# Would You Rely on an Eerie Agent?

## A Systematic Review of the Impact of the Uncanny Valley Effect on Trust in Human-Agent Interaction

Ahdiyeh Alipour [1], Tilo Hartmann [1], Maryam Alimardani [2]

[1] Department of Communication Science, Vrije Universiteit Amsterdam

[2] Department of Computer Science, Vrije Universiteit Amsterdam

## Author Note

Ahdiyeh Alipour https://orcid.org/0000-0003-0622-0269

Tilo Hartmann https://orcid.org/0000-0002-1862-7595

Maryam Alimardani https://orcid.org/0000-0003-3077-7657

We have no known conflict of interest to disclose.

Correspondence concerning this article should be addressed to Ahdiyeh Alipour, De Boelelaan 1105, 1081 HV Amsterdam. Email: a.alipour@vu.nl

[1] Ahdiyeh Alipour is the first and main author. Tilo Hartmann and Maryam Alimardani are listed as co-authors due to their active contributions throughout all stages of the work.

**Abstract**

Trust is a fundamental component of human-agent interaction. With the increasing presence of artificial agents in daily life, it is essential to understand how people perceive and trust these agents. One of the key challenges affecting this perception is the Uncanny Valley Effect (UVE), where increasingly human-like artificial beings can be perceived as eerie or repelling. Despite growing interest in trust and the UVE, existing research varies widely in terms of how these concepts are defined and operationalized. This inconsistency raises important questions about how and under what conditions the UVE influences trust in agents. A systematic understanding of their relationship is currently lacking. This review aims to examine the impact of the UVE on human trust in agents and to identify methodological patterns, limitations, and gaps in the existing empirical literature. Following PRISMA guidelines, a systematic search identified 53 empirical studies that investigated both UVE-related constructs and trust or trust-related outcomes. Studies were analyzed based on a structured set of categories, including types of agents and interactions, methodological and measurement approaches, and key findings. The results of our systematic review reveal that most studies rely on static images or hypothetical scenarios with limited real-time interaction, and the majority use subjective trust measures. This review offers a novel framework for classifying trust measurement approaches with regard to the best-practice criteria for empirically investigating the UVE. As the first systematic attempt to map the intersection of UVE and trust, this review contributes to a deeper understanding of their interplay and offers a foundation for future research.

**Would You Rely on an Eerie Agent? A Systematic Review of the Impact of the Uncanny Valley Effect on Trust in Human-Agent Interaction**

The agents are increasingly integrated into various domains of life, in different forms such as conversational social agents, avatars, chatbots, robots, intelligent virtual assistants, and voice-based technologies. The influx of these agents in important domains such as healthcare, education, and customer service raises significant questions about human-agent interaction (HAI). One central question in HAI research is to what extent agents are deemed by human users as reliable and trustworthy collaboration partners. Trust can be briefly defined as “the [human] operator’s confidence in the [AI-]machine’s capability (Muir & Moray, 1996). A key strategy followed by developers and designers is to create agents that might instigate trust by being more human-like in behavior and appearance. Indeed, research suggests that anthropomorphic features of agents (e.g., human voice, human-like appearance, human-like responses; Følstad & Brandtzæg, 2017) can enhance users’ compliance and trust in agents (Lee & Nass, 2010; Natarajan & Gombolay, 2020).

However, a greater anthropomorphic appearance might be a double-edged sword; while offering the potential to enhance trust, it also introduces the risk of triggering the Uncanny Valley Effect (UVE) (Mori, 1970). The UVE describes the curious effect that, in general, increasingly more human-like agents are perceived as increasingly more familiar and likable by human users, up to a certain threshold where agents that look and behave “almost but not quite human” start appearing strangely eerie, unfamiliar, and dislikable to users (see Figure 1).

**Figure 1**

*Mori's (1970/2012) hypothetical graph of the Uncanny Valley*

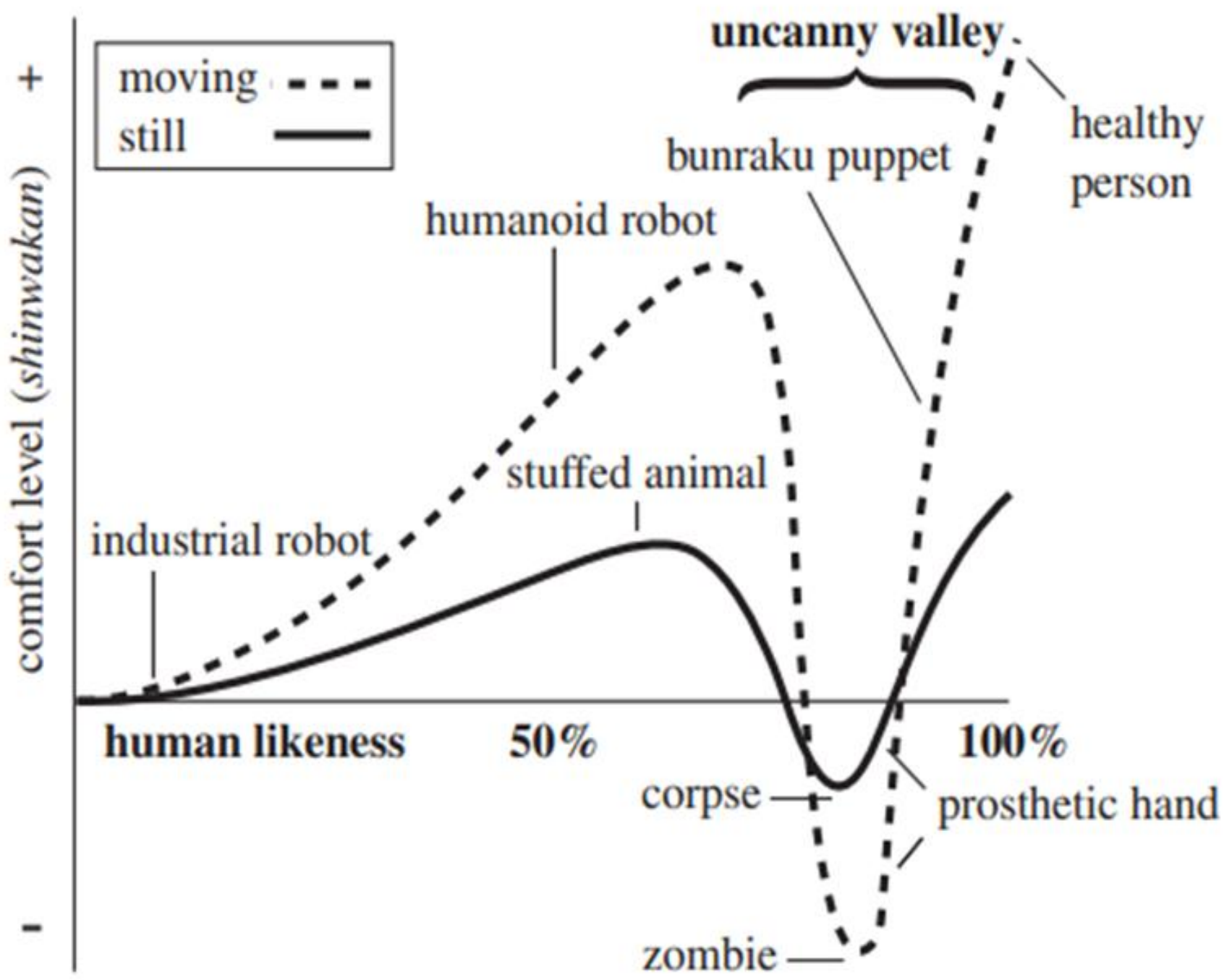

It should be noted that while the UVE is plausible, studies, even if applying different methodologies, have provided only mixed evidence for its existence to date. For example, it remains debatable if the agents' eeriness is directly linked to their "almost human-likeness" or if other factors, such as cues suggesting that a living being is (un)dead or sick, may contribute to discomfort (Wang et al., 2015). However, whatever the exact cause of an agent's uncanniness, if users do feel that an agent is uncanny, this might plausibly undermine their trust in and curb intentions to collaborate with the agent on a task. Accordingly, the UVE challenges the idea that greater human-likeness universally benefits trust.

A first unsystematic glance at the literature suggests that research on the UVE and its impact on trust is diverse and yields inconsistent findings. Studies adopted different methodologies when examining the relationship between human-likeness, the UVE, and trust (Gurung et al., 2023; Song & Shin, 2022). For instance, Song and Shin (2022) investigated the UVE among visually represented agents, while Gurung et al. (2023) examined the UVE in a context defined by incongruent visual and auditory representations of an agent. Studies also differ in their findings about the impact of the UVE on trust. A general problem of several studies that set out to examine the impact of the UVE on trust seems to be that they did not find any evidence of the UVE, or were unable to reproduce the UVE, in the first place. Hence, these studies fail to provide insights into how the UVE affects trust. For example, studies that focus specifically on audio features (Kühne et al., 2020), or human-realistic avatars in a Virtual Reality (VR) environment (Seymour et al., 2021) find that an appearance or speech that is closest to being human-like does not consistently provoke a sense of eeriness or reduce trust, defying the expectations of the UVE.

Despite these inconsistencies, at first glance, we noted more converging findings in the UVE and trust-related findings. Rheu et al. (2021) in a systematic review of trust-building factors in conversational agents, highlighted multiple studies (e.g., Shamekhi et al., 2018; Zanatto et al., 2016) showing that agents with a more attractive and human-like appearance can enhance trust, regardless of their expertise or reliability. However, while it is shown that human-like agents are generally more trusted, the level of detail in their human-like features yielded inconsistent results. More importantly, the review by Rheu et al. (2021) does not elaborate on the UVE and its impact on trust in almost human-like agents. Hence, the question remains if greater human-likeness indeed increases trust in a linear fashion, as the review suggests, or if this trend reverses for almost human-like agents, as the UVE would suggest.

To the best of our knowledge, no systematic review exists to date that illuminates how the UVE affects trust. Given the inconsistent findings related to anthropomorphism, UVE, and trust in the agents, a structured and systematic review of existing research is necessary to clarify how the UVE can affect trust. While some reviews have examined the UVE's mechanism and variables, such as literature review of its research (Zhang et al., 2020) or meta-analysis of its variables (Diel et al., 2021) or systematic review on reducing the effect (Stepp, 2022), their scope seldom extends to its consequences for trust. To tackle this research gap, the present literature review aims, by systematically analyzing empirical studies, to illuminate what we know about the empirical effects of an agent's uncanniness (caused by the UVE, i.e., of an almost human-like agent) on trust, and to determine whether and under what conditions the UVE impacts user trust in the agents. The findings can offer practical insights for future research, technical advancement, and design strategies aimed at fostering and maintaining trust in different kinds of agents. Accordingly, this systematic review addresses the general research question:

*How does the Uncanny Valley Effect impact humans' trust in an agent?*

We conducted a systematic review of empirical research illuminating the influence of the UVE on trust by following the PRISMA protocol. From an initial pool of 311 records, 107 studies were screened in full, and 53 were included for detailed analysis. Our findings reveal that trust in agents is a complex, context-dependent concept shaped by multiple factors. While many studies report that the UVE undermines trust, others found no significant impact, suggesting that the impact of the UVE on trust may depend on aspects such as the agent's modality, the congruence of visual, audio, and behavioral cues, and the interaction type between the human and the agent.

## How does the Uncanny Valley Effect of "Almost Truly Human-like Agents" Affect Trust?

In order to better understand the mechanisms underlying the impact of the UVE on trust in agents, it is important to first clarify what kinds of entities are referred to as agents. This lays the groundwork for exploring how trust operates in HAI, and how interdisciplinary insights from trust research and the UVE converge in the context of increasingly human-like agents.

### Agents

An agent can be broadly defined as a system that is capable of perceiving its environment, making decisions, and acting autonomously to achieve specific goals (Weiss, 2000). According to this book, unlike passive software programs or tools that only function upon user command, agents possess a degree of autonomy and responsiveness, enabling them to interact continuously and adaptively with users or other systems. The concept of agency arises from the need for intelligence to be embedded in a form that can operate in and react to a dynamic environment rather than existing solely as an abstract algorithm. This enables the agent to engage with its operational environment through perception, action, and adaptation, whether in the physical form (e.g., a robot) or simulated (e.g., a virtual avatar) (André & Pelachaud, 2010). Not everything that appears lifelike or automated qualifies as an agent. A video of a talking head or a scripted chatbot with no adaptive capacity may look agent-like but lack the autonomy, goal-directedness, or contextual awareness that defines true agency (Weiss, 2000).

Agents can be categorized in various ways. Embodied agents include physical robots and animated avatars, while non-embodied agents operate invisibly, such as recommendation systems or backend decision-making modules. Similarly, agents can be conversational (e.g., voice-based assistants like Alexa or Siri) or non-conversational, operating through actions or decisions without verbal interaction. Effective agents go beyond surface-level human-likeness: they exhibit communicative, emotional, and social behaviors that align with their roles and goals (Cassell, 2000). Despite their differences in form and function, these agents are better able to mimic human-to-human communication, i.e., perceive, act, adapt, and generate responses in real-time (André & Pelachaud, 2010). Human-like behavior and traits are expected to facilitate HAI, as users are more likely to perceive these agents as trustworthy and rely on them (Natarajan & Gombolay, 2020). However, the more important challenge lies in effectively gaining user trust through these design features.

## Trust in Agents

What is trust? A widely used definition stems from trust in interpersonal relationships by Mayer et al. (1995). According to this approach, trust is a combination of a person’s belief in another’s ability (their competence and effectiveness), benevolence (their intentions and motivations), and integrity (their reliability and adherence to ethical principles). Natarajan and Gombolay (2020) argue that trust refers to a positive belief and confidence in an agent’s ability to offer support and assistance, particularly in situations of uncertainty where individuals may feel vulnerable or at risk. The agents are often designed in a way that they supposedly should be held accountable for outcomes, even when users do not fully understand the logic behind them (Bach et al., 2024). In AI-assisted decision-making systems, trust is key to effectiveness and acceptance. When users trust an agent, they are more likely to rely on its recommendations, engage with it positively, and ultimately achieve better task outcomes (Vereschak et al., 2024).

In general, trust in an agent is influenced by three main factors, according to a recent systematic review of Campagna and Rehm (2024): human-related factors, environmental factors, and robot-related factors. Importantly, these factors dynamically change, and thus their interactive effect on trust also dynamically changes over time. This highlights that trust might not only be considered a momentary state, but that it also develops and fluctuates in time, paving a trust-trajectory (Bach et al., 2024). Human-related factors refer to a user's ability-based attributes (such as a user's prior experience, expertise, and competency) and personal characteristics (including demographics, attitude, and personality traits; Campagna & Rehm, 2024). Environmental factors include context, communication style, task type, and complexity. According to Campagna and Rehm (2024), robot-related characteristics have the greatest influence on trust, with 62% of studies emphasizing their roles. Robot-related factors include an agent's performance-based characteristics (such as behavior, reliability, predictability level of automation, and transparency) or attribute-based factors (like robot personality, robot type, and anthropomorphic features. The importance and strongest influence of robot characteristics and performance on perceived trust are also highlighted in a meta-analysis by Hancock et al. (2011). Their findings show that manipulating various aspects of the agent has the greatest impact on trust, more than any other factor, suggesting that designers should prioritize how to effectively adjust these characteristics to enhance trust.

## How an Agent's Human-Likeness Affects Trust

Human-likeness is often considered a key strategy in the design of agents to enhance user engagement and foster trust (Natarajan & Gombolay, 2020). Drawing on the Computers Are Social Actors (CASA) paradigm (Nass et al., 1994), users naturally interpret human-like systems as autonomous social entities and tend to apply social norms and expectations to their interactions with them. This effect occurs because human-like features reduce the cognitive effort required to initiate social interaction, making the agent feel more familiar and intuitive. As a result, users are more likely to view such agents as approachable, trustworthy, and emotionally intelligent (Lee & Nass, 2010; Natarajan & Gombolay, 2020). Furthermore, in support of the role of human-likeness in agents' interactions, the Modality, Agency, Interactivity, Navigation (MAIN) model proposed by Sundar (2008) suggests that human-likeness activates heuristics that influence users' perceptions of credibility. People are more inclined to trust entities that resemble real-world human cues, and thus, agents designed with human-like traits are often seen as more credible and effective communication partners.

Empirical studies show that integrating human-like features into an agent's appearance and behavior can foster more favorable user responses and increase trust (Natarajan & Gombolay, 2020; Salem et al., 2013). By way of example, Natarajan and Gombolay (2020) demonstrated that features (e.g., gestures, voice, expressiveness) significantly enhanced trust. In their experiment, participants interacted with robots (Pepper, Nao, Sawyer, Kuri) in embodied or virtual forms. Robots exhibited four behaviors: always correct, apologetic (admits mistakes), accountable (blames user), or indifferent. Participants completed a time-sensitive math quiz, relying on robot hints, while trust and anthropomorphism were measured via surveys. Their findings also highlighted that the users responded differently based on the robot's behavioral style, emphasizing the importance of aligning human-like behavior with social expectations. Similarly, Salem et al. (2013) investigated how a humanoid robot's use of hand and arm gestures influenced users' perceptions of its human-likeness, likability, and willingness to interact with it again. This study indicated that using both gestures and speech was perceived as more likable and human-like, and users were more willing to engage with them again, even when their gestures were only partially aligned with their verbal cues.

However, the research on HAI suggests that, on the one hand, human-like agents can elicit more trustworthy responses from users, on the other hand, human-likeness is not a one-size-fits-all solution for enhancing trust. The effectiveness of anthropomorphic features can be dependent on multiple factors, such as the context in which the agent is used and the content it delivers (Kulms & Kopp, 2019). For instance, Kulms and Kopp (2019) discovered that whether people perceived the agent as a computer, virtual entity, or human did not affect their actual trust in its advice during tasks. Instead, their trust was primarily influenced by the quality of the advice itself. However, participants tended to report feeling greater trust when the agent appeared more human-like, despite this not always aligning with their actual trust in the agent's advice during practical tasks. These findings underscore the complexity of trust and the numerous factors that can influence it in HAI.

Although several studies suggest that anthropomorphic features can enhance user trust, there is also evidence that excessive human-likeness may lead to discomfort in some contexts, highlighting the need to examine how the level of human-likeness relates to trust and whether it contributes to the potential triggering of the UVE. Importantly, overly human-like agents may lead people to develop unrealistic expectations that the agent can behave similarly to humans. Thus, this incongruence between reality and expectation results in diminished trust (de Graaf & Malle, 2017). Therefore, identifying the right balance of human-likeness requires investigating how the UVE can impact trust, which is necessary for designing agents that foster and maintain the desired level of user trust.

**The Uncanny Valley Effect**

The UVE describes a psychological phenomenon where emotional reactions to human-like robots follow a non-linear pattern (Mori, 1970/2012). As a replica becomes more human-like, people initially feel an increasing affinity, which is a sense of familiarity and attraction. However, when the replica nears but does not quite achieve full realism, emotional responses sharply drop into discomfort, eeriness, or even revulsion. This sudden dip is referred to as the "valley" in the curve mapping human-likeness against emotional response (see Figure 1). Mori (1970/2012) proposed that this aversive response persists until the replica becomes nearly indistinguishable from a real human, at which point emotional responses turn positive again. The UVE most commonly occurs with humanoid entities that closely mimic human appearance and behaviors, yet fall just short of full realism, thereby evoking an unsettling sense of almost-but-not-quite human.

Mori’s UVE (1970/2012) depicts the two-axis graph, with human-likeness on the horizontal axis and familiarity on the vertical axis. The human-likeness ranges from highly artificial forms like industrial robots to fully human. It is important to note that Mori's original Japanese terms, ‘bukimi’ and ‘shinwakan,’ used for describing the UVE, have multiple conceptual meanings. The term ‘bukimi’ is commonly translated as eeriness (Ho & MacDorman, 2010), though other negative related terms like creepiness and strangeness have also been used (Ho et al., 2008). However, the translation of Mori’s y-axis term (See Figure 1), shinwakan, is more complex. It is an unconventional Japanese word with no direct English equivalent (Bartneck et al., 2009), leading to varied translations including affinity, familiarity, warmth, and even likability, leading to interpretation as a positive effect.

While many things can evoke feelings of eeriness or discomfort, not all such reactions fall within the scope of the UVE. Eeriness caused by horror elements, death imagery, or culturally shaped fears (e.g., zombies or corpses) may resemble uncanny reactions but do not qualify as UVE (Kätsyri et al., 2015; Lay et al., 2016). It is not sufficient to claim that an entity belongs in the valley simply because it appears eerie, which introduces methodological circularity, in which something is assumed to be uncanny without proof (Lay et al., 2016). Mori's original examples, such as zombies and corpses, are problematic, as they mix human-likeness with other confounding factors (like morbidity), which may independently cause discomfort due to death cues, not uncanny human-likeness (Kätsyri et al., 2015; Zhang et al., 2020). Accordingly, creepiness or eeriness refers to a distinctive drop in affinity triggered by near-human qualities and thus must be defined and measured with care.

Regarding operationalization, there are minimal criteria that must be met to credibly claim that the UVE has been observed (Kätsyri et al., 2015; Lay et al., 2016). According to Lay et al. (2016), a study must use stimuli that span a continuum from artificial to human with a minimum of five points or ideally more (e.g., 0%, 25%, 50%, 75%, and 100% human-likeness), and must obtain independent ratings of human-likeness for these stimuli, rather than relying on assumptions. However, Kätsyri et al. (2015) propose a slightly less stringent requirement, suggesting that at least three levels of human-likeness are sufficient to test for a UVE pattern. Moreover, Lay et al. (2016) emphasize the importance of directly measuring eeriness to distinguish uncanny discomfort from general emotional responses. Similarly, Kätsyri et al. (2015) suggest a slightly more general interpretation, arguing that the negative affinity (e.g., eeriness) must be specifically demonstrated to be associated with almost-human agents, rather than with morbid or unfamiliar ones. Eventually, a valid demonstration of the UVE must indicate a clear non-linear trend, which is a valley-shaped dip in emotional response that occurs between 50% and 100% human-likeness (Lay et al., 2016). It is important to note that it is only valid if it is plotted against independently rated human-likeness and shows a significant deviation from a linear pattern (Kätsyri et al., 2015; Lay et al., 2016). For example, studies like Mathur and Reichling (2015) and MacDorman and Chattopadhyay (2016) met all core criteria. In contrast, studies such as McDonnell and Breidt (2010), which had no manipulation checks and eeriness ratings; Hanson (2005), which offered a qualitative, aesthetic perspective without empirical testing, and Gray and Wegner (2012), which tested only a single android and could not identify a valley pattern, were considered methodologically weak when they drew conclusions about the UVE without sufficient evidence.

Since the introduction of the UV, a considerable amount of research has been conducted to illuminate its existence, which can be categorized into perceptual and cognitive processing theories (MacDorman & Ishiguro, 2006; Wang et al., 2015). Perceptual processing, which is automatic and stimulus-driven, includes hypotheses like the Pathogen Avoidance hypothesis, Mortality Salience hypothesis, and Evolutionary Aesthetics hypothesis (MacDorman & Ishiguro, 2006). These indicate that uncanny human replicas may evoke thoughts of illness and mortality or be perceived as less appealing. For example, according to the Evolutionary Aesthetics hypothesis, feelings of uncanniness emerge when an entity’s appearance violates universal aesthetic expectations by falling short of perceived standards for human-likeness (MacDorman & Ishiguro, 2006). Cognitive processing theories, which scholars proposed at a later stage and are broader, include the Violation of Expectation hypothesis (MacDorman & Ishiguro, 2006), the Categorical Uncertainty hypothesis (Wang et al., 2015), the Perceptual Mismatch Hypothesis (MacDorman et al., 2009), and the Mind-Perception hypothesis (Gray & Wegner, 2012). These theories overlap in suggesting that perceptual inconsistencies in human replicas evoke an eerie sensation and emphasize the human-nonhuman distinction. For instance, the Perceptual Mismatch Hypothesis explains the UVE as a reaction to inconsistencies in human-likeness across different sensory cues within an artificial character (MacDorman et al., 2009). This mismatch, such as artificial eyes on an otherwise human-like face, can create a sense of unease, particularly when deviations from typical human norms are perceived. While Kätsyri et al. (2015) argue that the UVE is not a universal response to any near-human entities, they found strong empirical support for perceptual mismatch, as characters with inconsistent or atypical human-like features tend to elicit more negative affect.

Together, these theoretical perspectives highlight the complex and often inconsistent ways in which human-likeness can influence trust through the UVE. To clarify and synthesize the empirical evidence surrounding this relationship, we conducted a systematic literature review. The following section outlines our methodological approach.

# Method

In order to answer our research question, we conducted a systematic review following the PRISMA 2020 protocol (Page et al., 2021). This process included 1) developing a search strategy and identifying relevant keywords, 2) defining inclusion and exclusion criteria, and 3) conducting a screening and selection process. The results of the screening and selection are summarized in a PRISMA flow diagram to provide a clear schematic representation of our review process (see Figure 2).

**Search Strategy**

After defining our research question, we selected seven different electronic databases (Scopus, Web of Science, PubMed, APA PsycInfo, IEEE Xplore, ProQuest, and WorldCat) to account for the interdisciplinary nature of the research topic. Next, keywords were selected to capture studies using relevant terminology to the UVE and trust. The final search string that we applied was ("uncanny valley" OR "uncanny effect") AND (trust* OR believab* OR persuas* OR acceptab* OR acceptance OR competen* OR credib*). The inclusion of trust-related terms in the second part of the string (e.g., credibility, competence, persuasion, acceptance, etc.) was inspired by a literature review on trust measures in human-robot interaction (Campagna & Rehm, 2024), and was adopted to ensure comprehensive coverage of studies addressing various dimensions of trust. These constructs reflect agents' perceived capability, reliability, and users' willingness to accept recommendations or interact with agents, all of which are key determinants of trust.

In the databases, filters were applied to the search within titles, abstracts, keywords, subject headings, and index terms, with no restrictions on publication date. The search was further limited to specific peer-reviewed formats and sources, including journal articles, conference proceedings, book chapters, academic dissertations, and scholarly books. Additionally, only studies published in English were selected. The search was conducted on October 15, 2024, yielding an initial result of 641 studies, which were imported into EndNote 20. A total of 330 references were removed as duplicates or because they were master's theses or conference proceedings books (where keywords were identified in different chapters). PhD dissertations were included. This process left 311 studies for title and abstract screening.

**Inclusion and Exclusion Criteria**

We defined the following inclusion criteria that studies must have met to be included in our review:

- The study must offer minimal boundary conditions to be able to sufficiently prove that it indeed included and/or measured the UVE and not something else. While we did not apply the full set of measurement criteria proposed by Kätsyri et al. (2015) and Lay et al. (2016), we included studies that explicitly claimed to investigate the UVE and that met at least the minimum basic indicator.

  Operationalization of this criterion:

  - The study presented agents. Eligible agents included but were not limited to embodied agents (e.g., robots), chatbots, voice-based technologies, or even static images of agents. Furthermore, studies were eligible if at least one of the presented agents plausibly fell into the UV by being almost human-like. We included studies in which human-likeness or anthropomorphism was either manipulated directly or described as a central design feature of the agent. Importantly, we did not require studies to strictly vary human-likeness along a precise or objective scale but focused on whether agents were intended to appear (almost) human-like or were perceived as such by users, which allowed us to capture a broader range of studies.
  - The study assessed users' affinity response to an agent. To ensure comprehensive coverage of terms related to the UVE, we included both commonly used (e.g., likability) and measures (e.g., eeriness) frequently investigated in UVE studies, as identified by the meta-analysis of Diel et al. (2021). The best-practice measures of the user response underlying the UVE include the sensation that an agent felt eerie, creepy, uncanny, and threatening, as these directly capture the core experience of uneasiness. In contrast, other measures such as likability, attractiveness, familiarity, fear, disgust, anxiety, and discomfort (stemming from negative emotional reactions) are less effective and direct in explicitly capturing the UVE. We nevertheless included them to ensure a comprehensive review that accounts for both positive and negative affective measures of the UVE. To account for the variations of Mori's original Japanese terms, we included affinity and all its related meanings in our review.

- Combining these two aspects, we included studies if we found that minimal boundary conditions to infer the UVE were met. A study could meet these requirements based on compelling theoretical arguments (e.g., strong and convincing arguments, why the presented stimulus or stimuli fall into the UV) and/or empirical evidence (e.g., empirically replicating the UV). More specifically, studies with only one stimulus were included if they (a) applied an almost-human stimulus that plausibly fell into the range of stimuli within the UV (Ho & MacDorman, 2010) and (b) empirically showed that this stimulus produced "surprisingly low" affinity (or high eeriness) scores, or a similar outcome on any of the variables that we identified as validly representing users' affinity response. For example, Li et al. (2024) applied one stimulus (a virtual human; Miquela) that appeared to be highly yet not fully human-like, plausibly evoking uncanny perception. Similarly, Destephe et al. (2015) applied one stimulus (a humanoid robot; WABIAN-2R) to measure perceived humanness and eeriness on a 5-point scale, with the plotted curve resembling the UV.

- Next to these single-stimulus studies, studies that compared at least two stimulus conditions (e.g., experiments, comparing at least two stimuli, or two versions of a stimulus) were considered eligible if they met two key criteria: (a) they compared a stimulus that plausibly fell into the UV with a stimulus that plausibly did not, and (b) they empirically demonstrated the expected affinity-response pattern in users, based on a variable that adequately represents the UVE.

- The study must address trust or trust-related constructs (believability, persuasiveness, acceptance, competence, credibility), and measure these constructs directly or indirectly (e.g., through trust-related intentions or decisions such as intention to use/buy, interest in future interactions, or adherence to an agent's advice).
- The study must be based on data-driven human responses, provided by participants (no limitations on age, gender, or cultural context) who observed or directly engaged in interaction with agents during the experiment or reflected on prior experiences with such agents.
- The study has to include sufficient information to assess the relevance and quality of findings. This included detailed reporting of methodology and outcomes.

In addition to the above inclusion criteria, a number of exclusion criteria were considered, which are provided in Table 1.

**Table 1**

*Exclusion Criteria*

| **Exclusion Criteria** | |
|---|---|
| *Study Type* | • Theoretical<br>• Workshop<br>• Systematic review<br>• Meta-analysis |
| *Format and Study Source* | • Non-academic materials<br>• Magazines<br>• Newspapers<br>• Encyclopedias<br>• Websites<br>• The articles were not accessible in full text. |
| *Duplicate Author Studies* | • If multiple similar studies by the same authors were found, the studies with less relevance or impact were excluded, and studies with higher citation counts or journal publications were prioritized over conference proceedings. |
| *Language* | • The text was not available in English. |
| *Population* | • Non-human |

## Screening

In order to have a systematic screening, 311 studies were imported into Covidence, a widely used systematic review management tool (https://www.covidence.org). The first author conducted a two-stage screening process, starting with title/abstract screening followed by full-text screening. In the first stage, 197 studies were excluded based on the inclusion/exclusion criteria, leaving 114 studies for retrieval. Here, the second and third authors also screened a random selection of 80 studies to ensure the quality of the screening process and the clarity of the inclusion/exclusion criteria. Covidence's conflict resolution feature facilitated discussions between researchers, allowing disagreements to be resolved and the criteria to be refined for better comprehension. Seven studies could not be retrieved due to inaccessibility of the full text, leaving 107 studies eligible for the second stage of screening, i.e., full-text screening (see Figure 2).

Similar to the first stage, the first author initially screened the full text of 107 papers. To ensure screening quality, the second and third authors also assessed a set of 20 randomly selected studies. The inter-rater reliability, calculated using Fleiss' Kappa, was 0.61, which indicated substantial agreement among the three researchers (Landis & Koch, 1977). The studies with conflicting decisions were collectively discussed in a meeting to reach a consensus on eligibility for inclusion. Eventually, 54 studies were excluded in the full-text screening due to violation of one or more inclusion/exclusion criteria (see Figure 2 for a breakdown of the number of papers excluded under each criterion). This left a total of 53 papers that met all the inclusion/exclusion criteria and were included in the review.

The final 53 papers were moved to the data extraction stage, where the first author analyzed and summarized their content using predefined codes: theoretical research question, modality, type of agent, type of interaction, methods/design, manipulation, setting, participants, UVE measures, trust measures, results, and limitations (see Appendix A).

**Figure 2**

*PRISMA Flow Diagram*

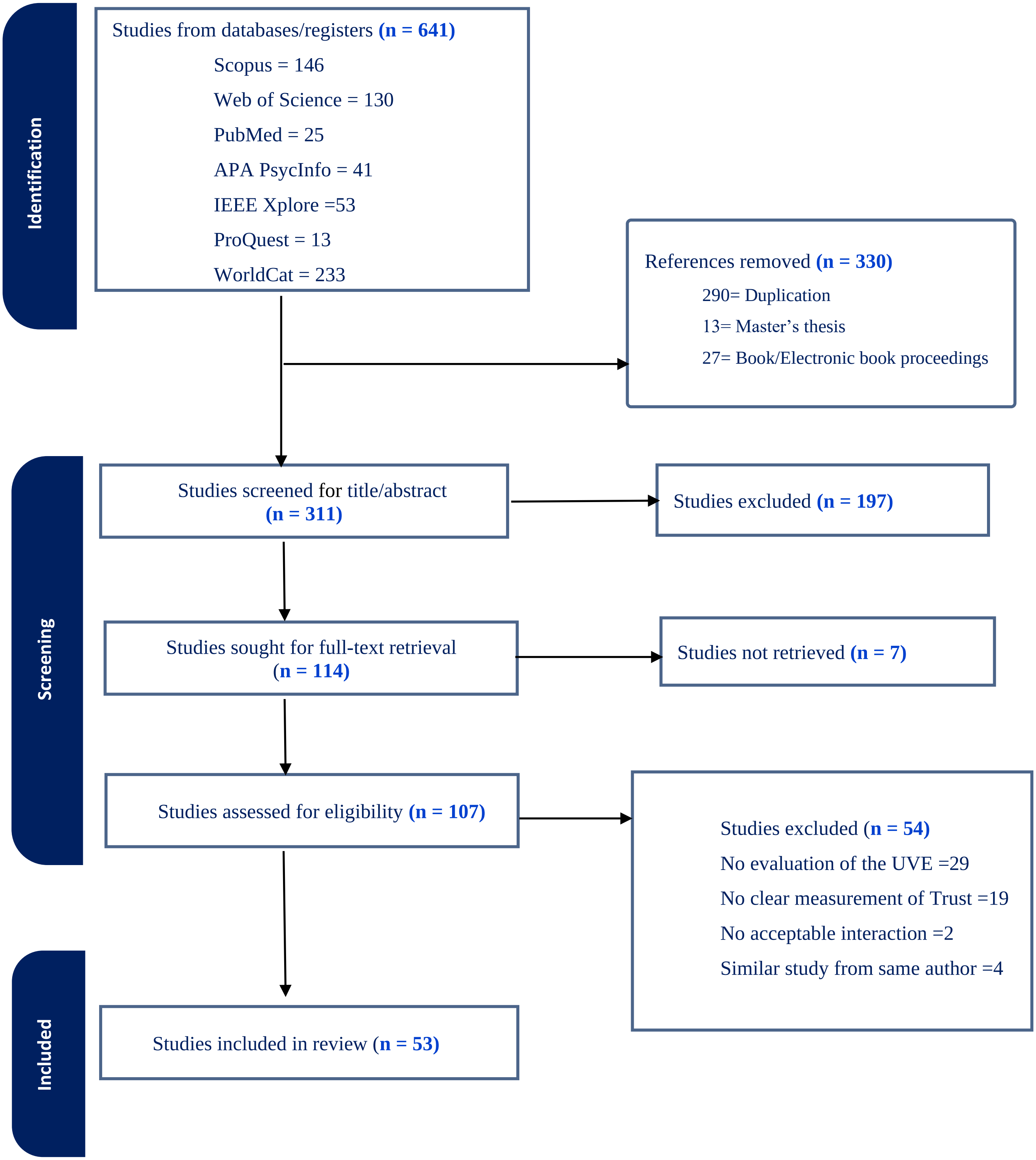

# Results

Appendix A provides a comprehensive overview of all 53 studies, along with a detailed summary of the extracted data from each paper. The findings for each data category are described below.

## Types of Agents and Encounters Examined in Reviewed Studies

### *Modality of stimuli*

In this review, modality refers to the primary sensory channel through which participants interacted with the agent. Four types of modalities were identified: 1) visual, involving images, animations, or visual interactions; 2) auditory, including spoken interactions, sounds, or auditory cues; 3) tactile/physical, which encompassed physical interactions such as touch or haptic feedback; and 4) textual, capturing interactions based on written or typed text.

The majority of studies examined agents represented in a visual modality, either alone or in combination with other modalities (see Figure 3). More specifically, 21 out of 53 studies examined visual-only representations, while 19 studies investigated agent representations in both visual and auditory modalities. Additionally, six studies explored visual and textual representations, two studies examined visual and tactile representations, and one study incorporated all modalities. Other modalities, for example, auditory, were less frequently studied. Only two studies focused solely on auditory representations of agents, while two studies combined auditory and textual modalities.

**Figure 3**

*Types of modalities in which agents were represented.*

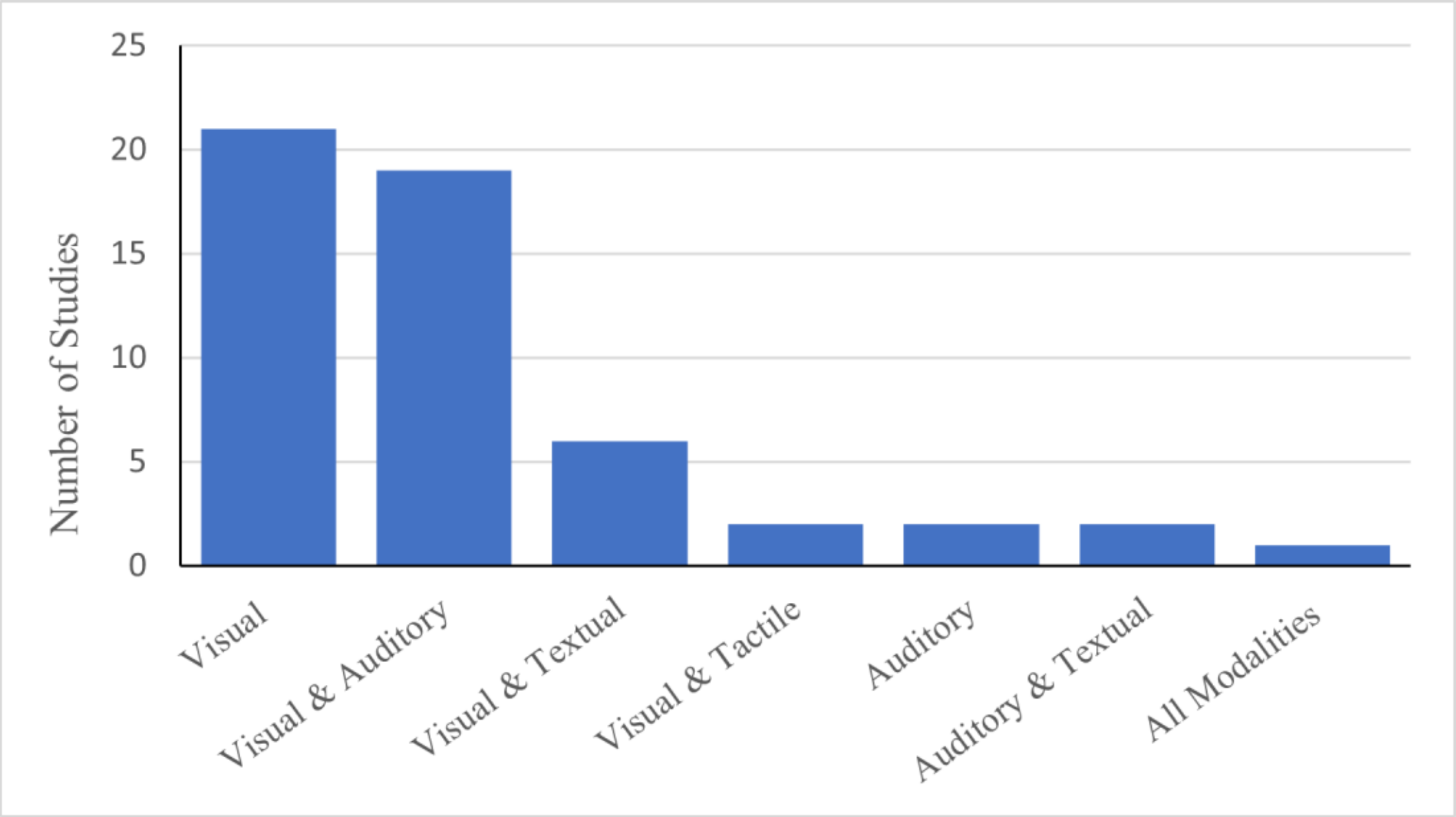

***Type of agent***

Across the studies included in this review, a wide range of *agent types* were used. These included robots (*n* = 25), virtual agents (*n* = 21), chatbots (*n* = 2), and voice-only agents (*n* = 2). The remaining three studies reflected on participants' prior experience with existing agents without exposing them to any agents during the study. For example, one study (Troshani et al., 2021) examined broader categories of AI systems, such as Siri, driverless cars, and smart robots. Another study (Pandey & Rai, 2023) explored AI-powered virtual assistants like Google Assistant or Alexa, while the third one (Lou et al., 2023) investigated the existing virtual influencers.

In terms of *representation format*, most studies focused only on static representations of agents (see Figure 4 for examples), meaning participants were exposed to static images or videos, rather than interacting with robots or virtual agents, etc., in real-time. Specifically, 14 studies used static images of robots, while one study used static images of virtual agents. Additionally, 11 studies used videos of virtual agents, and three studies employed videos of robots. One study incorporated videos and vignettes of virtual agents. Furthermore, one study combined static images and videos of robots, while another combined static images and videos of virtual agents, and one included static images and videos of film-based characters and CGI agents.

**Figure 4**

*The examples of static representations of agents*

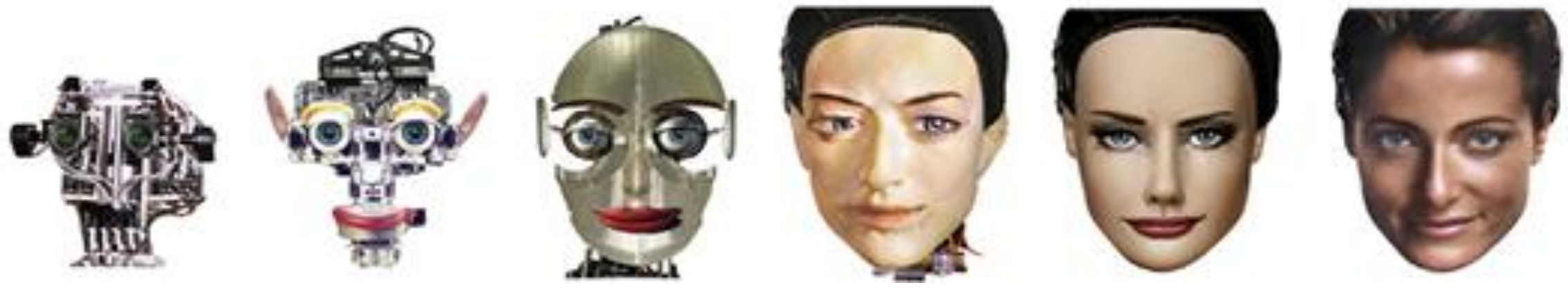

*Note.* These are the static images of agents that were used in the study by Abubshait et al. (2017, 2022) and in the composed stimulus set from Experiment 2A in Mathur and Reichling (2016).

By contrast, much fewer studies allowed participants to interact with a physically present or real-time-rendered agent. Seven studies focused on actual physical robots, while four studies explored virtual agents with real-time visual presence.

The chatbots ($n$ = 2) and voice-only agents ($n$ = 2) relied on interactive conversation, scripted transcripts, or pre-recorded dialogues. Lastly, three studies took into account participants' previous interactions with existing agents.

### *Type of interaction*

A substantial portion of studies (42 out of 53) examined only indirect interactions between users and agents, where participants observed and evaluated agents based on pre-recorded stimuli (e.g., static images, videos, or transcripts), without engaging in real-time interaction. In contrast, 10 studies explored direct interactions between participants and agents. Additionally, one study investigated both direct and indirect interactions between participants and agents.

## Methodologies Applied in Reviewed Studies

### *Methods and designs*

The greater part of the studies (42 out of 53) employed a quantitative research approach, using statistical analyses to examine the UVE and trust-related constructs. Most of these 42 quantitative studies were experiments featuring different conditions or versions of agents. Of all 42 quantitative studies, 41 were experimental studies, and only one employed a survey-based observational approach without experimental manipulation (Pandey & Rai, 2023). Of the 41 experimental studies, 28 studies employed a between-subjects experimental design, eight applied a within-subjects (pre-post) design, and five studies followed a mixed-design approach, incorporating both between-subjects and within-subjects factors.

An additional four studies used a qualitative approach, with three studies relying on semi-structured interviews and thematic analyses (Lin et al. ,2021; Lou et al., 2023; Thimm et al., 2024), and one study employed a focus group discussion to explore participants' experiences (Troshani et al., 2020).

An additional seven studies adopted a mixed-methods approach, integrating both quantitative and qualitative techniques. Within this group, one study combined a between-subjects design with qualitative methods of focus groups; one study employed a within-subjects design and included open-ended responses to gather additional insights; three studies incorporated both between- and within-subjects designs, supplementing their analyses with interviews or open-ended responses. Two studies employed non-experimental cross-sectional survey designs. One of these combined the survey with open-ended responses, while the other integrated an observational approach and semi-structured interviews.

### *Setting*

As shown in Figure 5, many studies (23 out of 53) examined general human-agent encounters without focusing on a specific applied setting. Agents in service settings such as hotels, restaurants, or customer service were examined in 10 studies. Five studies examined agents in a healthcare setting, and five in a social media setting. Other contexts, such as banking ($n$ = 3), virtual reality ($n$ = 3), education ($n$ = 1), industry ($n$ = 1), parenting ($n$ = 1), and design ($n$ = 1) were explored in fewer studies.

**Figure 5**

*Frequency of Context in Studies*

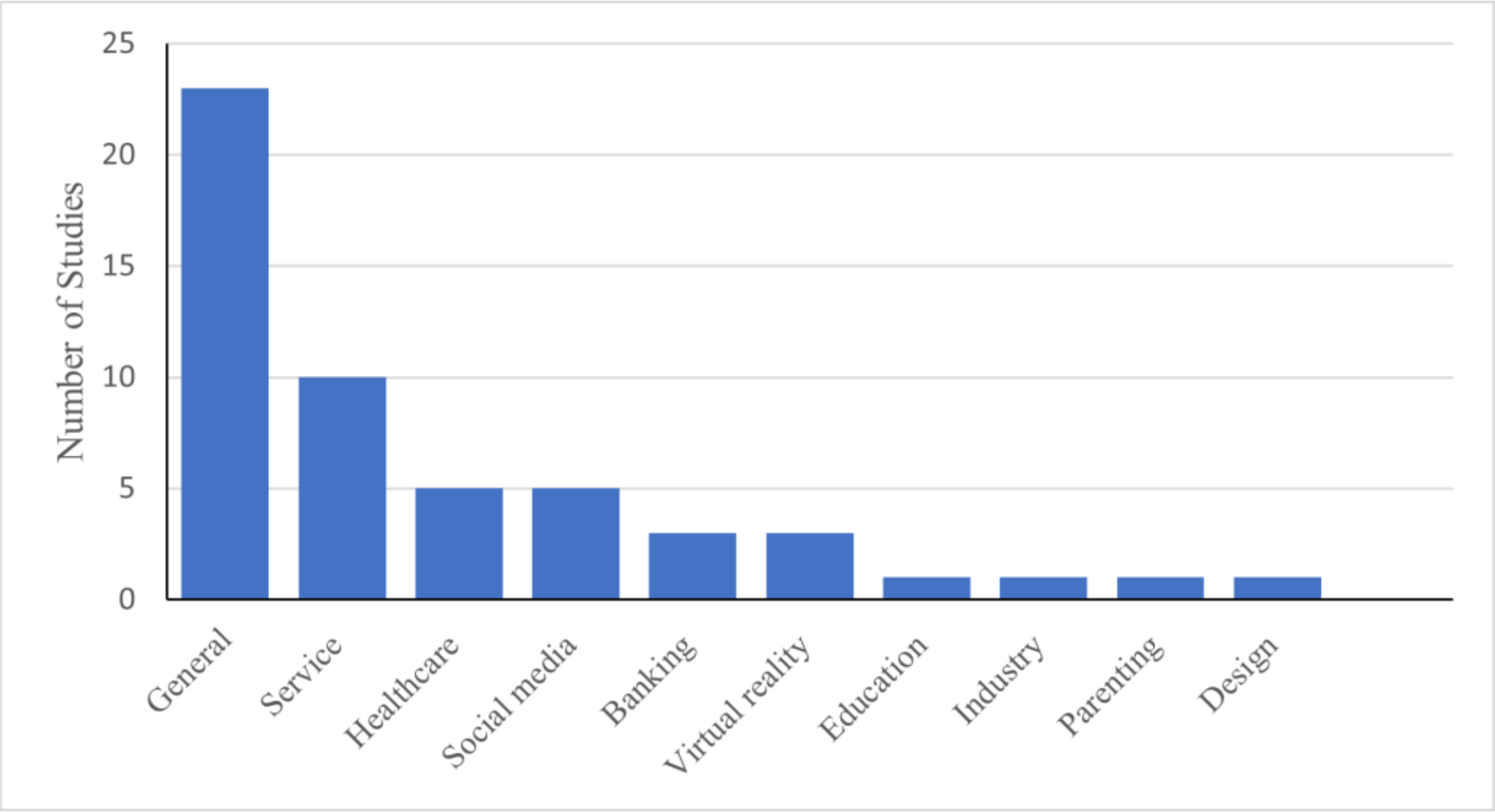

***Sample size***

Twenty-two out of the 53 studies recruited participants through university/school systems, such as campus advertisements or university websites, and thus featured student samples. Additionally, 15 studies recruited participants through online platforms (*n* = 7 Amazon MTurk; *n* = 4 Prolific; *n* = 4 other open online survey platforms). A total of eight studies recruited participants via email lists and social media, while other studies used the member list from the lab (*n* = 1), local institute (*n* = 1), or general invitations (*n* = 1). Three studies employed snowball sampling. One study specifically recruited conference attendees as participants. One study did not specify the participant recruitment approach.

Sample sizes varied widely across studies. A total of 14 studies had small sample sizes with fewer than 50 participants, including two studies with fewer than 20 and one study with fewer than 10 participants. Among studies with larger samples, eight studies included between 50 and 99 participants, while 14 studies had sample sizes ranging from 100 to 199 participants. Similarly, 10 studies recruited between 200 and 499 participants, and seven studies had sample sizes between 500 and 999 participants (see Figure 6).

**Figure 6**

*Distribution of Sample Size Rage Across Studies.*

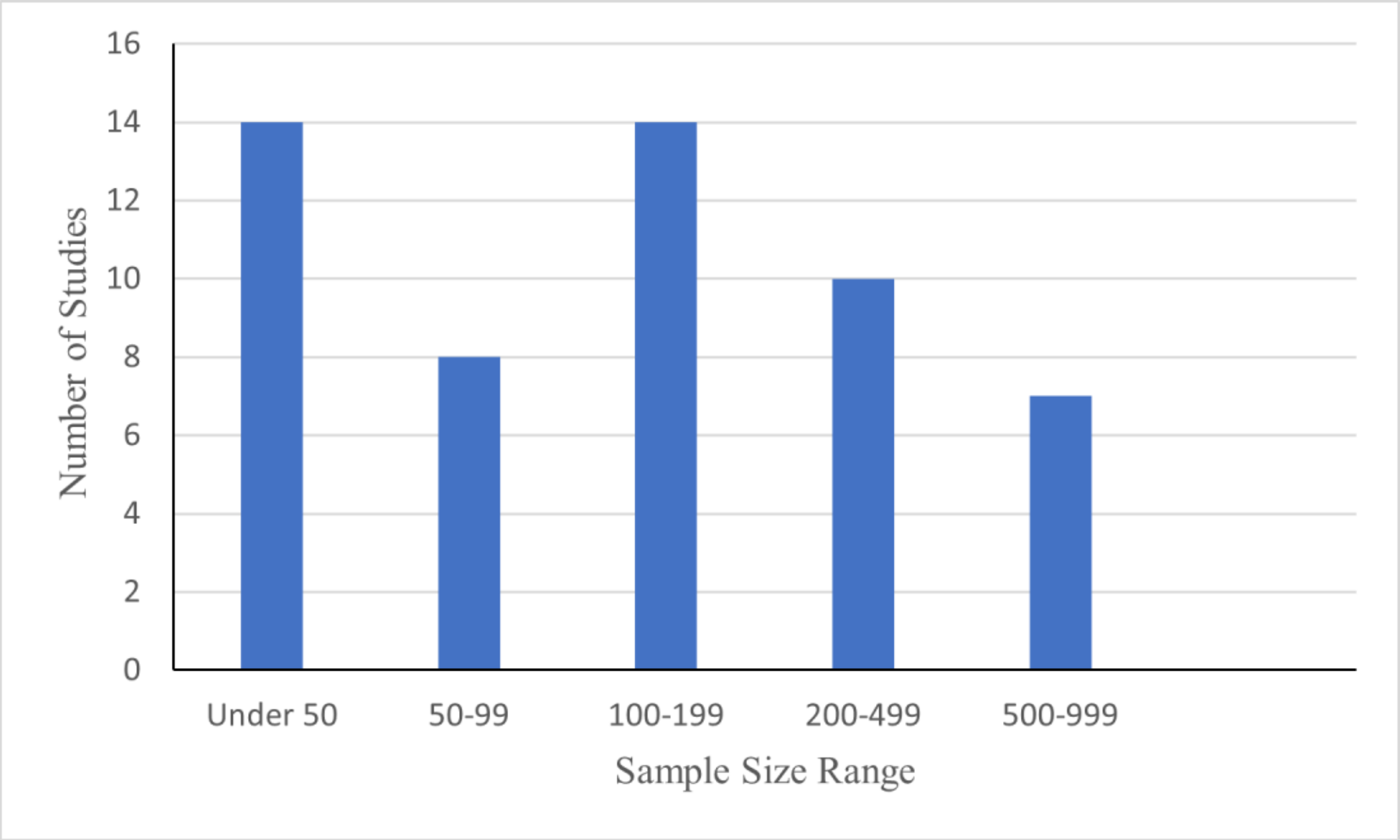

### *Culture*

The reviewed studies included participants from diverse cultural backgrounds. Sixteen studies involved U.S. participants. A total of 15 studies were conducted in Europe, with participants from Germany (*n* = 7), Austria (*n* = 2), the Netherlands (*n* = 2), Sweden (*n* = 2), Italy (*n* = 1), and England (*n* = 1). In Asia, 12 studies examined participants from different countries, including South Korea (*n* = 4), China (*n* = 2), Taiwan (*n* = 2), India (*n* = 2), Georgia (*n* = 1), and Singapore (*n* = 1). Additionally, one study included participants from both France and Japan, while another study involved participants from both Australia and Denmark. Another study specifically focused on New Zealanders.

Beyond country-specific samples, one study examined native English speakers across different regions, while three studies recruited international participants. Additionally, three studies did not impose geographical restrictions on participant selection (see Figure 7).

**Figure 7**

*Study Samples by Continent*

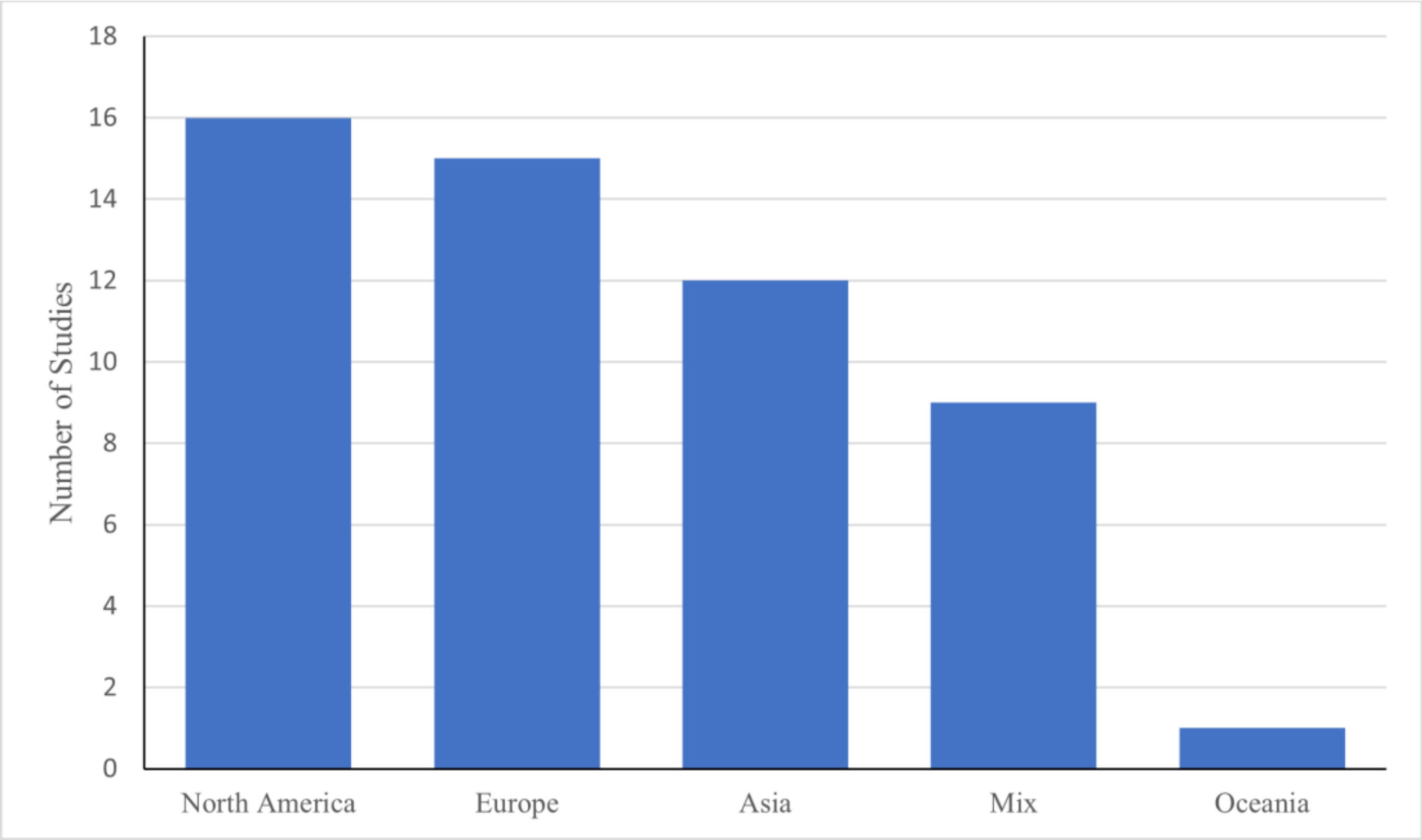

### *Gender*

Across the studies that reported gender ($n$ = 50), the mean number of female participants was 124.3 (median = 71.5), while the mean number of male participants was 102.7 (median = 61), indicating a slight skew toward larger female samples. Three studies did not specify the gender of their participants.

## Measurement Approaches in Reviewed Studies

### *UVE Measures*

We categorized the UVE measures based on users' responses collected by the studies. The collected measures were divided into two categories: measures of negative affinity (*n* = 50) and measures of positive affinity (*n* = 3). Since applying valid negative affinity measures is essential in UVE studies (Kätsyri et al., 2015; Lay et al., 2016), the first category was further divided into two subcategories. The first one included “best-practice” measures, which capture a distinct experience of uncanniness and have demonstrated larger-than-average effect sizes in Diel et al., (2021), specifically, eeriness and creepiness (*n* = 38). The second subcategory included “broader affective response” measures, which encompassed a wider range of emotional reactions but vary in effectiveness in capturing the UVE (*n* = 12; Diel et al., 2021). The second category (*n =3*) included studies that employed only valid positive (affinity) measures, such as emotional responses or reactions to the stimuli (e.g., familiarity or likability). A detailed breakdown of the UVE measures categories and their study counts is presented in Table 2.

**Table 2**

*Categorization of the UVE Measurement Approaches*

| Category | Subcategory | Measures | Number of Studies | Studies |
|---|---|---|---|---|
| **1.Negative Affinity** | **a.Best-practice Measures** | | | |
| | | Eeriness [a] | 7 | Aubel et al., 2022; Broadbent et al., 2013; Lou et al., 2023; Mal et al., 2024; Qiao & Eglin, 2011; Schreiberlmayr et al., 2023; Shin et al., 2019 |

| | | | |
|---|---|---|---|
| | Eeriness and Attractiveness | 9 | Cornelius et al., 2023; Destephe et al., 2015; Latoschik et al., 2017; Patel & MacDorman, 2015; Reuten et al., 2018; Stein & Ohler, 2018; Stein et al., 2020; Tastemirova et al., 2022; Weisman & Peña, 2021 |
| | Eeriness and Warmth | 5 | Dai & MacDorman, 2018; Dai & MacDorman, 2021; Grazzini et al., 2023; Kim et al., 2019; Pandey & Rai, 2023 |
| | Eeriness and Likability | 2 | Kim, 2022; Kühne et al., 2020 |
| | Eeriness and Familiarity | 1 | Song & Shin, 2022 |
| | Eeriness, comfort and Familiarity | 1 | Sharma & Vemuri, 2022 |
| | Eeriness and Affective response | 1 | Li et al., 2024 |
| | Eeriness and Discomfort [a] | 1 | Thimm et al., 2024 |
| | Eeriness and Threat | 1 | Stein et al., 2022 |
| | Eeriness and Empathy | 1 | MacDorman, 2019 |
| | Eeriness and Affinity | 1 | Seymour et al., 2021 |
| | Eerniness and Pleasentness | 1 | Schreibelmayr & Mara, 2022 |
| | Creepiness [a] | 4 | Kaate et al., 2023; Pinney et al., 2022; Troshani et al., 2020; Williams, et al., 2020 |
| | Creepiness and Likability | 3 | Abubshait et al., 2017; Abubshait et al., 2022; Mathur & Reichling, 2016 |
| **b.Other Negative Measures** | | | |
| | Discomfort | 1 | (Lin et al., 2022) |

| | | | | |
|---|---|---|---|---|
| | | Discomfort and Warmth | 1 | Paetzel-Prüsmann & Castellano ,2019 |
| | | Discomfort, Threat, warmth, and likability | 1 | Paetzel et al., 2020 |
| | | Threat | 1 | Liao et al., 2023 |
| | | Threat and Positive attitude [a] | 1 | Lin et al., 2021 |
| | | Human Identity Threat | 1 | Kim, 2023 |
| | | Fear | 1 | Nie et al., 2012 |
| | | Disgust and Likability | 1 | Tu, et al., 2020 |
| | | Disgust, Likability, and Familiarity | 1 | Chien et al., 2024 |
| | | Anxiety and Likability | 1 | Prakash & Rogers, 2015 |
| | | AI Anxiety | 1 | Mulcahy et al., 2024 |
| | | Favorability (uneasiness) | 1 | Jung et al., 2021 |
| **2.Positive Affinity** | **Affinity Measures** | | | |
| | | Familiarity [b] | 1 | Bae et al., 2024 |
| | | Familiarity and Likability [C] | 2 | Gurung et al., 2024; Rosenthal-von der Pütten et al., 2019 |

*Note.* [a] Indicates qualitative studies (Lin et al., 2021; Lou et al., 2023; Pinney et al., 2022; Thimm et al., 2024; Troshani et al., 2020) where UVE-related terms (e.g., eeriness, creepiness, and threat) were extracted from participants' statements and considered as UVE measures. [b] This study used the Familiarity Index, including Eeriness, Attractiveness, Pleasure, and Warmth. [C] Rosenthal-von der Pütten et al., (2019) employed uncanniness as one of the measures during the stimulus selection process in the pretest.

In addition, 40 out of 53 studies measured the perceived human-likeness and anthropomorphism of agents through participants' ratings. These ratings were obtained either in a manipulation check (*n* = 7) or independently in the main study *(n* = 33). The remaining studies (*n* = 13) did not involve any empirical manipulation or assessment; in these studies, the human-likeness of the agents was established by other means, such as findings from their own previous studies or relying on existing literature reviews.

### *Trust Measurement*

Studies were categorized into three groups based on the type of trust measurement that they employed (see Table 3). The first group (*n* = 24) consists of studies that directly and explicitly measured trust or trustworthiness empirically (e.g., based on self-report measures), either alone or in combination with other trust-related outcomes. The second group (*n* = 11) includes studies that assessed trust-related constructs, such as competence, acceptability, credibility, believability, and persuasiveness. The third group (*n* =18) contains studies that assessed trust-related intentions, such as adherence to an agent's advice, willingness to interact, or conformity to an agent's recommendation in a decision-making task (e.g., purchase or investment decisions) based on agents' recommendations. In the third group, some studies (*n* = 10) employed the intentions solely, and some studies (*n* = 8) combined with other trust-related constructs (see Table 3).

**Table 3**

*Categorization of Trust measure approaches*

| Type of Measures | Measures | Number of Studies | Studies |
|---|---|---|---|
| **Direct Measures of Trust** | | | |
| | Trust [a] | 11 | Abubshait et al., 2017; Abubshait et al., 2022; Jung et al., 2022; Latoschik et al., 2017; Mal et al., 2024; Mathur & Reichling, 2016; Nie et al., 2012; Pinney et al., 2022; Prakash & Rogers, 2015; Thimm et al., 2024; Troshani et al., 2020 |
| | Trustworthiness | 6 | Broadbent et al., 2013; Gurung et al., 2024; Kühne et al., 2020; Seymour et al., 2021; Tastemirova et al., 2022; Tu et al., 2020 |
| | Trustworthiness and Persuasiveness | 1 | Cornelius et al., 2023 |
| | Trust and Intentions | 3 | Song & Shin, 2022; Weisman & Peña, 2021; Williams, et al., 2020 |
| | Trustworthiness and Intentions | 2 | Kim, 2022; Shin et al. ,2019 |
| | Trust, Competence, and Intentions | 1 | Schreibelmayr et al., 2023 |
| **Trust-related Constructs** | | | |
| | Competence | 4 | Kim et al., 2019; Paetzel-Prüsmann & Castellano, 2019; Paetzel et al., 2020; Stein et al., 2022 |
| | Acceptance | 3 | Destephe et al., 2015; Schreibelmayr & Mara, 2022; Sharma & Vemuri, 2022 |
| | Credibility | 2 | Mulcahy et al., 2024; Stein & Ohler, 2018 |
| | Persuasion | 1 | MacDorman, 2019 |
| | Believability | 1 | Qiao & Eglin, 2011 |
| **Trust-related Intentions** | | | |
| | Intentions [a] | 8 | Chien et al., 2024; Kim, 2023; Li et al., 2024; Lin et al., 2021; Lin et al., 2022; Lou et al., 2023; Paetzel-Prüsmann & Castellano, 2019; Stein et al., 2020 |
| | Risk Perception and Behavior [b] | 1 | Aubel et al., 2022 |
| | Social Acceptance [c] | 1 | Reuten et al.,2018 |
| | Intentions and Competence | 5 | Dai & MacDorman, 2018; Dai & MacDorman, 2021; Grazzini et al., 2023; Liao et al., 2023; Pandey & Rai, 2023 |
| | Intentions and Credibility | 2 | Kaate et al., 2023; Patel & MacDorman, 2015 |
| | Intentions and Satisfaction | 1 | Bae et al., 2024 |

*Note.* [a] Indicates qualitative research studies (Lin et al., 2021; Lou et al., 2023; Pinney et al., 2022; Thimm et al., 2024; Troshani et al., 2020) where trust and intentions were extracted from participants' statements and considered as trust measures. [b] Involved safe and risky investment options, where perceived risk was interpreted as a lack of trust. [c] Assessed how comfortable participants would feel interacting with an agent in various futuristic contexts.

Across these three clusters, the majority of studies (*n* = 46) assessed trust or trust-related constructs/intentions solely through subjective self-report measures. These typically involved survey responses or structured trustworthiness scales, such as multi-item Likert scales, single-question measures, or qualitative designs. Among these, 14 studies combined a self-reported survey with self-reported behavioral intentions. In these studies, researchers were interested in how subjective trust or trust-related outcomes ratings relate to participants' stated intentions regarding future interactions or behaviors.

In contrast, a smaller number of studies (*n* = 7) employed objective trust measures, where trust or trust-related constructs were inferred from actual decisions, actions, or behavioral outcomes (e.g., trust games or decision-making tasks). These studies combined survey-based measures with observable behaviors (e.g., following advice or behavioral changes in game scenarios) to capture the dynamic and applied nature of trust. Of these, three studies employed a trust game (Abubshait et al., 2017; Abubshait et al., 2022; Mathur & Reichling, 2016), and one study assessed risk perception and risk behavior through an investment decision-making task (Aubel et al., 2022). Additionally, only one study measured behavioral intentions through the actual decision preferences in a choice task, such as participants' preference for humans over robots for gift selection (Rosenthal-von der Pütten et al., 2019). Notably, among the seven studies that measured trust through objective behavioral outcomes, most (*n* = 5) applied only indirect encounters in which participants made trust-related decisions based on hypothetical scenarios or static representations of agents, rather than engaging in real interactions with an agent. Only two studies (both assessed competence) applied direct interaction in which participants engaged with an agent in a real-time collaborative game (Paetzel-Prüsmann & Castellano, 2019; Paetzel et al., 2020). Hence, no study directly assessed trust objectively in a real interactive encounter with an agent.

**Findings on the Relationship Between UVE and Trust**

In the following, we review what the sampled studies reveal about the potential effects of the UVE on trust. In this section, we fully adhere to the interpretations of observations offered by the respective authors. That is, if the author(s) of a study claim to have observed the UVE or to have found a direct effect of the UVE on trust, we coded this as evidence, without further questioning the validity of such a claim. We refer to the discussion section for a more critical reflection on the methods and approaches that were applied to assess the UVE and its impact on trust.

### ***General Findings***

We first coded studies based on whether they provided empirical evidence about the relationship between the UVE and trust (see Figure 8 for a visualization of the coding categories). A total of 41 studies provided empirical evidence on the relationship between the UVE and trust or trust-related outcomes. We then coded these studies based on the type of effect reported. This included three types of effect: direct effect, conditional or indirect effect and no effect. It is important to note that studies could be coded multiple times if they reported more than one type of effect.

Studies were coded as reporting a “direct effect” if they showed a direct impact of the UVE on trust or trust-related outcomes. This includes both qualitative and quantitative-inferential analyses, such as significant correlations between the UVE and trust or a main/direct effect of a UVE-related experimental factor on trust or trust-related outcomes (e.g., in ANOVA or regression analyses). Among the 41 coded studies, 17 were categorized as showing a direct effect.

Studies were coded for reporting a “conditional/indirect effect” if they reported a moderation (i.e., conditional effects) or mediation (i.e., indirect effects) influencing the relationship between UVE-related patterns and trust or trust-related outcomes. Additionally, if the authors claimed that a factor influenced variation in UVE strength without formally testing it as a statistical moderator, we still coded the study as a conditional/indirect effect based on their interpretation. For qualitative studies, if the authors suggested that a factor or condition could impact the relationship between the UVE and trust, the study was included in this category. We coded 24 studies as showing a conditional/indirect effect.

"No effect" studies included studies in which the authors successfully induced the UVE but did not observe any effect on trust or trust-related outcomes. Among the 41 studies that empirically tested the relationship between the UVE and trust or trust-related outcomes, 10 were coded as showing no effect.

An additional 12 studies were coded as not empirically testing the relationship between UVE and trust or trust-related outcomes. Seven of these studies were coded as "failing to produce an effect" as they admittedly did not observe the UVE at all. Another five of these studies were coded as providing "inconclusive evidence for an effect", as they measured UVE and trust, or trust-related outcomes separately but did not test the relationship between them or relied on interpretative claims without empirical modeling. Instead, they discussed a possible connection between the two constructs based on existing literature.

**Figure 8**

*Overview of Coding Categories*

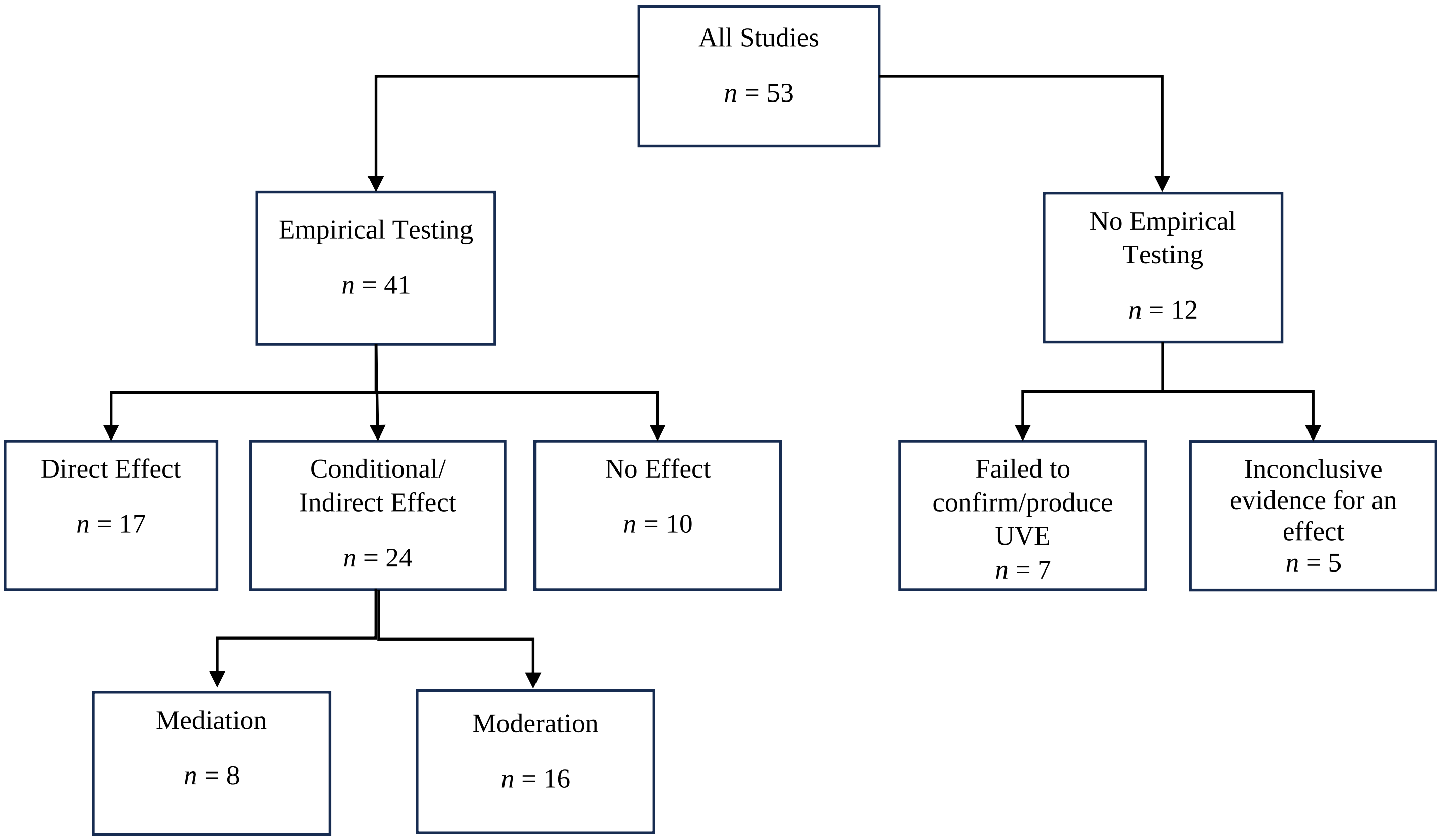

*Note.* Some studies were dual-coded under both direct and conditional/indirect effects (*n*=10); therefore, the total '*n*' under Empirical Testing exceeds the actual number of unique studies in this group.

### *Findings Grouped by Trust Assessments*

Because findings might plausibly vary with the way trust was operationalized and measured, to further scrutinize the existing evidence on whether the UVE affects trust, in the remainder of this section, we analyze findings grouped by the three types of trust measurement approaches that we previously identified. First, we review the findings of studies that measured trust directly; second, we review findings of studies that assessed trust-related constructs, such as competence or acceptability; and finally, we review the findings of studies that measured trust through trust-related intentions. Within each of the three trust assessment categories, we included some examples for each type of finding.

**Studies Directly Measuring Trust.** Out of the sampled 53 studies, 24 directly measured trust or trustworthiness by explicitly asking participants to rate the level of trust they placed in the agents. Figure 9 presents the coding results for all studies that directly measured trust.

**Figure 9**

*Coding of Studies with Direct Trust Measurement*

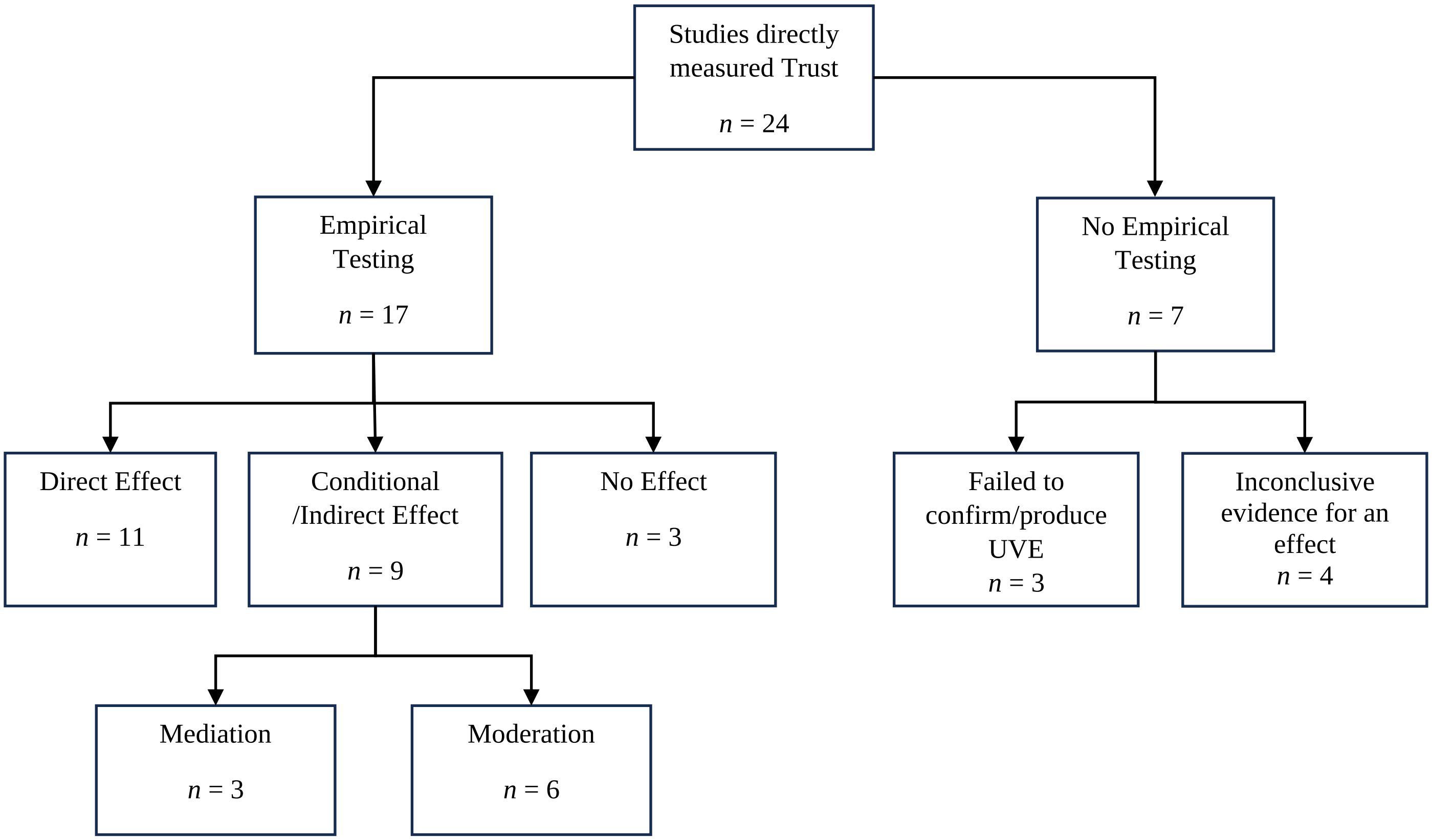

*Note.* Some studies were dual-coded under both direct and conditional/indirect effects (*n* = 6); therefore, the total '*n*' under Empirical Testing exceeds the actual number of unique studies in this group.

Of these, *11 provided a direct effect* of the impact of the UVE on trust. All of these studies reported that the UVE was induced, and the elicited affinity response directly diminished trust.

Perhaps the strongest study providing evidence of such a direct effect is offered by Mathur and Reichling (2016). This study examined how the UVE affects people's trust in robots based on their facial appearance. For this purpose, they designed a series of experiments in which participants were exposed to static 2D images of either real robot faces (wild-type robot, studies 1-3) or digitally composed robot faces (studies 3-5). In the first simple experiment, participants rated how human-like each face looked. In a second simple experiment, participants indicated how likable each face seemed (rated from –100 indicating less friendly, more unpleasant and creepier to +100 indicating more friendly, more pleasant and less creepy); this study replicated the typical uncanny valley curve. The third experiment was the main study. Participants played a trust "investment" game where they were exposed to one of the robots' faces and in which they had to choose how much (virtual) money (up to 100$) they would entrust to the robot. This study convincingly showed that trust follows a similar response curve as liking, with reduced trust in less likable robots that fall into the uncanny valley. Accordingly, users seem to trust robots in the UV less. In addition, the extent a robot's face was displaying negative or positive emotions was assessed as a moderator of this effect. However, the UV seemed to affect trust independent of the robot's displayed emotion. Studies 4 and 5 replicated these findings with a range of robot faces that were digitally composed, based on a number of extraneous parameters (e.g., eye size).

Of the studies that directly measured trust, 12 reported a conditional or an indirect effect. Among these nine studies reported a *moderation effect,* while three reported a *mediation effect.*

The nine studies reporting moderation effects found several factors that shaped or influenced the relationship between the UVE and trust (note that the values in parentheses represent the frequency with which each factor appeared across different studies): context of interaction (e.g., service vs. healthcare settings; $n = 4$); age (e.g., younger vs. older adults; $n = 2$); familiarity with the agent ($n = 2$); mismatch in aesthetic design or modality (e.g., realistic visuals with artificial voices; $n = 2$), perceived control during the interaction ($n = 1$); task type (e.g., emotional-social vs. mechanical; $n = 1$); pre-existing levels of interpersonal trust ($n = 1$).

A good example of a study that not only reported a direct effect but also a conditional effect is the study by Jung et al. (2021). This study was interested in understanding to what extent users trust agents that varied in human-likeness, depending on the context in which the agent offered a service (high-expertise context, like in tutoring, vs. low-expertise, like in hotel service). A sample of 505 participants were exposed to five different types of humanoid robot pictures with an increasingly human-like outer appearance, either presented as a tutor or a hotel clerk. The extent users felt “uneasy or unfriendly” when looking at the displayed robot was assessed as an affective response (uncanniness) to infer the UVE. Trust was measured using a single-item question: “Do you believe that the robot in this picture would be trustworthy (as a hotel receptionist/ tutor)?”. Confirming the UVE, the study found a U-curve pattern, with the almost human-like robot version making users most uneasy. Trust ratings in robots showed a dip for intermediate levels of human-likeness in both contexts, suggesting a direct effect of the UVE on trust. However, the study also found that this effect was conditional and qualified by the context of service. The undermining effect of an almost human-like robot on trust was alleviated (or weaker) in a high-expertise tutoring context as compared to a low-expertise hotel reception context. In summary, the study suggests that task expertise moderates the influence of robot appearance on trust, with an almost human-like agent's uncanniness undermining trust less strongly in high-expertise service contexts.

Another example of a study observing a moderation effect is provided by Tu et al. (2020). This study examined whether a robot's UVE varies across age groups, and whether it can impact likability, perceived trustworthiness, and acceptance of the robot's service and companionship. The study used an online questionnaire completed by 255 participants divided into three age groups: younger (18–39), middle-aged (40–59), and older adults (60–87). Each participant rated 16 robot faces (randomly drawn from a pool of 83). Results suggest a clear UVE pattern and drop in likability, perceived trustworthiness, and acceptance as robots became near-perfectly humanlike for young and middle-aged adults. In contrast, older adults did not show this effect. Instead, older participants displayed a linear positive relationship (without a dip) between a robot's human-likeness and trust, and similarly for likability and acceptance. Age, thus, significantly moderated the effect of a robot's human-likeness on trust.

In the three studies that reported *mediation effects,* perceived trustworthiness of the agent (Shin et al., 2019), trust (Song & Shin, 2022), and eeriness (Weisman & Peña, 2021) were identified as mediators. However, these studies did not report actual mediators of the UVE-trust relationship but reported mediators in the wider context of studying this relationship. For instance, Shin et al. (2019) investigated the mediating role of perceived trustworthiness in the relationship between the UVE (specifically, the feeling of eeriness induced by avatar realism and animacy) and the intention to befriend an avatar user in a virtual social networking service. Here, rather than being treated as an outcome, trust was examined as a mediator of the effect of the UVE on a trust-related outcome like befriending the avatar. Weisman and Peña (2021) examined the effect of doppelganger talking head avatars on affect-based trust and found that eeriness acted as a mediator. Here, rather than treating eeriness as a predictor in the UVE-trust relationship, it was treated as a mediator. Seeing a doppelganger AI avatar (a talking head featuring the participant's face) did not directly lower trust, it increased feelings of eeriness and discomfort, which in turn led to decreased trust in AI-generated messages.

Of the three studies that observed *no effect* of the UVE on direct measures of trust, a study by Kim (2022) can serve as an example. The study aimed to examine, among 119 participants, how an avatar’s realism (cartoonish vs. hyper-realistic) and conversational cues (introverted vs. extroverted) influence the avatar’s perceived personality, including trustworthiness, likability, and willingness to befriend. The study conducted a manipulation check and confirmed that hyper-realistic avatars were rated significantly higher on both uncanniness and freakiness than cartoonish avatars. Hence, greater realism did lead to greater eeriness, consistent with the UVE. However, the established UVE did not affect trust, as trustworthiness scores were statistically similar across both avatar types (either cartoonish or hyper-realistic), and no significant effect of realism was found on trust, likability, or willingness to befriend.

Finally, regarding studies that did not empirically evaluate the relationship between UVE and trust, *three studies were classified as failing* to produce an effect of the UVE, while *four studies were identified as inconclusive evidence* for an effect.

A study by Mal et al. (2024) can serve as a typical example of a study providing only inconclusive evidence for the UVE-trust relationship. The study investigated how immersion level (2D screen vs. VR) influences social judgments (trust and sympathy) of agents differing in human appearance (realistic vs. abstract). Participants were assigned to either a low-immersion (2D screen) ($n$ = 65) or high-immersion (VR) condition ($n$ = 26). Eeriness and trust ratings varied notably with the level of immersion. In low-immersion conditions, higher eeriness ratings were reported for realistic virtual humans compared to abstract ones, suggesting a possible UVE. However, this effect was not observed in fully immersive VR, where no significant difference in eeriness was found. Regarding trust, there were no significant differences between the two types of virtual humans in the 2D setting, but in VR, trust ratings were significantly higher for realistic virtual humans compared to abstract humans. The authors speculated that high immersion may enhance trust in more human-like agents, while trust might have been undermined by the eeriness of realistic virtual humans presented on a 2D screen. However, this remains hypothetical, as no correlation or testing between eeriness and trust was provided.

**Studies Applying Trust-Related Constructs.** Among the 53 studies sampled in this review, 11 studies examined the impact of UVE on trust-related constructs, such as competence, acceptability, credibility, believability, and persuasion. Figure 10 displays the coding of studies that measured trust-related constructs.

**Figure 10**

*Coding of Studies with Trust-Related Constructs Measures*

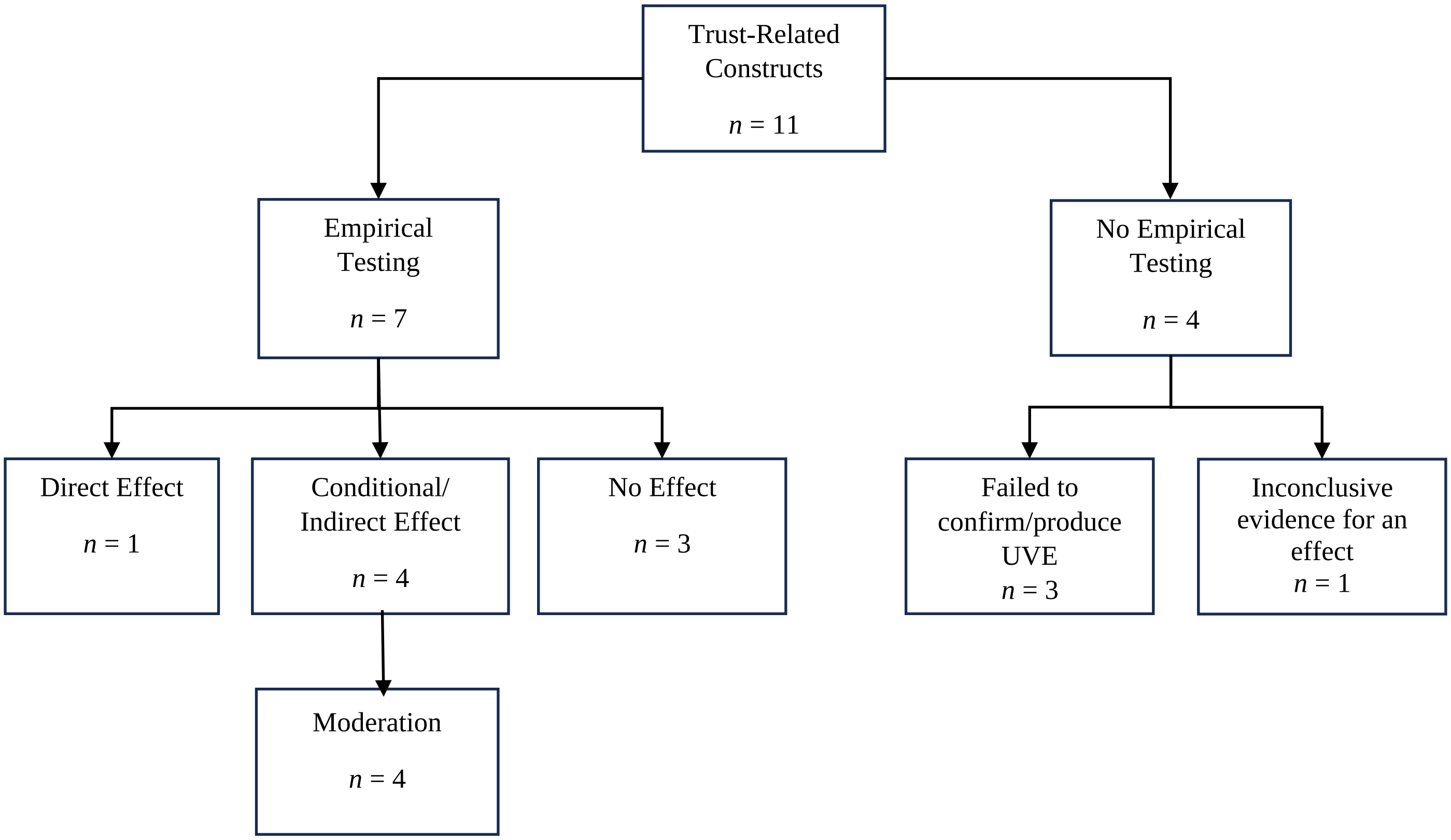

*Note.* One study was dual-coded under both direct and conditional/indirect effects; therefore, the total '*n*' under Empirical Testing exceeds the actual number of unique studies in this group.

Only one study provided a *direct effect* of the UVE on trust-related constructs namely acceptance of the agent. Four studies reported *a conditional effect* where the relationship between the UVE and trust-related constructs (e.g., competence, acceptability, and credibility) was found to be explained by moderating roles of initial impressions, interaction frequency, familiarity with the agent (e.g., Tintin), task type (e.g., cognitive vs. creative roles), and both verbal and visual forms of anthropomorphism (reflecting an asymmetrical effect of anthropomorphic design).

A study by Sharma and Vemuri (2022) was coded twice as reporting a direct effect and a conditional effect for the UVE-acceptance relationship. The study examined how human-likeness in digital avatars influenced users' acceptance, comfort, eeriness and emotional responses across different social and professional roles. Across three experiments, combining eye-tracking and pupil dilation with subjective ratings, participants evaluated static images and videos of film-based characters and CGI agents, with varying levels of human-likeness. The results showed that while very human-like avatars (e.g., Dexter, Siren) were rated as more comfortable, pupil dilation elicited during human-like movements (e.g., talking or smiling), revealed eeriness. The study reported a negative significant correlation between comfort and eeriness, providing support for UVE. Familiarity was associated with lower eeriness, as agents from well-known contexts (e.g., Tintin) elicited less eeriness despite their cartoonish designs. Moreover, this study highlighted that acceptance was highly role-dependent, since avatars with higher perceived human-likeness were accepted for roles requiring cognitive skills (e.g., IT) but resisted for roles demanding empathy or creativity (e.g., judges, artists).

Another study by Paetzel and Castellano (2019) investigated the relationship between the UVE and perceived competence through direct interactions with a Furhat robot during collaborative games. This study demonstrated that the impact of the UVE was not fixed but varied depending on the level of exposure or repeated interaction, making it an example of a conditional effect. In the study, ratings of uncanny feeling (warmth and discomfort), human-likeness and competence were collected at three time points; after an initial impression, after the robot introduced itself (using a two-minute pre-recorded video), and following an interactive session involving a 20-question game. The results demonstrated that applying a morphed facial texture to the robot (intended to elicit the UVE) initially caused higher discomfort, but the perceptions of competence improved after participants ($n$ = 48) interacted with the robot. Uncanny feelings significantly decreased after the robot introduced itself, in comparison to the first impression. However, there was no significant difference during the subsequent interaction.

These results suggest that the robot's appearance shaped participants' early perceptions of both uncanniness and competence. However, as interaction with the robot increased, perceived competence improved, and feelings of uncanniness diminished. In other words, the initial uncanny impressions associated with the robot’s appearance had a potential indirect negative influence on early perceptions of competence, but these effects were subsequently mitigated and then improved through repeated interaction. Moreover, qualitative feedback also revealed shifting perceptions after interactions, in which many participants initially described the robot as "*dead or not real*", but after engaging with it, they perceived it as more "*alive and relatable*". Altogether, these findings of this study suggest that interaction with a robot can play a determinant role in mitigating the effects of the UVE, leading to a boost in perceived competence over time.

Finally, three studies reported *no effect* of the UVE on the trust-related constructs.

A study by Dai and MacDorman (2021) shows a case of no effect, as the negative UVE did not impact competence or behavioral intentions. The study aimed to investigate how different levels of visual realism in a virtual doctor, combined with bedside manner (e.g., using a friendly tone, showing empathy, making eye contact), influence adherence intentions and perceived competence in a virtual health consultation. The study used a 2×5 between-subjects design, manipulating bedside manner (good vs. poor) and visual depiction of the doctor (five realism levels: cartoon, bigeye, swapped, 3D animation, and real). Participants' responses ($n$ = 441) were assessed using measures of eeriness, competence, warmth, and adherence intention to medical advice. Eeriness ratings were collected both before (pretest) and after (posttest) the virtual consultation to establish a baseline measurement independent of the doctor's subsequent behavior (i.e., bedside manner) during the consultation. Competence, warmth, and adherence intention were only collected after the interaction. The pretest revealed a clear UVE pattern, as the 3D-animated versions were rated significantly eerier than cartoon or real depictions. Nevertheless, posttest results showed that eeriness did not significantly impact adherence intentions or competence. Instead, bedside manner had a significant direct effect on both warmth and adherence intention. Increased warmth, in turn, had a significant direct effect on perceived competence, and competence affected adherence intention; thus, creating an indirect pathway from bedside manner to competence through warmth, regardless of realism levels.

Finally, regarding studies that did not empirically evaluate the relationship between UVE and trust, three studies were classified *as failing to produce* an effect of the UVE, while one study yielded *inconclusive evidence* for an effect.

Concerning the single study that led to inconclusive evidence for an effect, the study by Qiao and Eglin (2011) investigated how different aspects of facial behavior realism (head movement, facial expression, and eye movement) affect perceptions of eeriness, believability and accurate behavior recognition in a computer-generated (CG) human head. The study used a within-subjects factorial design in which 34 participants viewed video clips of avatars displaying different combinations of motion types. This study found that the inclusion of facial expressions had a significant effect on increasing eeriness, while head movement reduced eeriness and significantly improved believability. The combination of facial expression and head movement produced the highest believability ratings. Their findings indicated that motion plays an important role in mitigating eeriness, making avatars appear more natural and believable to participants. The study did not formally test the UVE as a causal structure and did not check whether the increased eeriness had a direct or mediating effect on believability. Therefore, while eeriness and believability were measured, the relationship between them remained untested.

**Studies Measuring Trust-related Intentions.** A noteworthy portion of the reviewed literature (18 studies) assessed trust indirectly by measuring intentions, either alone (*n* = 10) or in combination with other trust-related constructs (*n* = 8). Figure 11 summarizes the coding of studies that measured trust-related intentions.

**Figure 11**

*Coding of Studies with Trust-Related Intentions Measures*

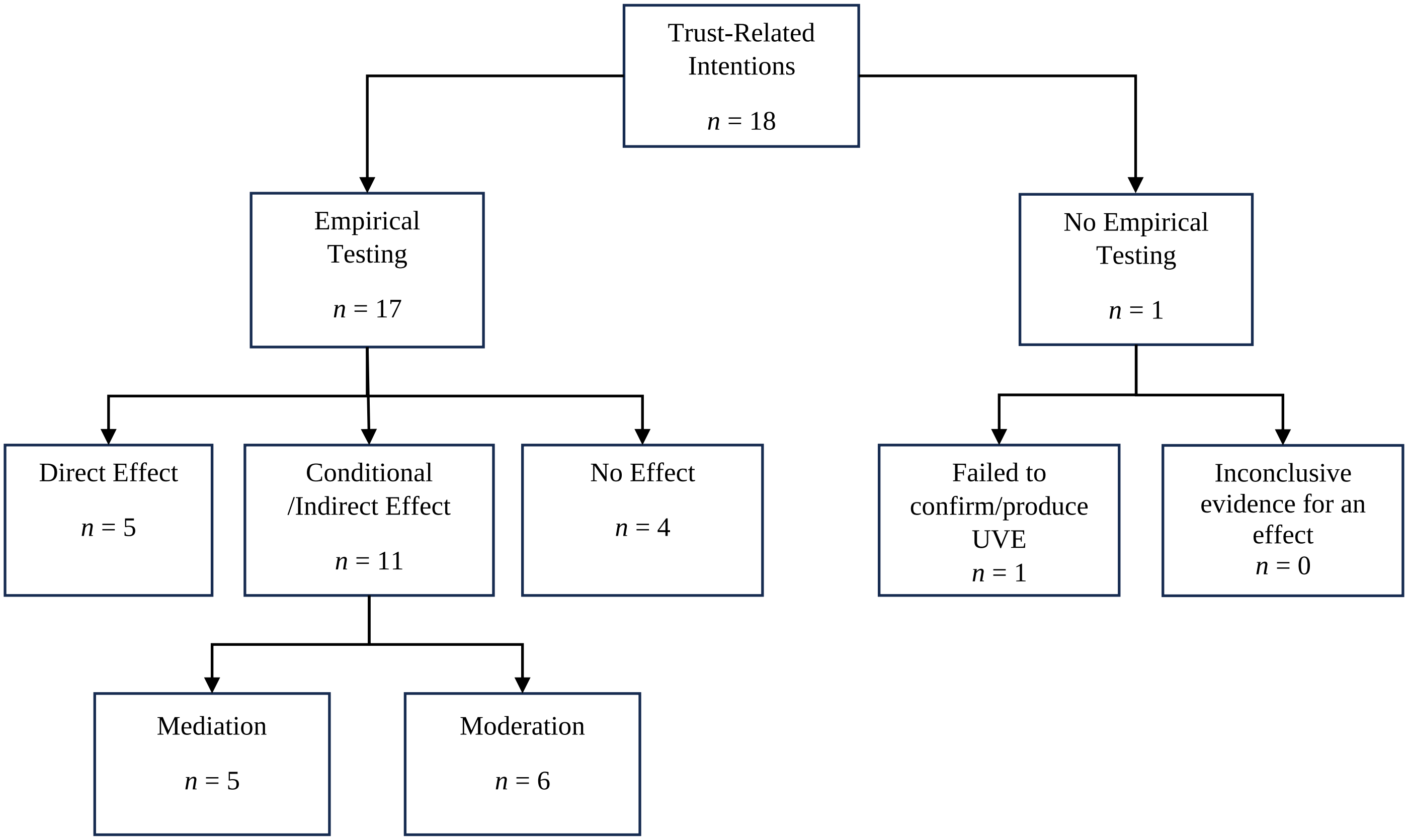

*Note.* Some studies (*n* = 3) were dual-coded under both direct and conditional/indirect effects; therefore, the total '*n*' under Empirical Testing exceeds the actual number of unique studies in this group.

Among 18 studies, *five studies provided a direct effect*, supporting the impact of UVE on intentions.

A study by Reuten et al. (2018) provided a direct effect, showing that the UVE directly reduced participants' willingness to engage with robots in social contexts. The study used a within-subjects repeated-measures design in which participants viewed static images of robots and humans displaying emotional expressions. Participants (*n* = 40) rated each agent on several dimensions (eeriness, naturalness, and attractiveness) and indicated how comfortable they would feel interacting with the agent in various imagined future scenarios (e.g., healthcare, service roles). Simultaneously, the researchers used pupillometry as a physiological marker of uncanniness. The results showed that robots that appeared nearly human-like scored highest on uncanniness and lowest on naturalness and attractiveness, confirming the UVE. Additionally, pupil dilation was weaker for uncanny robots, suggesting reduced emotional engagement. The study indicated that the UVE negatively impacted participants' willingness to engage in social interactions with uncanny robots, as these robots were consistently rated as less comfortable for interaction across all tasks.

Eleven studies provided a *conditional/indirect effect of the UVE on intentions.* Across these studies, five studies reported indirect effects of the UVE on intentions and introduced the following as mediators in their results: disconfirmed expectations (Grazzini et al., 2023), identity threat (Kim, 2023), emotional expressions (Li et al., 2024), competence and threat (Liao et al., 2023), and familiarity and satisfaction (Bae et al., 2024). However, it should be noted that these studies did not all explicitly treat the UVE as a formal predictor or did not run mediation analyses that frame trust-related intentions as the direct outcome of UVE. Instead, they often examined mediators in the broader context of UVE-related variables and trust-related outcomes. Therefore, these variables were reported as a mechanism that linked the agent-related UVE patterns to trust-related intentions.

The study by Bae et al. (2024) provided a clear example of an indirect effect of the UVE on intentions. This study examined how ‘visual’ and ‘behavioral’ human-likeness in meta-humans (hyper-realistic digital humans with advanced behavior) influenced user acceptance, satisfaction, and purchase intention, through the UVE. Using videos of meta-humans and digital humans (conventional photorealistic avatars), they collected survey data from 280 participants and applied structural equation modeling (PLS-SEM). The study indicated significantly higher user acceptance for meta-humans since meta-humans surpassed digital humans in user acceptance, human-likeness (both visual and behavioral), and purchase intention. Moreover, the result demonstrated that the impact of the UVE was indirect, mediated by familiarity and satisfaction. Familiarity did not directly influence user acceptance but instead mediated satisfaction, which in turn significantly shaped purchase intentions. Additionally, the study found that the UVE emerged only for digital humans' appearance (i.e., visual human-likeness), whereas meta-humans avoided visual eeriness but continued to struggle with behavioral realism.

Out of the 11 studies, six identified factors that contributed to the relationship between the UVE and trust-related intentions through a *moderating* role. The following factors were reported: age (Chien et al., 2024); context of use, perceived agency, and intelligence (Lin et al., 2021); task type (social vs. mechanical tasks) (Lin et al., 2022); curated flaws (i.e., Small, intentional imperfections added to virtual influencers to make them look more natural and relatable) and user self-justification (i.e., the cognitive process when followers convince themselves that real people are managing the virtual influencer’s account, helping them feel more comfortable continuing to engage despite any uncanniness) (Lou et al., 2023); personality traits (Aubel et al., 2022); and individual differences in neural responses (Rosenthal-von der Pütten et al., 2019).

A study by Rosenthal-von der Pütten et al. (2019) can serve as an example that was coded twice, reporting both a direct and conditional effect of UVE on trust-related intentions. This study aimed to explore how the UVE influences humans' behavioral decisions and their underlying neural mechanisms. The study collected fMRI data from participants (*n* = 21) who viewed pictures of six categories of agents and completed two main tasks: a rating task (assessing likability, familiarity, and human-likeness of the agents) and a decision-making task (choosing whether they preferred to receive a gift from a human or an artificial agent). The agents ranged from non-humanlike mechanoid and humanoid robots to highly humanlike android robots and artificial humans as well as real humans with and without physical impairments. The study provided both behavioral and neurological evidence supporting the UVE, showing that highly human-like artificial agents elicited the strongest UVE responses. The study found UV reactions significantly influenced the decisions participants made during the task. During the decision-making task, preferences were shaped by likability, familiarity, and human-likeness of the agent, as participants' ratings on these measures were significant predictors for their choices on whom to receive a gift from, with a general tendency to prefer humans over artificial agents. Individual differences in the UVE reactions were reflected in the neural activity of several brain regions and explained the strength of behavioral choices, neural responses, and variations in the UVE strength across participants.

Four studies were classified as reporting *no effect* of the UVE on trust-related intentions.

A study by Stein et al. (2020) is among the studies that provided evidence of no effect of the UVE on trust-related intentions. The study aimed to examine how virtual agents' visual realism (embodiment) and cognitive sophistication (mental prowess) influence users' perceptions and behavioral intentions. The study presented participants (*n* = 134) with videos and vignettes of agents that varied in appearance (human-like rendering vs. text interface) and mind complexity (simple algorithm vs. complex AI), and measured responses including eeriness, perceived usefulness, and interest in future interaction. The results demonstrated a significant interaction effect on eeriness, where agents with a human-like body paired with simple algorithms elicited notably higher eeriness, aligning with the UVE. However, this increase in eeriness did not significantly reduce behavioral intention. Neither embodiment nor mind complexity had a direct effect on participants' willingness to interact with the agent again. Interestingly, despite higher eeriness ratings, human-like agents were still perceived as more useful. Hence, the UVE was present but did not impair behavioral intention.

Finally, regarding studies that did not empirically assess the relationship between UVE and trust-related intentions, one study was classified as *failing to produce* an effect of the UVE, while no studies were identified as *inconclusive evidence* for an effect.

## Discussion

In this systematic review, we examined how the UVE impacts trust by synthesizing findings from 53 studies.

Our review identifies multiple factors, under various conditions, that contribute to the way elicited response from the UVE influences trust. In addition to classifying the key findings, we also provide a comprehensive overview of the methodological approaches across the studies. We coded and analyzed important methodological aspects, including the type of agent and interaction, study design, sample size, context, and culture, as well as the varied measurements of the UVE and trust employed by the studies. Together, the results underline the multifaceted nature of both the UVE and trust and underscore the need for a more in-depth and methodologically grounded investigation into their relationship.

**Findings and interpretations**

In our review, we organized the findings based on the type of trust assessment evaluated in each study: direct trust measures, trust-related constructs, and trust-related intentions. Among the 24 studies that directly measured trust, 17 empirically tested the relationship between the UVE and trust. Of these, 11 studies reported a direct effect, nine reported a conditional/indirect effect, and three reported no significant effect. Regarding trust-related constructs (e.g., competence, acceptance, believability, persuasion, and credibility), 11 studies evaluated these outcomes, and seven of them empirically tested their relationship with the UVE. Within this group, one study reported a direct effect, four reported a conditional/indirect effect, and three reported no effect. Finally, 18 studies assessed trust-related intentions (e.g., willingness to purchase or to interact with the agent in the future), with 17 studies providing an empirical test. Among these, five reported a direct effect of the UVE on trust-related intentions, 11 reported a conditional/indirect effect, and four reported no effect (see Figure 9-11).

Taken together, the findings across all types of trust suggest the relationship between the UVE and trust is often not straightforward. Among the 41 studies that empirically tested the impact of UVE on trust or trust-related outcomes, 17 studies reported a direct effect. However, a large portion (more than half, $n$ =24), reported a conditional/indirect effect, indicating that the relationship between UVE and trust can be either mitigated or amplified by additional factors (see Figure 8). These patterns suggest that the effect of UVE on trust or trust-related outcomes is complex and dependent on external factors that appear to play a determinant role in shaping user responses. These factors can be categorized into three overarching groups:

- User-related factors: These factors relate to demographic characteristics, individual traits, and cognitive processes of users. They include age, personality traits, individual differences in neural responses, initial impressions, prior levels of interpersonal trust, familiarity and satisfaction with the agent, disconfirmed expectations, perceived identity threat, and user self-justification.

- Agent-related factors: These refer to design elements, capabilities, and expressive behaviors of the agent. Examples involve design mismatches, agent autonomy, physical warmth, competence, agency, intelligence, agent's emotional expression, curated flaws, eeriness and trustworthiness of the agent.
- Environment-related factors: These involve the contextual and situational conditions in which users interact with the agent. This category encompasses the context of use, task type, and interaction frequency.

The identified factors are broadly consistent with those reported to influence trust in agents, as outlined in Campagna and Rehm's (2024) systematic review, which categorized them into human-related factors, environmental factors, and robot-related factors. Notably, a meta-analysis by Hancock et al. (2011) also emphasized the particularly strong influence of robot characteristics and performance on perceived trust, concluding that agent-related manipulations have a greater effect on trust than any other category.

Among the identified factors, some appeared more frequently across the included studies, indicating more influence on the relationship between the UVE and trust. These commonly mentioned factors included: familiarity with the agent ($n$ = 3), users' age ($n$ = 3), personality characteristics ($n$ = 3), asymmetrical effect of anthropomorphic design cues ($n$ = 3), and context of use and task type ($n$ = 6).

Notably, the impact of UVE on trust appears to vary depending on the context of use since task type and role expectations shape whether humans embrace or resist the agents. As shown in studies by Jung et al. (2021), Sharma and Vemuri (2022), and Lin et al. (2022), trust was less influenced by agent appearance in high-expertise or cognitively demanding roles (e.g., tutoring or IT support), where functionality was prioritized over form. In these settings, humanoid designs, despite their human-likeness, can gain trust. However, the preference for human-likeness was task-dependent. High human-likeness tended to be avoided in roles requiring empathy or creativity, such as in judicial or artistic tasks (Sharma & Vemuri, 2022). Similarly, Lin et al. (2022) found that highly human-like robots reduced trust and usability in mechanical tasks (e.g., room cleaning), whereas moderately human-like robots were generally preferred across both emotional-social and mechanical domains. Additionally, studies on parental and healthcare acceptance (e.g., Lin et al., 2021; Troshani et al., 2020) emphasized that trust is strongly tied to perceived usefulness and contextual integration. In these domains, users evaluated agents based on how well they fit into daily routines and fulfilled specific roles. For example, storytelling robots were accepted as parental aides when their function aligned with real-life parenting needs (Lin et al., 2021), while AI in healthcare was more trusted for routine, low-risk tasks than for emotionally sensitive decisions (Troshani et al., 2020).

These findings related to context and task characteristics are also in line with previous research. Roesler et al. (2020) emphasize that industrial and social contexts play distinct roles in shaping trust, noting that the increasing use of collaborative robots (cobots) has shifted perceptions from viewing robots as solely autonomous tools to seeing them as responsive team members. In technical environments, robots were perceived as more reliable when presented with a more functional design, compared to only anthropomorphic alternatives. Furthermore, the role of task difficulty in shaping trust and decision-making has been highlighted by Christoforakos et al. (2021) and Herse et al. (2023). Their findings show that participants changed their initial decisions more frequently when confronted with difficult stimuli, likely due to reduced confidence in their own judgments. This suggests that the nature and complexity of a task can influence users' willingness to rely on an agent's recommendations.

Moreover, competence has not only been identified as a key factor influencing the relationship between UVE and trust-related outcomes, but it has also been highlighted in several studies as a potential buffer against the negative effects of eeriness. For instance, MacDorman (2019) found that the potential negative effect of eeriness did not weaken persuasion when participants perceived the computer-animated doctor as competent. Similarly, Dai and MacDorman (2018, 2021) reported that greater perceived competence enhanced persuasiveness and adherence intentions, whereas eeriness had no significant post-test effect. Instead, warmth and competence emerged as the key predictors of adherence, independent of the agent’s realism. Supporting this, Patel and MacDorman (2015) concluded that advice content had a stronger influence on compliance than the appearance or perceived trustworthiness of the agent, further highlighting how competence and advice quality may override the discomfort associated with the UVE.

Thus, it is possible to conclude that the interplay of context, task characteristics, and competence of the agent can act as a shield to protect the negative feelings, which typically stem from the UVE. In other words, the effectiveness of highly anthropomorphic designs appears contingent on the context of use and the nature of the task. This interpretation is further supported by a recent study by Alimardani et al. (2024). The study found that while a voice and appearance mismatch influenced participants’ perception of anthropomorphism, it had no significant effect on trust. The authors suggested that the agent's recommendation quality and perceived competence played a more decisive role in fostering trust than its human-likeness.

However, as our review also showed, findings across studies remain inconsistent, making it difficult to draw definitive conclusions about when and how exposure to highly human-like agents leads to the UVE and influences trust in them. Even in the case of competence, which is often considered a positive trust cue, the relationship with the UVE is not straightforward. For example, Grazzini et al. (2023) found that presenting highly human-like robots with an emphasis on competence intensified the negative effects of disconfirmed expectations, leading to reduced customer attitudes and behavioral intentions. Similarly, Liao et al. (2023) showed that while increased anthropomorphism enhanced perceived competence, it also elevated perceived threat, which ultimately decreased acceptance of the agent. This suggests that while human-likeness may initially boost competence perceptions, it can also backfire by heightening expectations and triggering a psychologically negative response. These inconsistencies may stem, in part, from the unclear conceptualization of the UVE itself, including variations in how it is defined, operationalized, and measured across studies. These variations lead to divergent methodologies and limitations in UVE research.

## Limitations of the Existing Literature

In reviewing the limitations of the included studies, we organized them according to the broader framework of research validity, including external validity, internal validity, and construct validity. Under each of these validities, we identified existing limitations observed in the studies. Since each type of validity addresses a different aspect of research quality, this approach provides a better understanding of the methodological shortcomings and highlights key areas for improvement in future research (see Table 4 for a full summary of the limitations across studies).

***External validity***

Two key types of limitations were identified across the included studies: (a) ecological validity and (b) sample representativeness. These limitations refer to the extent to which study findings can be generalized to other populations, settings, or times.

**Ecological validity limitations.** One of the prominent limitations identified in the existing investigation of the UVE-trust relationship was the nature of interactions between participants and agents. Nearly 80% of the studies (41 out of 53) employed indirect encounters, where participants passively viewed static images or videos rather than engaging in real-world and dynamic interactions with the agents. The absence of naturalistic interactions raises concerns about ecological validity and limits the generalizability of findings to real-world settings, where users typically engage with agents in reciprocal and socially embedded environments.

According to Campagna and Rehm (2024), trust in human–agent interaction is a cognitive evaluation that emerges during engagement between the user and the agent. This definition implies that some level of active interaction is required, beyond merely observing a static image. Trust is not formed instantly; it develops gradually through repeated engagement, where users have the opportunity to assess the system’s integrity and reliability (Campagna & Rehm, 2024). Therefore, indirect or one-time passive exposure is insufficient to capture this dynamic process. Without interaction, participants cannot observe the agent’s behavior over time, evaluate its task performance, or experience how it responds in varied contexts.

**Sample-related limitations.** Sample size plays an important role in determining the statistical power and precision of a study’s findings. Among the included 53 studies, only seven explicitly reported a sample size justification based on power analysis. Furthermore, 14 studies included relatively small samples (fewer than 50 participants). In addition, seven studies recruited their samples from narrow populations (e.g., a specific academic or professional group), which may have distorted findings and reduced generalizability due to potential bias. For instance, Seymour et al. (2021) recruited participants from SIGGRAPH, which is a professional design conference. Similarly, Liao et al. (2023) involved only university students rather than actual industrial workers in a study investigating collaborative robots. One study (Mal et al., 2024) also exhibited a highly imbalanced sample size across conditions (e.g., 2D: 65; VR: 26), which can introduce bias and reduce the interpretability of between-group comparisons.

### *Internal Validity*

Many issues related to stimuli and experimental procedures were identified across the reviewed studies, which raised concerns about internal validity and whether causal relationships could be reliably inferred from reported designs.

**Limitations in Stimuli Design and Use.** Many studies (*n* = 34) exhibited limitations in the design and variety of stimuli in the included studies. Four studies relied on only a single stimulus, such as a video of the WABIAN-2R humanoid robot (Destephe et al., 2015), a static image of the NAO V5 robot (Stein et al., 2022), a single virtual influencer (Miquela) (Li et al., 2024), and the RoboSapien RS-Media with limited social behaviors (Nie et al., 2012). Additionally, 19 studies used a limited range of stimuli, often involving only two variations. This constraint reduces the ability to represent a continuum of human-likeness and limits the applicability of findings across different agent types. It also impairs the ability to capture the non-linear nature of the UVE, making it difficult to clearly position the stimuli as falling within, before, or beyond the valley. Consequently, these limitations weaken support for hypotheses predicting a U-shaped pattern in participants' responses. Moreover, three studies (Lou et al., 2023; Pandey & Rai, 2023; Troshani et al., 2021) based their stimulus selection on participants’ prior experience with the agents and their perceived level of human-likeness. As a result, the spectrum of human-likeness was not systematically manipulated.

Furthermore, four studies had limitations in stimulus design that hindered the accurate measurement of the UVE. For example, Kaate et al. (2023) compared three persona modalities (static image, text-based narrative, and AI-generated video), making it difficult to isolate the effect of human-likeness from the medium itself. Similarly, Sharma and Vemuri (2022) lacked live-action versions of highly human-like CGI avatars (e.g., ‘Dexter’, ‘Siren’) in comparison to other stimuli used in the study, limiting the ability to position these stimuli along the UV curve.

Gender bias in stimuli was observed in four studies, where only female agents were tested (Bae et al., 2024; Kim, 2022; Shin et al., 2019; Schreibelmayr & Mara, 2022). A robot’s gendered appearance can introduce bias, as it has been shown to influence both affective and cognitive evaluations, including how people interact with the agent and the extent to which they perceive it as trustworthy (Bernotat et al., 2021).

**Design Flaws and Lack of Experimental Control.** Several studies (*n = 23*) demonstrated potentially confounding variables that may have affected the interpretation of their findings as they were not systematically controlled. Some of these confounding variables were participant-related. For example, differences in demographic characteristics between two groups (e.g., age, recruitment method) may have impacted how their responses were compared and interpreted (Abubshait et al., 2022). In addition, when participants’ prior experience with financial AI systems was not taken into account, it may have influenced participants’ responses and the development of trust toward the AI-powered virtual assistant in the banking setting (Schreibelmayr et al., 2023).

Some other confounds were stimulus-related. For instance, measures of eeriness were confounded with perceptions of warmth and competence, making it difficult to isolate the specific contribution of the UVE (Dia & MacDorman, 2018; 2021). The use of celebrity avatars introduced possible biases related to attractiveness or perceived expertise (Song & Shin, 2022). Additionally, stimuli created using internet-sourced robot faces may suffer from inconsistent and biased representation (Mathur & Reichling, 2016). Also, Sharma and Vemuri (2022) relied on commercially available animated clips, which limited the researchers' ability to control cinematographic elements such as lighting, background, and facial or body movement. Finally, the findings by Seymour et al. (2021), such as higher trust or affinity for digital avatars in VR conditions, may have been influenced by confounding factors like the host's persona or speaking style, not just the level of realism. Since this study did not include a control group or random assignment, internal comparisons were limited, which made it harder to isolate causal effects.

Furthermore, ineffective or unsuccessful manipulations were reported in five studies as a threat to internal validity. For example, Dai and MacDorman (2018) reported a failed manipulation check for eeriness, and their follow-up study (2021) was underpowered to detect the effect of the agent's visual representation on adherence intention. Similarly, Kim (2022) found that the manipulation of eeriness was not effective, and Aubel et al. (2022) noted that their manipulation of the agent's anthropomorphism led to non-significant perception differences between medium and high anthropomorphism levels. In Li et al. (2024), emotion manipulations (e.g., lust) were inconsistently interpreted, especially by male participants. These examples highlight the importance of effective manipulation design and pretesting to ensure causal conclusions can be confidently drawn.

### ***Construct validity***

Across the reviewed studies, two primary types of construct validity limitations were identified, both related to the extent to which studies accurately measured the theoretical concepts they aimed to assess.

**Measurement issues.** Twenty studies reported concerns regarding measurement issues and the robustness of their approach. For example, some studies relied on single-item scales, which may not sufficiently capture complex constructs such as trust or eeriness. In one such case, Chien et al. (2024) asked participants to rate Disgust, Likability, Familiarity, and Acceptance through a single-item measure. Similarly, Paetzel and Castellano (2019) experienced a limited range of discomfort ratings, as the morphed agent in this study could receive a maximum rating of 3.43 on a 7-point scale, showing at most mild uncanny feelings. Reuten et al. (2018) did not include separate measures for uncanniness and human-likeness across robotic emotional expressions. Although it was a deliberate methodological choice to avoid potential order effects, this limitation reduced the clarity of their findings and restricted their ability to explore how different emotional expressions might influence key perceptions related to the UVE. Moreover, Liao et al. (2023) employed a cross-sectional design that may have introduced common method bias, as all measures were collected simultaneously using the same method. Additional limitations involved the incomplete or ambiguous operationalization of constructs. For instance, Broadbent et al. (2013) measured eeriness in only two of three conditions, limiting comparability across conditions. Additionally, Kim et al. (2019) defined anthropomorphism levels based solely on researcher interpretation, which may not align with participants' subjective perceptions. Similarly, Tu et al. (2020) did not define terms "human-like" or "non-humanlike", so participants rated based on their own judgment, which may have introduced subjective variability. Finally, Troshani et al. (2021) introduced potential moderator bias, as all focus group discussions were led by the same moderator, increasing the risk of groupthink.

**Self-report biases.** As 46 studies relied on self-reported evaluations, they are susceptible to self-report biases such as social desirability, response tendencies, or participant misinterpretation of constructs. These limitations challenge the validity of conclusions drawn about affective responses for trust or trust-related constructs/intentions. Additionally, self-report measures may not capture automatic or unconscious reactions, which are especially relevant in HAI. For instance, participants may consciously report trust but unconsciously exhibit hesitation or discomfort, indicating the need for integration of behavioral or physiological data to reliably capture UVE-related responses in future research.

**Table 4**

*Distribution of Limitations Identified Across the 53 Included Studies.*

| Research Validity | Limitation Category | Specific Issues | Number of studies |
|---|---|---|---|
| External Validity | | | |
| | Ecological validity limitations | Indirect encounters | 41 |
| | Sample-related limitations | No justification based on power analysis | 46 |
| | | narrow samples | 7 |
| | | Imbalanced sample size across conditions | 1 |
| Internal Validity | | | |
| | Limitations in Stimuli Design and Use | Use of a single stimulus | 4 |
| | | Limited range of stimuli | 19 |
| | | Prior experience with the agents | 3 |
| | | Limitations in stimulus design | 4 |
| | | Gender bias in stimuli | 4 |
| | Design Flaws and Lack of Experimental Control | Confounding variables | 23 |
| | | Ineffective or unsuccessful manipulation checks | 5 |
| Construct Validity | | | |
| | Measurement issues | Robustness of measures | 20 |
| | Self-report biases | Over-reliance on subjective evaluations | 46 |

**Research Gaps and Recommendations for Future Research**

Despite the valuable insights from the included studies, the aforementioned limitations highlight important research gaps that require more rigorous and comprehensive investigation. Given the growing interest in trust within HAI and inconsistencies in findings regarding the UVE and its impact on trust, future research should address the following areas to advance theoretical and practical understanding (see Table 5 for a summary of recommendations).

First and foremost, the nature of interaction in most studies remains indirect, often relying on static images, pre-recorded videos, or hypothetical scenarios. Such methods may not reflect how trust develops during real-time and dynamic interactions. In addition, longitudinal approaches, which could demonstrate how trust evolves over time, are rarely employed. Only two studies by Paetzel and Castellano (2019) and Paetzel et al. (2020) examined perceptions of competence and eeriness across multiple sessions. While the 2019 study found that perceived competence improved and uncanny feelings decreased after repeated interaction, the 2020 study observed that competence was established early (within the first two minutes) and remained stable across sessions. The 2019 study involved a single 20-minute interaction, whereas the 2020 study included three sessions with intervals of no interaction in between. This inconsistency highlights the need for more systematic longitudinal studies that directly assess, particularly trust trajectories in repeated, real-time interactions.

Therefore, we recommend that future research shift toward interactive, real-time paradigms and adopt longitudinal designs to examine how trust is formed, maintained, and potentially repaired over time. The recent systematic reviews (Bach et al., 2024; Campagna & Rehm, 2024) have suggested that trust may gradually increase with repeated exposure, as users adjust their expectations and become more familiar with the agent. However, these studies also highlight that trust is fragile, and negative incidents or feelings, such as a malfunction, can break it, though it may be recovered through subsequent positive experiences. Therefore, future research should explore whether uncanny feelings experienced in early interactions impact trust and continue to influence it during and after real-time, multi-session engagements.

The second major gap concerns the assessment of trust, which in most reviewed studies is predominantly subjective, relying on self-reported measures. While subjective measures offer valuable insight into users' perceptions, they often fail to capture the dynamic, behavioral, and evolving nature of trust formation in HAI. As Campagna and Rehm (2024) and Hancock et al. (2011) emphasize, relying solely on subjective trust measures poses several limitations. First, such measures offer only a fleeting snapshot of a dynamic process, failing to capture how trust is established, eroded, or restored throughout the interaction. Second, they are prone to post-hoc rationalization, where participants interpret trust-related items inconsistently across contexts, potentially leading to mismatches between self-reported and actual trust levels. Third, there is often a misalignment between subjective responses and observable behavior, as users may report high trust while acting otherwise. To address these limitations, researchers such as Campagna and Rehm (2024) and Bach et al. (2024) advocate for incorporating objective trust measures, including trust games, decision-making tasks, behavioral indicators (e.g., advice following, task delegation), and physiological signals (e.g., HRV, EEG, pupil size, skin conductance) that capture moment-to-moment trust fluctuations. Therefore, future research should combine subjective and objective trust measures, as this integration would enable real-time monitoring, support more holistic and reliable trust assessments, and facilitate the development of trustworthy interface designs, particularly in efforts to address uncanny feelings and better understand how the UVE influences trust.

The next gap concerns the lack of consistency in how the UVE is measured. Inconsistencies in implementation and measurement across studies contribute to variability in research approaches and challenges in interpretation. Therefore, it is important to establish an approximate consensus, based on existing literature (Kätsyri et al., 2015; Lay et al., 2016), to follow best-practice criteria. We propose that future investigations use a sufficient number of stimuli that span a continuum of human-likeness, ideally across three to five or more points, to reveal the non-linear nature of the UVE. Studies should also measure both negative responses (e.g., eeriness) and positive responses (e.g., familiarity, likability), and apply valid statistical tests to demonstrate a valley-shaped pattern in affinity response.

Additionally, we encourage recruiting a sufficient number of participants, justified through statistical power analysis, while avoiding narrow samples (e.g., samples composed solely of one specialized group) and ensuring balanced group sizes across conditions. This approach will promote diversity in the sample by covering a range of characteristics such as gender, age, cultural background, and prior experience with agents. Furthermore, we advise against using stimuli representing only one gender (e.g., females), in order to avoid potential biased responses. Finally, we recommend systematically controlling for confounding variables to the greatest extent feasible.

Finally, the lack of studies that systematically model the mechanisms linking the UVE to trust or trust-related outcomes highlights the need for a new direction in future research. Past research often reported circumstantial associations whereas an imperative direction for future research would be to treat the UVE as a predictor and trust as an outcome to investigate underlying mechanisms that explain how or why the UVE influences trust. This may involve identifying and testing possible various mediating factors, such as emotional discomfort, perceived competence, perceived warmth, anthropomorphism, cognitive load, or threat perception. These variables may explain the psychological pathways through which uncanny experiences shape trust formation. A deeper and more accurate understanding of these mechanisms could enable researchers and designers to target the roots of mistrust, develop more trustworthy agents, and ultimately enhance the quality of HAI.

In summary, the current body of research is marked by inconsistent operationalization of key constructs related to the UVE and trust or trust-related outcomes, reliance on static or hypothetical interactions, subjective measurement of trust, a lack of attention to the dynamic nature of trust, and the underlying mechanisms linking the UVE to trust outcomes. These methodological inconsistencies limit the ability to draw robust conclusions about how uncanny feelings influence trust and underscore the indispensable need for further research.

**Table 5**

*List of Research Gaps and Recommendations for Future Studies on UVE and Trust*

| Gap | Recommendation |
|---|---|
| Lack of direct and repeated interaction | • Use dynamic, real-time interaction<br>• Conduct longitudinal studies to track trust formation, maintenance, breach and repair across multiple sessions |
| Overreliance on self-report measures | • Combine subjective surveys with objective behavioral outcomes<br>• Implement real-time trust monitoring through interaction data or physiological signals |
| Inconsistencies in implementation and measurement of the UVE | • Manipulate a continuum of human-likeness across at least 3–5 levels to account for non-linearity of the UV curve.<br>• Measure both positive and negative responses using validated scales<br>• Conduct statistical tests to show a valley-shaped pattern in affinity response |
| Inadequate Statistical Planning and Control | • Justify sample sizes through power analysis<br>• Carefully select samples to minimize potential biases and ensure diversity across key characteristics (e.g., gender, age, cultural background)<br>• Systematically control for confounding variables |
| Underlying Mechanisms of UVE-Trust relationship | • Consider the UVE as a predictor and trust as an outcome to uncover underlying mechanisms though the mediation Model UVE → Mediator → Trust |

## Limitations of the Current Paper

While this review offers a structured and comprehensive synthesis of research on the UVE and trust, a few limitations should be acknowledged. First, the inclusion criteria were intentionally strict, focusing only on studies that explicitly used the term "Uncanny Valley " or "Uncanny Effect" in order to emphasize empirical studies that directly measured the UVE. This narrow focus may have excluded studies that addressed similar phenomena using alternate terms or indirect designs. Conversely, trust-related constructs and intentions were included more broadly, in line with existing literature, to capture a comprehensive view of trust in HAI. However, this resulted in variation in how trust was operationalized across studies, which limited the ability to make direct comparisons and may have reduced the clarity of findings specifically related to trust, as its nature may differ from other related constructs and intentions. Finally, the review included only English-language, peer-reviewed articles and dissertations to ensure quality and academic rigor; however, this may have introduced publication bias by excluding relevant studies published in other languages, as well as gray literature or industry reports.

Nevertheless, this review represents the first systematic effort to bridge research on the UVE and trust in HAI. Beyond summarizing key patterns and methodological challenges, it introduces a novel framework for categorizing trust measurement approaches and addressing the gaps in the prior research. Additionally, it offers practical guidelines and recommendations for improving the design, operationalization, and interpretation of UVE-trust studies. By doing so, it contributes not only to theoretical clarity but also equips future researchers with a concrete foundation to conduct more ecologically valid and methodologically sound research. Ultimately, this review helps shift the field toward more precise, meaningful investigations of how human-likeness influences trust, offering insight for both HAI scholars and system designers seeking to build trustworthy agents.

## Conclusion

The purpose of this systematic review was to investigate the relationship between the UVE and human trust in agents. Our systematic search following PRISMA protocol and using pre-defined inclusion and exclusion criteria identified 53 studies that examined both the UVE and trust or trust-related constructs and intentions. The analysis revealed that the nature of both UVE and trust within HAI is multilayered and complex. Although the UVE remains a theoretically supported effect that may occur under specific conditions, inconsistencies in its induction and measurement have led to variation in findings across studies. Moreover, the lack of attention to the dynamic nature of trust, the type of interaction, and the methods used to assess trust limits the ability to draw robust conclusions about the general effect of the UVE on trust. This review aimed to support ongoing research by highlighting the research gaps and offering recommendations that address these limitations. By building on the patterns and gaps identified here, future studies can deepen our understanding of whether and how uncanny feelings influence trust and contribute to the development of more effective, trustworthy human–agent systems.

# Appendix A

## Comprehensive Overview of All Studies

| Reference | Theoretical Research Question | Methods/designs | Manipulation and stimulus | Context | Type of the Agent | Modality: * | Type of Interaction * | Participants | UVE Measurement | Trust Measurement | Results | Limitations |
|---|---|---|---|---|---|---|---|---|---|---|---|---|
| 1. (Abubshait et al., 2017)<br><br>“Seeing human: Do individual differences modulate the Uncanny Valley?” | RQ1: Does the Uncanny Valley pattern exist in trust, likability, and mind attribution ratings?<br>RQ2: Do individual differences in empathy, social reasoning, and mentalizing abilities moderate the Uncanny Valley pattern? | Between-subjects experimental design using nested model comparisons | Six wild-type agents with varying degrees of humanness, ranging from machine-like to nearly human-like | General human-agent interaction | Static images of robots | 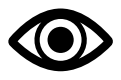 | Indirect | Sample Size: 64<br>-----------------<br>Age: M=35.5<br>-----------------<br>Gender: 33 females<br>-----------------<br>Culture: No geographical restrictions | Likability/creepiness Along with a polynomial model (5th level) to fit the ratings. | Trust game | - Evidence for UV patterns in trust, likability, and mind attribution.<br>- Trust followed the UV pattern, with a drop in trust for agents in the mid-range of humanness<br>-Individual differences did not moderate this relationship | - Static images limit ecological validity, no dynamic or real-world interaction<br>- Restricted generalizability due to the use of online participants and a single experimental design<br>- Individual differences may require alternative measures |
| 2. (Abubshait et al., 2022)<br><br>“Loneliness During COVID-19 Influences Mind and Likeability Ratings in the Uncanny Valley” | Does loneliness, induced by COVID-19 restrictions, influence the Uncanny Valley pattern in ratings of trust, likability, and mind perception of robots with varying levels of human-likeness? | Between-group comparisons (COVID-19 group vs. pre-COVID-19 group) using retrospective survey | Six wild-type agents with varying degrees of humanness, ranging from machine-like to nearly human-like | General human-agent interaction | Static images of robots | 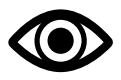 | Indirect | Sample size: 69<br>-----------------<br>Age: M=20.5<br>-----------------<br>Gender: 46 females<br>-----------------<br>Culture: American | Likability/creepiness Along with a polynomial model (5th level) to fit the ratings. | Trust game | - Evidence for the UV pattern was found for mind perception, likability, and trust in both pre-COVID-19 and COVID-19 groups. The COVID-19 group displayed less pronounced UV patterns for mind | - Static images limit ecological validity, no dynamic or real-world interaction.<br>- Lack of loneliness data for the pre-COVID-19 group.<br>- Potential confounds from differences in demographic characteristics between the |
| | | | | | | | | | | | perception and likability.<br>- Trust ratings followed UV patterns but showed no significant differences between the two groups | two groups (age and recruitment) |
| 3.<br>(Aubel et al., 2022)<br><br>“Risk Perception and Risk Behavior in Response to Service Robot Anthropomorphism in Banking” | How does service robot anthropomorphism (low, medium, high, human) influence risk perception and behavior via uncanniness, and how are these effects moderated by technology readiness and behavioral inhibition? | Online between-subjects experiment | Four anthropomorphism conditions: low (Ethon 2), medium (Pepper), high (Erica), and human (real human image labeled as a robot). Stimuli were validated in previous research. | Banking | Static images of robots | 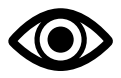 | Indirect | Sample Size: 293<br>----------------<br>Gender: females 45.4%<br>-----------------<br>Age: M=26.2<br>-----------------<br>Culture: International | uncanniness (e.g., uneasy, creeped out) | -Indirectly linked to risk perception (safe and risky investment options), assessed through psychological, risk (I would worry a lot when using services from this bank), functional risk (When using services from this bank, I would worry about how reliable they would be, Privacy risk, and time risk measures.<br>-Also, Risk Behavior (investment decision-making task) was measured<br>-Perceived risk can be interpreted as a "Lack of Trust" | - Medium anthropomorphism (Pepper) evoked the least uncanniness and lowest risk perceptions.<br>- High anthropomorphism (Erica) increased uncanniness, which mediated higher risk perceptions (e.g., privacy, psychological, time risks).<br>- Risk behavior was not significantly affected by anthropomorphism, except for moderation effects (e.g., high behavioral inhibition increased risk-taking for low anthropomorphism)<br>- Uncanniness influenced risk perception, but | - No direct interaction with<br>- Results specific to banking and may not generalize to other domains (e.g., healthcare, tourism).<br>- Small differences between medium and high anthropomorphism levels in manipulation checks |
| | | | | | | | | | | | not actual financial decision-making. | |
| 4. (Bae et al., 2024)<br><br>“Effects of meta-human characteristics on user acceptance: from the perspective of uncanny valley theory” | RQ1. In light of the uncanny valley effect, does the use of meta-humans increase user acceptance when compared to digital humans?<br>RQ2. What are the factors that influence overall user acceptance of these two technologies? | Between-subjects experimental design using online surveys | and videos of meta-humans (hyper-realistic digital humans with advanced behavior) and digital humans (conventional photorealistic avatars), with differences in appearance and behavior | Service | Videos of virtual avatars | 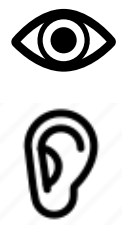 | Indirect | Sample Size: 280 participants (139 viewed digital humans; 141 viewed meta-humans)<br>----------------<br>Gender: 51% male<br>----------------<br>Age: 20s to 50s<br>-----------------<br>Culture: Korean | Humanness and familiarity | Indirectly measured trust through user satisfaction and purchase intention | - Meta-humans surpassed digital humans in user acceptance, human likeness (visual and behavioral), and purchase intention.<br>- The UVE was evident for digital humans (appearance), but meta-humans overcame it visually while still struggling behaviorally<br>- familiarity does not directly affect user acceptance, but mediates satisfaction<br>-User satisfaction significantly impacts purchase intention<br>-The UVE was confirmed between human likeness and both user familiarity and user satisfaction | - Lack of dynamic interactions; only pre-recorded videos were used.<br>- No consideration of user diversity in evaluating meta-humans across cultural or social contexts.<br>- Behavioral likeness was less advanced compared to visual likeness, limiting full acceptance<br>- The effect of the gender of the digital humans or meta-humans |
| 5. (Broadbent et al., 2013)<br><br>“Robots with Display Screens: A Robot with a More Humanlike Face Display Is Perceived to Have More Mind and a Better Personality” | How does the level of human-likeness in robot facial displays influence perceptions of the robot's mind, personality, and eeriness, as well as the participant’s blood pressure? | Within-subjects experimental design | The robot's display screen alternated between a human-like face, a silver metallic face, and a no-face (text display), with each condition presented in a randomized order, measuring blood pressure. | Healthcare setting | Healthcare robot (Peoplebot) | 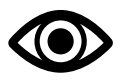 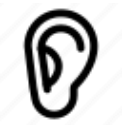 | Direct | Sample Size: 30<br>-----------------<br>Gender:<br>14 females<br>-----------------<br>Age:<br>M=22.5<br>-----------------<br>Students and staff from New Zealand | Measured eeriness using a visual analog scale, asking participants to rate "How eerie did the robot’s face look?" for each condition | Trust is assessed through personality traits (trustworthy factor derived from oppositional trait ratings such as dishonest-honest, unstable-persistent) | - The human-like face was rated least eerie and most sociable and amiable.<br>- The silver face was rated more eerie than the human-like face, confirming the UVE<br>- No-face display was rated the least human-like and sociable.<br>- eeriness negatively correlated with perceptions of trustworthiness, amiable and sociable traits, only for human-like faces<br>- trustworthiness remained relatively high across all face types | - Small sample size limits generalizability The study was restricted to a single robot and three face types; broader variations were not tested.<br>- Eeriness was measured for only two of the three conditions (human-like and silver face)<br>-It is not clear whether the participants were rating the faces on the display screens or the robots as a whole |
| 6. (Chien et al., 2024)<br><br>“Exploring the Developmental Aspects of the Uncanny Valley Effect on Children's | How does the Uncanny Valley Effect develop in children, and what are the age and gender differences in their preferences for robot appearance and interactions? | 2 (grade: 1st and 4th) × 2 (gender: boy and girl) between-subject design Observational study with quantitative surveys | 12 robot face images spanning a range of humanness levels (mechanical to near-human). | Education | Static images of robots | 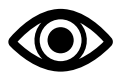 | Indirect | Sample Size: 162 children<br>-----------------<br>Age:<br>1st-grade boys/girls M=6.75<br>4th-grader boys/girls M=9.61<br>-----------------<br>Gender:<br>67 boys | Measured likability, disgust, humanness, and familiarity ratings | Measured indirectly through acceptance (intention to play). | - UVE-like pattern observed in 4th-grade girls, where likability and intention to play decreased and then increased with rising humanness.<br>- No UVE observed in | - Limited ecological validity; static images may not fully capture children’s perceptions of the dynamics and interactions of robots.<br>- No verbal comments about the robot images |
| Preferences for Robot Appearance” | | | | | | | | ----------------- Culture: Taiwanese | | | 1st-graders or 4th-grade boys.<br>- Positive correlation between likability and acceptance across all groups. | - Limited understanding of disgust for 1st-grade children.<br>All measures were asked through a single item |
| 7. (Cornelius et al., 2023)<br><br>“Significance of Visual Realism - Eeriness, Credibility, and Persuasiveness of Virtual Influencers” | RQ1.What is the optimum level of visual realism acceptable to people among realistic digital characters like Virtual Influencers (VIs)?<br>RQ2.Does the Uncanny Valley Effect exist for virtual influencers, and how does it impact perceptions of their eeriness, credibility, and persuasiveness? | Between-subjects factorial design. | Four levels of visual realism were used: cartoon, imperfect human, human-like, and human (pre-tested for humanness scores). | Social media | Virtual influencers | 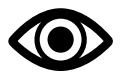 | Indirect | Sample Size: 368<br>-----------------<br>Gender: 49.2% female<br>-----------------<br>Age 18–40 years<br>-----------------<br>Culture: American | Measured eeriness along with polynomial modeling to identify UVE patterns | Measured through trustworthiness and persuasiveness (intention to follow, share, and donate) | - Imperfectly realistic VIs were rated as the eeriest and least credible, triggering the UVE<br>- No significant difference in eeriness between imperfectly realistic and realistic VIs, but imperfect VIs exhibited higher eeriness compared to less realistic VIs.<br>- Less realistic VIs were rated as most persuasive.<br>- Less realistic and highly realistic VIs were perceived as more trustworthy<br>- Even if people do not feel eerie, they may still distrust uncanny digital | - Lack of dynamic interaction: Only static images were used.<br>- Generalizability is limited to social media contexts.<br>- Eeriness perceptions might not fully explain the UVE, as attractiveness and trustworthiness played a more critical role in this study.<br>-Self-reported measures for persuasiveness |
| | | | | | | | | | | | characters, showing that UVE affects trust and persuasiveness beyond just eeriness. | |
| 8. (Dai & MacDorman, 2018)<br><br>"The doctor's digital double: How warmth, competence, and animation promote adherence intention" | How do warmth, competence, and animation realism affect adherence intention, enjoyment, and perceptions of eeriness in virtual doctor-patient consultations? | 2 × 2 × 2 between-subjects posttest-only experiment. | Three factors were manipulated:<br>- Warmth: Doctor's bedside manner (good vs. poor)<br>- Competence: Doctor's outcome (fellowship vs. malpractice)<br>- Realism: Depiction as either a human actor or a computer-animated digital double<br>-Hypothetical virtual consultations to examine patient adherence to treatment advice in clinical contexts. | Healthcare setting | Videos of virtual avatars | 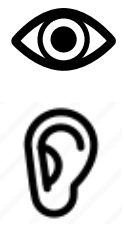 | Indirect | Sample Size: 738<br>-----------------<br>Gender: 73% female<br>-----------------<br>Age 18–82<br>-----------------<br>Culture: American | Eeriness | Indirectly measured through adherence intention and also, as part of the warmth index, which included competence and trustworthiness subscales. | - eeriness ratings for the animated doctor were not significantly higher than for the real doctor (No expected UVE pattern)<br>-Warmth (good bedside manner) and competence (fellowship) increased adherence intention and consultation enjoyment.<br>- The computer-animated doctor was rated as less eerie than expected and increased adherence intention and warmth compared to the human actor.<br>- Eeriness negatively correlated with adherence intention, but its effects were mitigated by the | - Eeriness measures may have been confounded by the doctor's bedside manner and competence<br>- The study used to role-play participants, limiting generalizability to actual patient populations.<br>- used semantic differential scales with visual analog bars, with no option for a neutral response, which may have pushed participants into favoring one response or the other<br>- a potential threat to internal validity, because the manipulation check for Eeriness, was unsuccessful |
| | | | | | | | | | | | perception of warmth<br>-A high-competence source was more persuasive than a low-competence source | |
| 9.<br>(Dai & MacDorman, 2021)<br><br>"Creepy, but Persuasive: In a Virtual Consultation, Physician Bedside Manner, rather than the Uncanny Valley, Predicts Adherence" | How does a physician's bedside manner and level of realism influence adherence intention and behavior change through warmth, competence, and eeriness? | 2 × 5 between-groups factorial design with pretest and posttest for eeriness. | Five levels of realism were manipulated for the virtual physician (Dr. Richards): cartoon, bigeye, swapped, animation, and real. Bedside manner was manipulated as either high warmth (good) or low warmth (poor). - Participants assumed the role of a patient in consultation regarding potential diabetes and provided adherence responses | Healthcare setting | Videos of virtual avatars | 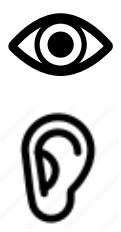 | Indirect | Sample Size: 441<br>---------------<br>Gender: 72% female<br>----------------<br>Age: 18–70<br>-----------------<br>Culture: American | Eeriness | Indirectly measured through adherence intention and warmth, and competence index. | - The UVE was observed in pretest eeriness ratings, with 3D-animated doctors rated eeriness more than real or cartoon versions.<br>The UVE impact on adherence intention and behavior change was nonsignificant.<br>-Posttest eeriness ratings were significantly affected by bedside manner<br>- Bedside manner significantly increased ratings of warmth, competence, and adherence intention, regardless of depiction.<br>- Eeriness had no significant posttest impact. | - The study was underpowered (small effect on pretest eeriness) to detect the effects of depiction on adherence intention.<br>- Behavior change was measured through self-reported scales, which may be biased.<br>-The measurement of eeriness was confounded with other factors such as warmth and competence, or patients' emotional responses. |
| 10. (Destephe et al., 2015)<br><br>“Walking in the uncanny valley: Importance of the attractiveness on the acceptance of a robot as a working partner” | How do robot movement and attractiveness intensity influence perceptions of the Uncanny Valley Effect and acceptance of robots as work partners in different cultures? | Cross-cultural experimental study(between-experiment) | -Two walking conditions: Happy, Sad<br>-Two emotional intensity levels for walking: Natural vs. Exaggerated<br>-Humanoid robot WABIAN-2R presented in videos. | General human-agent interaction in different cultural backgrounds | Video of Humanoid robot (WABIAN-2R) | 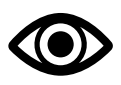 | Indirect | Sample Size: 69 (47French, 22Japanes)<br>-----------------<br>Age: 21-81<br>-----------------<br>Gender:32 females<br>-----------------<br>Culture: French vs. Japanese | Measured through perceived humanness, eeriness, and attractiveness. | Indirectly assessed through robot acceptability ratings (willingness to accept the robot in workplaces) | _UVE confirmed: Moderately humanlike movements were rated as eeriest. they plot Attractiveness and Familiarity scores against the Perceived Humanness score.<br>-A valley-like effect was observed in familiarity and attractiveness scores<br>- The Attitude toward robots was the main influence of the UVE?<br>- Cultural Differences: Japanese participants preferred naturally expressed emotions, while French participants showed no preference<br>- Acceptance: The robot’s attractiveness, rather than eeriness, was the main predictor of workplace acceptability | - Used pre-recorded videos rather than real-time human-robot interaction.<br>- Only examined one humanoid robot model |
| 11. | How do robots’ human-likeness, warmth, and | Mixed methods: One focus group (qualitative) and | Human-likeness manipulated | Service settings | Videos Social robots | 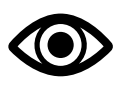 | Indirect | Sample Size: Study 1=15 | Measured perceived eeriness | Indirectly measured through | - Robots high in human-likeness | - Limited to controlled lab settings; lacks |
| (Grazzini et al., 2023)<br><br>“Dashed expectations in service experiences Effects of robots human-likeness on customers’ responses” | competence influence customers' responses in service settings, and what role do the perceived eeriness, and disconfirmed expectations play in mediating these effects? | two experimental lab studies<br>Study1: focus group<br>Study2: A one-way (human-likeness: high vs. low) experimental design with 2 mediators (i.e., perceived eeriness and disconfirmed expectations)<br>Study 3: A 2 (human-likeness: high vs. low) × 2(warmth vs. competence) experimental design with 1 mediator (i.e., disconfirmed expectations) | as high vs. low (Studies 2 and 3).<br>Warmth vs. competence manipulation in Study 3.<br>Stimuli included videos of robots interacting in service contexts (e.g., hotel check-ins) | (e.g., hotel reception) | | 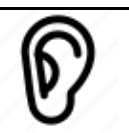 | | Study 2=140<br>Study 3=139<br>---------------<br>Age:<br>Study 1=M35<br>Study 2=M26<br>Study 3=M26<br>-----------------<br>Gender: Female<br>Study 1=60%<br>Study 2=45%<br>Study 3=45%<br>-----------------<br>Culture:<br>From Italy with prior hotel booking experience | ratings (3-item scale: Uneasy–Unnerved–Creeped out; from Gray & Wegner, 2012). | attitudes and behavioral intentions (e.g., willingness to interact or purchase) | combined with a focus on competence led to higher disconfirmed expectations and reduced customer attitudes and behavioral responses.<br>- Warmth moderated this effect, reducing negative responses for high human-like robots.<br>- Competence alone did not mitigate the negative effects of the UVE, in fact, high human-likeness robots focused on competence heightened the negative impact of disconfirmed expectations.<br>- Disconfirmed expectations were a stronger predictor mediator of negative responses | real-world interaction data.<br>- Focus on specific service context (hotels) limits generalizability<br>-No measurement of actual customer behaviors in the field |
| 12. (Gurung et al., 2023)<br><br>“The Uncanny Effect of | How do speech (human vs. machine-like) and appearance (real human vs. virtual avatars) influence | Within-subjects experimental design | - Three appearances: real human (Stelarc), high-fidelity avatar (PH V2), and low- | General human-agent interaction | Virtual avatars (PH V1 and PH V2) and a real human (Stelarc) | 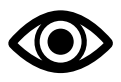<br>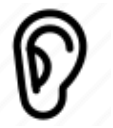 | Indirect | Sample Size: 49<br>-----------------<br>Gender:<br>28 females<br>-----------------<br>Age: | Explored through selected items from the different questionnaires by impression | Directly measured via participants’ ratings of trustworthiness for each | - Human voice and appearance (Stelarc speaking) received the highest ratings | - Small sample size limits generalizability<br>- Short exposure to stimuli (10 seconds per |
| Speech: The Impact of Appearance and Speaking on Impression Formation in Human-Robot Interactions” | impressions (trustworthiness, likability, human-likeness, etc.) in human-robot interactions? | | fidelity avatar (PH V1). - Two speech conditions: presence and absence of human voice (real Stelarc) vs. machine-like voice (PH V2) and more machine-like voice (PH V1). - 6 Videos presented consistent phrases and head movements. | | | | | 23-70 ----------------- Culture: Australian and Denmark | ratings on human-likeness, familiarity, and likability, | video condition. | across all dimensions (e.g., trustworthiness, realism, human-likeness). - Machine-like voices significantly decreased positive impressions of avatars, especially for the lower-fidelity avatar (PH V1). -Machine-like voices negatively impacted trust -The combination of human-like appearance with machine-like voices reduced the convincing quality of avatars, causing the UVE -familiarity and human-like effects of the speaking voice and visual perceptions can play a strong role in forming the UVE and the UVE of speech | video) may not fully capture long-term impressions. - Results specific to video-based interactions without direct or dynamic engagement. |
| 13. (Jung et al., 2021) | RQ1. Does the context of human | Between-subjects experimental survey design | Five humanoid robot images | Service contexts: Hotel | Static image of robots | 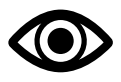 | Indirect | Sample Size: 505 ----------------- | Measured human likeness, | Trust was measured using a single | - Favorability and trust ratings | - Single-item scales for each factor may |
| "Users' Affective and Cognitive Responses to Humanoid Robots in Different Expertise Service Contexts | interactions with humanoid robots affect the uncanny valley? RQ2. How is trust in humanoid robots reflected in the affective responses presented in the uncanny valley? RQ3. Do the contexts of interactions with humanoid robots influence humans' trust in robots? | | with varying degrees of human-likeness, ranging from highly mechanical to human-like, were presented to participants randomly. | reception (low expertise) vs. tutoring (high expertise) | | | | Age: M=39.3 ----------------- Culture: Korean | favorability(uneasiness) | question: Do you believe that the robot in this picture would be trustworthy as a hotel receptionist staff (or tutor)? | followed a U-curve pattern, consistent with the UVE (Robots with intermediate levels of human likeness were rated less favorably)<br>- Trust was higher for humanoid robots in the tutoring (high expertise) context than in the hotel reception context.<br>- Favorability positively influenced trust across all robots and contexts.<br>- The depth of the UVE (i.e., differences in favorability across the stages) decreased trust<br>- Attitudes are less influenced by the appearance of the robot when the task requires a higher level of expertise | limit measurement robustness.<br>- Static images lack ecological validity compared to dynamic robot interactions.<br>- Cultural bias due to the homogeneous sample from South Korea |
| 14. (Kaate et al., 2023) "The realness of | RQ1. How does persona modality affect designers' perceptions of the persona? | Within-subjects (repeated measures) experimental design. | Three persona modalities were used: deepfake (AI-generated video), classic (static photo | Design context | Static images and videos of avatars | 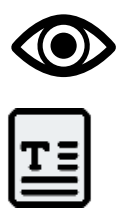 | Indirect | Sample Size: 90 ----------------- Age: M=33.1 ----------------- Gender: 33.7% female | Explicitly measured using a three-item Likert scale for eeriness | Measured indirectly via credibility and willingness to use | - Deepfake personas elicited a significantly higher UVE than classic | - Results may not generalize to other professional contexts beyond design tasks. |
| fakes: Primary evidence of the effect of deepfake personas on user perceptions in a design task “ | RQ2. How does persona modality affect design task performance?<br>RQ3. Is there an uncanny valley effect associated with deepfake personas? | | and text), and narrative (text only). Stimuli included three personas (James, Susan, Fiona), each presented in one of the three modalities. | | | | | -----------------<br>Culture: English-speaking professionals (UK, US, Ireland, Canada, Australia) | | | and narrative personas.<br>- Deepfake personas were rated lower in empathy, credibility, clarity, willingness to use, and immersion.<br>- Design task performance (time spent) was not affected by modality<br>-The UVE directly undermined credibility and willingness to use deepfake personas. | - No real interaction<br>- Limited to three personas; findings may vary with different persona designs or applications.<br>-The study used one particular deepfake generation service (Synthesia). Other services or technologies might yield different results.<br>-The persona's dwell time was impacted by the length of deepfake videos |
| 15. (Kim et al., 2019)<br><br>“Eliza in the uncanny valley: anthropomorphizing consumer robots increases their perceived warmth but decreases liking | How does anthropomorphism in consumer robots (via appearance and behavior) influence perceptions of warmth, competence, and uncanniness, and how does it affect consumer attitudes? | Four separate between-subjects experiments.<br>Study1: 3(robot human-likeness) ×2 (trait dimension)<br>Study2: 2 (anthropomorphized behavior: High vs. low) ×2 (trait dimension: warmth vs. competence)<br>Study3: 3 (anthropomorphism: more human-like robot vs. less human-like robot; plus, a human control | Anthropomorphism was manipulated through appearance (human-like vs. non-human-like robots) and behavior (e.g., emotionality, gestures). Stimuli included images and videos of real robots (e.g., Pepper, Nadine) and designer-rendered faces for comparison. | Service contexts | Static images and videos of robots | 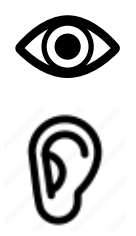 | Indirect | Sample size:<br>Study 1: 106<br>Study 2: 112<br>Study 3: 205<br>Study 4: 484<br>----------------<br>Gender: female<br>Study 1: 35%<br>Study 2: 41%<br>Study 3: 44%<br>Study 4: 41%<br>-----------------<br>Age: 21-50<br>-----------------<br>Culture: No geographical restrictions | Measured using a 3-item uncanniness scale | Measured indirectly through warmth and competence ratings and consumer attitudes toward robots in specific roles (e.g., helper, service agent) | - Anthropomorphism increased warmth but not competence perceptions.<br>- Robots with high anthropomorphism were rated higher in uncanniness, leading to lower overall liking.<br>- Uncanniness mediated the negative relationship between warmth and consumer attitude | - Results may not generalize beyond controlled lab settings.<br>- Limited to a few specific robots and scenarios, restricting broader applicability.<br>- Anthropomorphism levels were defined by researchers; participants’ personal perceptions might differ.<br>- a lack of consideration of the features |
| | | condition) ×2 (trait dimensions: Competence vs. warmth)<br>Study4:<br>2 (target: human vs. consumer robot) ×2 (dimensions: competence vs. warmth) ×3 (levels: high vs. med vs. low) | | | | | | | | | - While anthropomorphism enhanced warmth perceptions, it did not positively influence attitudes or liking. Instead, higher warmth created discomfort | of human-likeness such as body parts, gestures, and voice qualities |
| 16. (Kim, 2022)<br><br>“Whom do you want to be friends with: An extroverted or an introverted avatar? Impacts of the uncanny valley effect and conversational cues” | RQ1. How does the hyper-realism of avatars affect perceived trustworthiness, likability, and willingness to befriend?<br>RQ2. How do conversational cues (introverted vs. extroverted) influence the perceived traits (personality judgments and willingness to befriend) of avatars?<br>RQ3. To what extent are those conversational cues impacted by realism or uncanny valley effects? | 2 × 2 between-subjects online experimental design | Four avatar types combine realism (cartoonish vs. hyper-realistic) and conversational cues (introverted vs. extroverted). | Social media | Videos of virtual avatars | 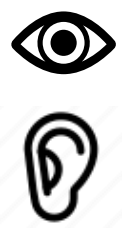 | Indirect | Sample Size: 119<br>-----------------<br>Gender: 59% female<br>-----------------<br>Age: M=37.83<br>-----------------<br>Culture: American | Eeriness and likability | Directly measured trustworthiness and willingness to be friends. | - Eeriness was higher for hyper-realistic avatars compared to cartoonish avatars.<br>- Extroverted conversational cues increased willingness to befriend hyper-realistic avatars but not cartoonish avatars.<br>- Extroverted conversational cues reduced eeriness for hyper-realistic avatars, while introverted cues increased it.<br>- Trustworthiness and likability mediated befriending decisions.<br>- Trustworthiness ratings were not significantly | - Limited stimuli (one avatar model) restricts generalizability<br>- No dynamic or real-time interactions; participants only evaluated pre-recorded videos.<br>- The sample was not diverse in terms of race and cultural background.<br>-The avatars were modeled after a female model; gender stereotypes may impact the result.<br>-the manipulation of eeriness was not effective |
| | | | | | | | | | | | affected by the avatar's realism (was more influenced by conversational cues) | |
| 17. (Kim et al., 2023)<br><br>"Is it the best for barista robots to serve like humans? A multidimensional anthropomorphism perspective" | RQ1. How do the multidimensional anthropomorphic features (visual and autonomy) of service robots interactively shape consumer intention to use?<br>RQ2. How do human identity threats and consumer resistance mediate the effects of these features on consumer intention? | between-subjects factorial experimental vignette design 3(low, mid, and high physical dimension) × 2(low, high psychological dimension | Visual anthropomorphism (low: mechanized robot, medium: humanoid robot, high: android robot) and autonomy (low: limited tasks under human supervision, high: fully autonomous) were manipulated using video and audio stimuli | Café service context | Videos of robots | 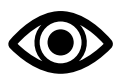<br>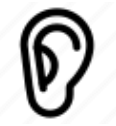 | Indirect | Sample Size: 369<br>-----------------<br>Gender: %50 female<br>-----------------<br>Age: 20–49<br>-----------------<br>Culture: from South Korea, all office workers with prior exposure to robotic café environments. | Measured Human Identity Threats (e.g., I feel that this service robot threatens my existence)<br>They inferred from negative responses (e.g., lower intention to use) to highly human-like robots with high autonomy. | Indirectly measured through consumer acceptance and intention to use service robots. | - Medium visual anthropomorphism combined with high autonomy yielded the highest intention to use.<br>- High visual anthropomorphism (android robot) resulted in the lowest intention to use, especially when paired with high autonomy (UVE)<br>- Human identity threats and consumer resistance mediated the effects.<br>- High anthropomorphism (visual or autonomy) triggered identity threats | - Limited to video simulations; no real-world robot interactions.<br>- Single geographic region may limit generalizability<br>- Short-term, transactional service scenarios do not account for long-term relational effects. |
| 18. (Kühne et al., 2020)<br><br>"The Human Takes It All: Humanlike | How does human-likeness in synthesized voices influence perceptions of eeriness, likability, and speaker trustworthiness, | Mixed methods: Quantitative study with qualitative feedback (open-ended responses). Within-subjects experimental survey design | Three types of voices: synthesized (Watson, IBM), humanoid (Sophia, Hanson Robotics), and | General human-robot interaction | Synthesized voices | 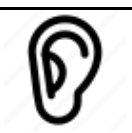 | Indirect | Sample Size: 95<br>-----------------<br>Gender: 62 females<br>-----------------<br>Age: M= 27.5<br>----------------- | Measured human likeness, likability, and eeriness of the speaker (e.g., How eerie (creepy) did you find the | Assessed the prosody's trustworthiness.<br>Qualitative: the artificial voice was not honest and fake | - No UVE<br>-Human voices were rated as least eerie and most likable and trustworthy.<br>- Synthesized voices were | - Limited stimuli: Only three voices (one per category) were tested.<br>- Auditory-only stimuli lack multimodal |
| Synthesized Voices Are Perceived as Less Eerie and More Likable. Evidence From a Subjective Rating Study” | and what role do listener characteristics (e.g., personality traits, gender) play in voice evaluation? | | human (native English speaker). Audio clips were 1–11 seconds long and included verbal phrases about social interactions. | | | | | Culture: German and non-native English speakers from the University of Potsdam | speaker?) Qualitative: The artificial voice was too perfect and odd | | less likable and more eerie than human voices but rated higher by female participants. - Human-likeness negatively correlated with eeriness, contrary to typical UVE predictions. | context, such as visual or embodied interactions. - Listener English language proficiency was not controlled through formal testing. -eeriness we measured through a single item |
| 19. (Latoschik et al., 2017) “The effect of avatar realism in immersive social virtual realities” | How do avatar realism and anthropomorphic human-like appearance (wooden mannequin vs. photogrammetry-based human-like avatars) affect uncanny valley experiences, virtual body ownership, and social interactions in immersive virtual realities? | 2 (Own Avatar Appearance: Wooden mannequin vs. photogrammetry-generated realistic avatar) × 2 (Other’s Avatar Appearance: Wooden mannequin vs. realistic avatar) within-subjects experimental design | Two avatar types were manipulated: - Wooden Mannequin: Gender-neutral, abstract humanoid. - Realistic Avatars: Photogrammetry-based male/female avatars with detailed features, matching participant gender. -Participants engaged in social interactions within a virtual environment using Oculus Rift HMD and motion capture for full-body tracking | Immersive virtual reality | Virtual avatars | 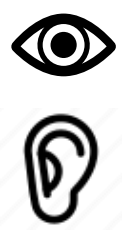 | Direct | Sample Size: 20 ----------------- Gender: 11 females ----------------- Age: M=20.25 ----------------- Culture: German students | Measured using a 3-component scale: humanness, eeriness, and attractiveness | Measured trust directly via three items | - Photogrammetry-based realistic avatars were rated more human-like but slightly eerier than wooden mannequins (Not significant: limited evidence for UVE). - Wooden mannequins scored higher in attractiveness (Not significant). - Interaction partner’s avatar realism influenced participants’ self-perception and virtual body ownership. - Realistic avatars for their own | - Small sample size limits statistical power. - No dynamic facial or gesture tracking, limiting avatar expressiveness. - Static pre-recorded behaviors (waving) may reduce the ecological validity of social interaction measures. |
| | | | | | | | | | | | appearance enhanced virtual body ownership, with participants reporting higher acceptance of the avatar as their own body compared to wooden mannequins.<br>-No significant result for trust (Avatar realism did not influence trust) | |
| 20. (Li et al., 2024)<br><br>“Social Media Users’ Affective, Attitudinal, and Behavioral Responses to Virtual Human Emotions” | Do media users’ (a) affective responses toward the social media posts, (b) attitudes toward the virtual human, and (c) behavioral intentions significantly differ among emotional cues with approach tendencies, including happiness, sadness, and lust, compared with no emotion? | 4 (emotion: happiness, sadness, lust, no emotion) × 1 between-subjects experimental design. | Four experimental conditions featuring a virtual influencer (Miquela) with emotions conveyed through Instagram posts (photo + caption) | Social media | Virtual influencer | 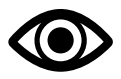 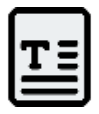 | Indirect | Sample Size: 362<br>----------------<br>Gender: 49% female<br>----------------<br>Age: Generation Z and Millennials (aged 18–40)<br>----------------<br>Culture: U.S.-based Instagram users | Measured using a 4-item eeriness | Indirectly measured through attitudes toward the virtual human (e.g., pleasantness, favorability) and behavioral intentions (e.g., I intend to follow this virtual human on Instagram) | - Lust elicited the strongest affective, attitudinal, and behavioral responses, followed by happiness, sadness, and no emotion.<br>- Eeriness moderated the relationship between affective responses and behavioral intentions<br>-Higher eeriness reduced positive outcomes (both attitudes toward the virtual human and behavioral intentions) | - Limited to one virtual influencer, findings may not generalize to other virtual humans.<br>- Emotion manipulations (e.g., lust) were sometimes misidentified, especially by male participants.<br>- No dynamic or interactive components; participants only viewed static posts<br>- It is possible that the setting of the photo in the “no emotion” condition (i.e., posing under a |
| | | | | | | | | | | | | clear blue sky; traveling) connotated a happy impression |
| 21. (Liao et al., 2023)<br><br>“Research on the acceptance of collaborative robots for the industry 5.0 era – The mediating effect of perceived competence and the moderating effect of robot use self-efficacy” | How does the anthropomorphism of collaborative robots (cobots) affect their acceptance in the workplace, and what role do perceive competence and perceived threat play in this relationship? | Between-subjects experimental vignette methodology | - Three levels of robot anthropomorphism:<br>1) Low: One-armed cobot (UR).<br>2) Medium: Two-armed cobot (Yumi).<br>3) High: Humanoid cobot (Baxter, with a "face").<br>- cobot images and videos | Industry | Static image of robots | 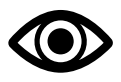 | Indirect | Sample Size: 300<br>-----------------<br>Gender: 150 Females<br>-----------------<br>Culture: College students in China | - Anthropomorphism manipulation aligns with UVE principles and was measured.<br>-Perceived threat (capturing concerns about job displacement and identity loss due to cobots e.g., Technological advances in the field of collaborative robotics are threatening what makes. us human.) | Measured indirectly through perceived competence (robot’s ability to perform tasks) and acceptance of cobots (willingness to work with them) | - The UVE confirmed that higher anthropomorphism increased perceived threat, reducing cobot acceptance<br>-Higher anthropomorphism increased perceived competence, but this also raised perceived threat, leading to lower acceptance<br>-Perceived competence increased perceived threat, which then reduced acceptance, confirming a serial mediation effect<br>-Employees with higher robot-use self-efficacy were less likely to feel threatened by competent cobots (moderation effect) | - No direct measurement of eeriness (UVE inferred through perceived threat).<br>- Cross-sectional studies may also lead to common method bias. This is because. the measurement methods used are influenced by the same factors.<br>- Focus on industrial settings (results may not generalize to other domains like healthcare or social robotics)<br>- Participants were students, not industrial workers, limiting workplace realism<br>- all Chinese<br>-No real interaction |
| 22. (Lin et al., 2021)<br><br>"Parental Acceptance of Children's Storytelling Robots: A Projection of the Uncanny Valley of AI" | RQ1. To what extent do parents accept children's storytelling robots in the home?<br>RQ2. How do parents?<br>envision these robots in the future?<br>RQ3. What aspects of robots could support or hinder their acceptance among parents? | Qualitative exploratory: Semi-structured interviews and qualitative thematic analysis.<br>(A form of speculative design that opens up discussions on the use of emerging technologies and their ethical and social implications) | Two storytelling robots with distinct designs: *Luka* (autonomous, expressive, zoomorphic with human voice) and *Trobo* (humanoid, remote-controlled with robotic voice) These robots were used as probes to facilitate discussion and envision future storytelling robots | Parenting | Robots | 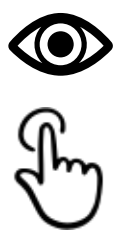 | Direct | Sample Size: 18<br>-----------------<br>Gender: mothers 77%<br>-----------------<br>Age: M=32<br>-----------------<br>Culture: American | The UVE was explored qualitatively through cognitive dissonance (parental reflections on the human-likeness and intelligence of the robots)<br>-Parental felt threatened and scared by the level of humanlike appearance linked to intelligence and perceived agency in robots (Highly autonomous, emotionally capable robots) | Not explicitly measured; inferred through the theme code "attitudes toward using robots" and subthemes "parental concerns." about preference, reliability, and safety of the robots in order to *accept* the storytelling robots.<br>(The study aimed to find the predictors of parental acceptance of these robots)<br>Qualitative: Parents indicated that robots would be more child-friendly than televisions, tablets, smartphones, and other devices because a robot's predetermined content was controllable and trustworthy. M4 explained, "I would kind of trust a pre. programmed thing made for kids, whereas | - Parents valued storytelling robots as stress-reducing aids but feared robots might replace parent-child bonding.<br>- Human-like intelligence elicited cognitive dissonance and concerns (parents struggled with the contradiction of robots appearing "too human-like" yet failing to fulfill human expectations), consistent with the UVE.<br>- Context of use, perceived agency, and intelligence were key acceptance factors.<br>- Parents showed conditional acceptance (repetitive tasks) for storytelling robots | - Qualitative study with a small sample size limits generalizability<br>-limited samples<br>- Findings based on speculative design fiction rather than real-world applications.<br>- Study focused on parents, excluding children's direct feedback |
| | | | | | | | | | | something like YouTube, they could go down a wrong path and see things that they should not be seeing." | | |
| 23. (Lin et al., 2022)<br><br>"Promotors or inhibitors? Role of task type on the effect of humanoid service robots on consumers' use intention" | RQ1. How does the uncanny valley effect of anthropomorphic robots' impact consumers' use intention in hotel settings?<br>RQ2. Does the type of task (mechanical vs. emotional-social) moderate the relationship between anthropomorphic robots and consumers' use intention?<br>RQ3. How do consumers' sense of discomfort and task attraction mediate the relationship between anthropomorphic robots and consumers' use intention, and how does task type moderate these mediating effects? | Mixed design:<br>Study 1: Single-factor between-subjects design (3 levels of anthropomorphism robot type: *high*: HARs, *moderate*: MARs, *low*: LARs)<br>Study 2: 3 (anthropomorphism: HARs, MARs, LARs) × 2 (robot type × task type). Within-subjects factorial design<br>Study 3: 3 × 2 (Anthropomorphism × Task type) between-subjects | Photographs of six service robots (two HARs, two MARs, two LARs) with standardized backgrounds. Task type manipulated as emotional-social (e.g., handling complaints) vs. mechanical (e.g., room cleaning). | Hotel service context | Static image of robots | | Indirect | Sample size:<br>Study1:138<br>Study2:108<br>Study3:618<br>-----------------<br>Age:<br>Study1:16-24<br>Study2:25-34<br>Study3:25-44<br>-----------------<br>Gender:<br>Female<br>Study1:60.5%<br>Study2:61.1%<br>Study3:57.3%<br>-----------------<br>Culture:<br>Chinese | Measured anthropomorphic appearance and discomfort which refers to eeriness and threat (e.g., If this robot served you, you would feel uneasy) | Indirectly measured through consumers' use intention and task attraction ratings | - HARs elicited significantly higher discomfort and lower task attraction, and intention of use compared to MARs or LARs.<br>- Emotional-social tasks mitigated the aversive effects of HARs on discomfort and use intention.<br>- MARs were most preferred for both task types, confirming the UVE<br>- The UVE reduced task attraction and use intention for HARs, especially for mechanical tasks | - Static images lack ecological validity compared to dynamic robot interactions.<br>- Limited to the hospitality context; findings may not generalize to other industries.<br>- Task manipulations were simplified to binary categories (emotional-social vs. mechanical)<br>-No focus on social presence that may affect the consumers' use intention |
| 24. (Lou et al., 2023)<br><br>"Authentically Fake? | RQ1. What are the primary motivations for following virtual influencers (VIs)? | Exploratory qualitative semi-structured interview design with deductive | Not manipulated; The study categorizes VIs based on their level of | Social media | Virtual influencers | | Indirect | Sample Size:<br>26<br>-----------------<br>Age:<br>20-28<br>----------------- | Explored qualitatively through the theme "Perceptions of Vis | Indirectly measured through themes "Purchase intentions," | - Most participants found VIs uncanny when human-likeness was | - Small, homogeneous sample limits generalizability (using snowball sampling) |
| How Consumers Respond to the Influence of Virtual Influencers ” | RQ2. Whether and how do followers react to the uncanniness of VIs? RQ3. Which factors could alleviate followers perceived uncanniness of VIs? RQ4. How do VIs drive marketing effectiveness (i.e., brand awareness, brand image, and purchase intention) and why? | and inductive thematic analysis | human-likeness into anime-like VIs and human-like VIs. Participants described their experiences with real-life VIs, such as Lil Miquela, Shudu, and Imma.gram, based on their social media interactions. | | | | | Gender: 65% female ----------------- Culture: Singaporeans | Uncanny & eerie,” participants’ reflections on eeriness and discomfort (e.g., “There is a bit of that uncanny valley feeling… It looks very human, but it is not, and you can clearly tell it is not.”) | “Authenticity fake,” “similarity,” and “parasocial relations “with Vis (e.g., “Absolutely not. … Like if they are attempting to influence people to purchase something, or they are endorsing something. It is a bit hard because they are not real.”) | too realistic, but also described them as authentically fake. - Curated flaws (e.g., imperfections in appearance) and self-justification (the belief that humans were behind Vis) reduced eeriness. - VIs were effective for brand awareness but not for purchase intention. -Almost all the interviewees indicated that they trusted VIs less. - Weak parasocial relations with VIs limited their ability to build trust and influence purchase intentions | - Results depend on subjective accounts of a small number of well-known VIs. - Focused on followers' perceptions; insights from creators or marketers were not included. -Not based on the real interaction |
| 25. (MacDorman, 2019) “In the Uncanny Valley, Transportation Predicts Narrative Enjoyment | How does the uncanny valley impact emotional empathy, narrative enjoyment, and persuasion in interactive narratives, and how do these effects differ | 2×2×2 between-groups posttest-only experiment (Participants assumed the role of a patient in a virtual consultation scenario with a doctor (real or computer- | - Three manipulated variables: Character (hero vs. villain), Ending (happy vs. tragic), and Depiction (real human | healthcare | Videos of virtual avatars | 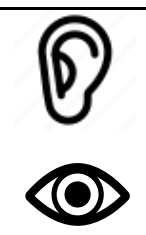 | Indirect | Sample Size: 738 ----------------- Age: 18-82 ----------------- Gender: 73% females ----------------- Culture: | The computer-animated doctor used in this study had been rated as significantly more eerie than the real human counterpart in | Evaluated indirectly via persuasion (adherence to the doctor’s advice) | - Computer-animated consultation was more persuasive and enjoyable than the real consultation due to greater transportation (the UVE did not negatively | - Did not separate the positive effects of animation quality from the UV - Short narrative duration may not capture long-term empathy or |
| More Than Empathy, but Only for the Tragic Hero” | between heroes and villains? | animated) as hero or villain) | vs. computer-animated).<br>- Scenarios included spoken dialogues and visual depictions of the doctor as a hero or villain, concluding in either a just (hero happy, villain tragic) or unjust (hero tragic, villain happy) ending. | | | | | U.S. university students | an earlier study. | | impact persuasion and may have even enhanced it)<br>-Cognitive perspective-taking (the ability to reason about the doctor's mental state) was a significant predictor of persuasion in all four computer-animated consultations The UVE negatively impacted emotional empathy for the computer-animated hero, especially in tragic endings, compared to the real hero.<br>- The UVE's negative impact on emotional empathy did not strongly influence persuasion when the computer-animated doctor was perceived as competent | enjoyment dynamics.<br>- Results focused on healthcare narratives, limiting generalizability to other genres |
| 26. (Mal et al., 2024)<br><br>“From 2D Screens to | How does the degree of immersion (low/high) affect the perceived plausibility of | 2 × 2 mixed *experimental* design. degree of immersion (low/high) as a | Two conditions of immersion (low: *2D* screens; high: *VR*) and two | Virtual reality and desktop environme nts | Videos of virtual avatars | 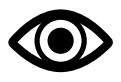 | Indirect | Sample Size: 91 (65 in 2D condition, 26 in VR)<br>-----------------<br>Age: | Explicitly measured humanness and eeriness as the UVE index | Measured using a trust subscale on social judgments of | - Realistic virtual humans were rated significantly higher in humanness | - No interaction with virtual humans; results limited to non-interactive settings. |
| VR: Exploring the Effect of Immersion on the Plausibility of Virtual Humans “ | virtual humans with varying realism and anthropomorphism? | between-subject factor and virtual human realism (abstract/realistic) as a within-subject factor | types of virtual humans (*realistic*: photogrammetry with life-like body movements vs. *abstract*: Anthropomorphic, wooden mannequin-like figures) with consistent pre-generated animations including idle and waving motions. | | | | | M=21.52<br>----------------<br>Gender:<br>53 females<br>----------------<br>Culture:<br>All undergraduate students from the University of Würzburg in Germany | | virtual humans | and plausibility than abstract figures.<br>- A possible UVE was observed for low immersion but not for high immersion.<br>- In VR, trust and sympathy ratings increased for realistic humans, but in 2D no significant difference<br>-VR mitigates the UVE, improving trust and reducing eeriness for realistic virtual humans. | - Imbalanced sample sizes between conditions (2D: 65; VR: 26).<br>- Limited range of stimuli (six realistic, four abstract).<br>- The online 2D setting lacked control over the participant environment. |
| 27. (Mathur & Reichling, 2016)<br><br>“Navigating a Social World with Robot Partners: A Quantitative Cartography of the Uncanny Valley” | RQ1. Does the Uncanny Valley Effect exist in human reactions to android robots??<br>RQ2. How does the UVE influence humans' willingness to trust robots as social partners? | Mixed design:<br>- Study 1A: Observational study between-subjects assessing the mechano-humanness of 80 robot faces.<br>- Study 1B: Cross-sectional between-subjects survey evaluating likability ratings.<br>- Study 1C: Trust-motivated behavior between-subjects assessed through a one-time investment game. | 80 wild-type robot faces (Experiments 1A–1C) and 6 digitally composed faces | General human-robot interaction | Static image of robots | 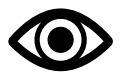 | Indirect | *Sample size:*<br>Study1A:66<br>Study1B:342<br>Study1C:334<br><br>Study 2A:52<br>Study 2 B:98<br><br>Replication Study2A:105<br>Replication Study2B:98<br>---------------<br>*Median age:*<br>30–32 years across sub-experiments A<br>32–44 years across sub-experiment B | Measured likability/creepiness along with a polynomial model<br>Also, Mechano-humanness scores are derived from explicit ratings of human vs. mechanical resemblance. | Trust-motivated behavior was assessed using a game-theory investment game. | - Likability followed a classic UVE pattern<br>- Trust mirrored the UVE pattern but was less pronounced and sensitive to facial characteristics.<br>- The UVE significantly reduced trust-motivated behavior, but this effect was more pronounced for wild-type | -Overemphasis on visual cues(face); limited ecological validity (potentially overlooking other aspects like voice, or behavioral cues, although explored emotion as a moderator)<br>- Single-shot trust game; lacks longitudinal context (May not fully reflect |
| | | - Study 2: Controlled within-subjects experiment using six digitally composed faces.<br>-Replication study 2A and 2B: Similar within-subjects design | | | | | | --------------<br>*Gender:* female<br>1A :56%<br>2A: 60%<br>1B:40%<br>2B:78%<br>1C: 42%<br>---------------<br>*Culture:* The majority of Caucasians recruited from the United States | | | robots (Experiment 1C) than for the controlled stimulus set (Experiment 2B) | trust dynamics in real-world interactions with robots)<br>- Trust results were not robustly replicated in the follow-up experiment using male-faced stimuli<br>- Results may not generalize to robots outside the mechano-humanness spectrum used in the study.<br>-the corpus of robot face images available through an Internet search may be a biased representation |
| 28. (Mulcahy et al., 2024)<br><br>"Avoiding Excessive AI Service Agent Anthropomorphism: Examining Its Role in Delivering Bad News" | RQ1. How do verbal and visual anthropomorphisms impact consumer subjective well-being and co-creation behaviors?<br>RQ2. What roles do AI anxiety and trust play in mediating these relationships, particularly in the context of bad news delivery? | Mixed-methods design with three experimental studies.<br>- Study 1: Between-subjects design testing verbal anthropomorphism (present vs. absent).<br>- Study 2: 2 × 2 factorial between-subjects design testing verbal (present vs. absent) and visual anthropomorphism (robot vs. human). | - Study 1: Text-based chatbot responses with and without verbal anthropomorphic cues (e.g., manipulated with empathy, affective language, first-person pronouns).<br>- Study 2: Chatbot responses paired with either a human-like avatar or a | Banking | AI-service agents(chatbot) | 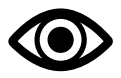<br>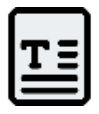 | Direct | Sample size:<br>Study 1: 242<br>Study 2: 211<br>Study 3: 194<br>---------------<br>Gender: 64.5% male<br>---------------<br>Age:<br>Study1: M = 37.42<br>Study2: M = 30.00<br>Study3: M=33.50<br>---------------<br>Culture: U.S. consumers | Explicitly tested using AI anxiety ratings as a proxy for discomfort related to excessive anthropomorphism. | Measured as trust in message credibility (e.g., honest/dishonest, credible/not credible). | - Verbal anthropomorphism reduced AI anxiety, increased trust in the message, and positively impacted participants' subjective well-being<br>- High verbal and high visual anthropomorphism triggered AI anxiety, consistent with the UVE. | - not dynamic interactions (pre-set dialogue)<br>- Limited to the banking context, findings may not generalize to other domains.<br>- The study did not measure changes in pre-existing AI anxiety levels (e.g., pre- and post-test designs). |
| | | - Study 3: Extension of Study 2, testing co-creation behavior as the outcome. | robot-like avatar.<br>- Study 3: Same manipulations as Study 2, with co-creation behavior measured<br>- Participants imagined receiving bad news (e.g., loan rejection) from an AI service agent in text-based or visual communicatio n. | | | | | | | | - Verbal anthropomorp hism paired with a robot avatar increased co-creation behavior.<br>- High anthropomorp hic cues (human avatar + verbal anthropomorp hism) reduced willingness to engage in co-creation, mediated by heightened AI anxiety<br>- Visual anthropomorp hism alone did not increase trust | |
| 29. (Nie et al., 2012)<br><br>"Can You Hold My Hand? Physical Warmth in Human-Robot Interaction" | Does physical warmth (via handholding) influence perceived friendship, trust, and human-likeness in human-robot interaction, and how does it impact emotional regulation in frightening scenarios? | Between-subjects experimental design. | Three conditions of robot hand temperature:<br>- Warm Hand: A robot hand with a thin hot pad maintained at 40°C.<br>- Cold Hand: The robot's unaltered mechanical hand.<br>- No Handholding: No physical interaction with the robot.<br>-Participants watched a five-minute | General human-robot interaction | RoboSapien RS-Media humanoid robot (36 cm tall) with a mechanical appearance and limited social behaviors. | 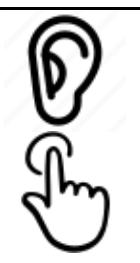 | Direct | Sample Size: 39<br>---------------<br>Age: M=26.1<br>---------------<br>Culture: Recruited from a private university in Seoul, South Korea. | Measured human - likeness. and fear of the robot | Measured using a multi-item trust scale assessing reliability, credibility, and immediacy (e.g., "How much do you trust this robot?") | - Warm hand conditions increased perceived friendship, trust, and human-likeness compared to cold and no handholding conditions.<br>- Warm touch heightened fright toward the robot, aligning with UVE (touch can exacerbate the UVE)<br>- While the warm touch increased | - Small sample size limits generalizability<br>- The robotic appearance limited realism and may have amplified uncanny effects.<br>- Limited to one robot model, lacking diversity in stimuli. |
| | | | horror movie clip (Grudge) while holding or not holding a robot hand, | | | | | | | | trust, it also triggered feelings of discomfort<br>- Fright from the robot was higher than from the movie in the warm conditions. | |
| 30. (Paetzel & Castellano, 2019)<br><br>"Let Me Get to Know You Better: Can Interactions Help to Overcome Uncanny Feelings?" | RQ1. How does the perception of a robot embodiment change when participants are gradually exposed to its capabilities?<br>RQ2. How can different textures project onto a blended embodiment create both likable and uncanny perceptions of the platform? | Mixed design: 3 × 4 (Level of Exposure × Embodiment)<br>-Between-subjects: Participants interacted with one of four robot facial textures (humanlike, machinelike, morph, sick).<br>-Within-subjects: Participants evaluated the robot after three stages of exposure (appearance, introduction, interaction)<br>-a semi-structured interview: Capture participants' experiences with the robot and how their perceptions of it changed over time. | Four facial textures for the embodiment robot:<br>- Humanlike: Based on a human photograph.<br>- Machinelike: Geometric facial features.<br>- Morph: A blend of humanlike and machinelike.<br>- Sick: Altered humanlike with pale/yellowish skin and redness around the eyes.<br>3 levels of Exposure:<br>- Stage 1: First impression (robot uncovered, no movements).<br>-Stage 2: Robot introduced | Laboratory setting: Playing a 20-question interactive game. | Blended embodiment humanoid robot (Furhat), capable of facial expressions and head movements | 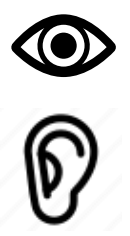 | Direct | Sample Size: 48<br>---------------<br>Age: M=26<br>---------------<br>Gender: 13 females<br>---------------<br>Culture: Computer Science Sweden students at Uppsala University | Explicitly measured using Godspeed anthropomorphism subscales (5 items) and discomfort ratings<br>(The combination of the robot's warmth and discomfort is used to judge people's uncanny feelings) | Measured through competence and social presence ratings (e.g., "The robot appears trustworthy and capable"). | - The morph texture was rated as significantly more uncanny than other textures, consistent with UVE<br>- Machinelike Texture rated less eerie than the morph but more discomforting than the humanlike face.<br>- Discomfort decreased significantly after the introductory phase and remained low after the interactive phase.<br>- Perceptions of competence and human-likeness increased over time<br>-perceived competence can help mitigate the effects of | - Small, homogeneous sample (primarily Computer Science students).<br>- Short interaction duration (approx. 20 minutes); findings may differ for long-term interactions.<br>- Limited range of discomfort ratings; stimuli elicited mild uncanny feelings at most |
| | | | itself with speech and facial movements.<br>-Stage 3: Participants played an interactive 20-question game with the robot. | | | | | | | | uncanniness over time.<br>- By Stage 3, the morph and machinelike textures were rated as more competent than the humanlike texture (Participants expected more humanlike behavior from the humanlike texture but found its performance disappointing when it did not meet those expectations)<br>-Qualitative: Many noted that the robot felt "dead" or "not real" during the first impression, but became more "alive" and relatable after the interaction stages. | |
| 31.<br>(Paetzel et al., 2020)<br><br>“The persistence of first impressions: The effect of repeated interactions on the perception | RQ1. How does a collaborative game interaction influence people’s first impression of a social robot?<br>RQ 2. How do repeated interactions influence the perception of a social robot? | 3×3 mixed experimental design. Participants engaged in a collaborative game with embodiment (humanlike, mechanical, and morph) as between-subject and repeated interactions | Manipulated 3 robot facial textures:<br>- Humanlike: Realistic human photograph projection.<br>- Mechanical: Geometric and metallic robot-like features. | Laboratory setting: Collaborative human-robot interaction using a geography game | Blended embodiment robot (Furhat) | 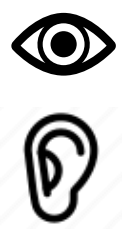 | Direct | Sample Size: 49<br>---------------<br>Age: M=24.78<br>---------------<br>Gender: 13 females<br>---------------<br>Culture: Computer Science Sweden students at | Explicitly measured perception of the robot using Godspeed anthropomorphism, likability, and perceived threat subscales. Also, measured | Measured through competence subscale from robotic social attributes | - Initial impressions improved after gameplay: anthropomorphism, likability, and comfort<br>- Repeated interactions stabilized anthropomorphism and likability but | - Small, homogeneous sample (primarily students in Sweden).<br>- Limited to three sessions; findings may not generalize to longer-term interactions.<br>- The morph texture was less |
| of a social robot” | RQ 3. How accurate are people in recalling their perception of a robot over multiple days of zero exposure?<br>RQ 4. Does the robot’s level of anthropomorphism influence people’s perception of it in repeated interactions?<br>RQ 5. Can a robot with mismatching cues recover from initial negative impressions through repeated interactions? | (sessions S1, S2, and S3) as within-subject factors. Between each session, participants had three to ten days of zero exposure | - Morph: Blend of humanlike and mechanical traits, designed to evoke mismatched perceptions<br>There are levels of interaction session:<br>S1: First impression and game interaction<br>S2: After 3–10 days of zero exposure<br>S3: After 3–10 days of zero exposure. | | | | | Uppsala University | discomfort subscale from robotic social attributes | | reduced discomfort and threat, but differences between robot embodiments remained constant, indicating a persistence of the UVE<br>- The morph robot elicited higher initial discomfort and threat, which persisted despite repeated interactions<br>- Familiarization alone did not significantly improve perceptions of uncanny robots.<br>- Perceived competence was higher for humanlike robots and stabilized after the first impression<br>- The perceived competence of the robot was determined very early on, within the first two minutes of interaction, and remained stable across all sessions (competence | distinguishable from the mechanical face, which may have confounded results. |
| | | | | | | | | | | | is judged more on its initial appearance and general interaction modalities than on its performance or interaction content)<br>-People's initial biases against robots remain unchanged over time. | |
| 32. (Pandey & Rai, 2023)<br><br>"Consumer Adoption of AI-Powered Virtual Assistants (AIVA): An Integrated Model Based on the SEM–ANN Approach" | How do various antecedents (e.g., anthropomorphism, effort expectancy) and consequences (e.g., trust, loyalty) of AI-powered virtual assistant (AIVA) adoption interact, and how does the Uncanny Valley Effect moderate these relationships? | Cross-sectional survey with moderated mediation analysis using Structural Equation Modelling (SEM) and Artificial Neural Network (ANN) | No direct manipulation: participants responded to survey items assessing AIVAs they had previously used (based on their prior interactions). Stimuli were embedded in survey items representing real-world experiences. | General consumer adoption of AIVAs (e.g., in service, entertainment, and productivity contexts). | AI-powered virtual assistants, including Google Assistant, Alexa, and Siri. | 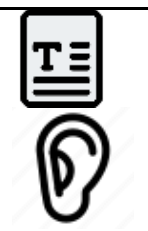 | Indirect | Sample Size: 412<br>---------------<br>Age: Millennials (18–36)<br>---------------<br>Culture: Indians | Explicitly measured Anthropomorphism and uncanniness subscale (e.g., "I often find AIVA as eerie," "AIVA seems unnatural") | Measured through perceived competence, warmth, and loyalty ratings (e.g., "I intend to keep using AIVA in the future"). | - The UVE was not significant<br>- Anthropomorphism positively influenced AIVA adoption but did not induce uncanniness<br>-AIVA adoption had a positive influence on perceived competence<br>- The UVE did not significantly moderate warmth-related adoption outcomes<br>- Higher competence and warmth significantly increased consumer adoption and | - Limited to millennials in India; findings may not generalize globally.<br>- no direct interaction with AIVAs.<br>- Survey-based responses are susceptible to self-report biases. |
| | | | | | | | | | | | loyalty toward AIVAs<br>- anthropomorphic traits (e.g., friendliness, sociability) can enhance trust and drive adoption without concerns of uncanniness because the study showed that the UVE did not significantly moderate the relationship between adoption and warmth | |
| 33. (Patel & MacDorman, 2015)<br><br>"Sending an Avatar to Do a Human's Job: Compliance with Authority Persists Despite the Uncanny Valley" | How does the uncanny valley effect influence compliance with advice from an authority figure when conveyed via computer-animated avatars compared to real human recordings? | 2 × 2 factorial between-subjects experimental design | - Depiction: digitally recorded real human vs. computer-animated avatar.<br>- Motion Quality: Smooth vs. jerky motion.<br>- ethical advice: Go condition (advocating disclosure of sensitive information) vs. no-go condition (advocating non-disclosure) | Healthcare | Videos of virtual avatars | 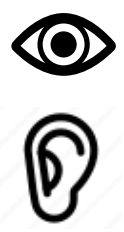 | Indirect | Sample Size: 426<br>---------------<br>Gender: 64% female<br>---------------<br>Age: M=22<br>---------------<br>Culture: undergraduate students recruited from U.S. midwestern universities | Measured explicitly using an eeriness and humanness subscale | Measured credibility through competence, trustworthiness, and goodwill; also, compliance was measured | - Animated avatars were rated as eerier and less human-like compared to real humans.<br>- Compliance with authority was not significantly affected by UVE, as both real and animated depictions were equally persuasive<br>- No significant difference in trustworthiness or competence between the human and avatar | - No direct interaction with avatars<br>-Sample population: undergraduate students from a single university<br>- Limited realism of stimuli: pre-recorded videos and the speaker's fixed identity, the framing of the narrative, and the assumption that in-study behavior maps directly to real-life behavior<br>- Ethical advice scenario may not replicate real-world |
| | | | | | | | | | | | depictions, despite higher eeriness ratings for the avatar<br>- The content of the advice influenced compliance more than the source's trustworthines s or depiction | consequences or emotional stakes<br>-The limitations in validity:<br>- self-reported<br>- The study assumes compliance is driven by the authority of the source but does not explore other moderating factors, such as emotional connection. participants' prior experience with avatars or virtual agents, task context |
| 34. (Pinney et al., 2022)<br><br>"Human-Robot Interaction : The Impact of Robotic Aesthetics on Anticipate d Human Trust" | How do robotic aesthetics (e.g., facial design and color schemes) influence anticipated trust in robots? | Mixed methods: Quantitative (cross-sectional online survey) and qualitative (open-ended responses) | Image of robot manipulated:<br>- Facial Features: Cartoon, blurred, human-like, realistic eyes.<br>- Color Schemes: Pink, orange, blue, yellow.<br>- Chest Screen Imagery: Contrasting facial and chest screen content. | General human-robot interaction | Static image of robots (Canbot U03) | 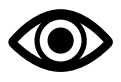 | Indirect | Sample Size: 74<br>---------------<br>Gender: 50 females<br>---------------<br>Age: 16+<br>---------------<br>Culture: Globally recruited from diverse regions | Measured through creepy feeling ratings associated with anthropomorp hic and blurred features. | Measured trust ratings (e.g., How trustworthy do you find this robot?) and qualitative feedback solely on its appearance. | - Visualization with no modifications was found to have a substantially higher percentage of trust in those with past experience<br>- Blurred facial features and mismatched facial/chest screen content elicited discomfort and reduced trust, aligning with UVE predictions.<br>- Cartoon faces were | - Static images lacked ecological validity for real-world interactions.<br>- Small sample size limits generalizability<br>- No dynamic behaviors or real-time interactions were tested<br>- Limited stimuli |
| | | | | | | | | | | | rated as more trustworthy and less threatening than human-like faces (avoiding the UVE)<br>- Human-like features (e.g., realistic eyes) were described as creepy and intimidating<br>- Warm colors (e.g., yellow, pink) were associated with positive emotions and increased trust | |
| 35. (Prakash & Rogers, 2015)<br><br>"Why Some Humanoid Faces Are Perceived More Positively Than Others: Effects of Human-Likeness and Task" | RQ1. How do people's perceptions of robot faces vary for a range of human-likeness in facial appearance?<br>RQ2. Do perceptions of robots of different levels of human-likeness vary across tasks?<br>RQ3. Do younger and older adults differ in their perceptions of robots? | Mixed methods: Quantitative and qualitative study using ratings and interviews.<br>2 (age: younger vs. older) × 3 (human-likeness: robot, mixed, human) × 4 (task) mixed design<br>age group = between subjects<br>human-likeness and task = within subjects | - Human-likeness: Robots with three levels of appearance (robotic, mixed human-robot, human-like) for each of 4 faces<br>- Tasks: Four scenarios (personal care, chores, social, and decision-making) | Service | Static image of robots | 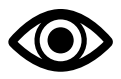 | Indirect | Sample Size: 32 youngers 32 elders<br>---------------<br>Age: Youngers: M=20.16 elders: M=70.09<br>---------------<br>Gender: 37 females<br>---------------<br>Culture: Georgian | Explored through likability and anxiety ratings across human-likeness levels.<br>Qualitative: Code of human likeness and Aesthetics/design features | Measured via trust ratings for each robot face in different task contexts (e.g., How much would you trust this robot for the task?) Also, perceived usefulness was measured as it is one of the main variables in the technology acceptance model.<br>Qualitative: code of personality (trustworthy and friendly) | -The UVE: Mixed human-robot faces were rated significantly lower in likability, trust, and perceived usefulness compared to robotic and human-like faces (older adults mentioned "alien-like")<br>- Age Differences: The UVE was most pronounced for older adults. They preferred human-like faces more than younger | - Relied on static images without dynamic interaction.<br>- Limited sample size for generalizing age-related differences.<br>- Mixed appearances may have been aesthetically inconsistent, influencing perceptions beyond human-likeness<br>-Anxiety was negatively correlated with perceived usefulness, trust, and likability, and the data were not included in |
| | | | | | | | | | | | adults, who favored robotic appearances.<br>- Human-like faces were preferred for social and decision-making tasks, while robotic faces were favored for chores and personal care tasks.<br>- Human faces were trusted the most for tasks involving emotional or social interactions<br>-Trust and perceived usefulness were positively correlated across age groups and tasks | the main qualitative analysis because the term was confusing for participants |
| 36. (Qiao & Eglin, 2011)<br><br>"Accurate Behaviour and Believability of Computer-Generated Images of Human Head" | How do variations in facial expressions, head movement, and eye movement of a computer-generated (CG) human head affect perceived eeriness, believability, and accurate behavior? | 2 x Eye Movement 2 x Facial Expression 2 x Head Movements within-subjects factorial design.<br>. | Eight videos with no speech of the CG human head without hair, textures, or voice manipulation:<br>- Facial Expressions: Moving vs. non-moving expressions.<br>- Head Movement: Moving head vs. static. | Audience perception of non-verbal behaviors in a controlled experimental setting | Videos of virtual avatars | 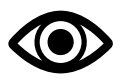 | Indirect | Sample Size: 34<br>--------------<br>Gender:<br>6 females<br>---------------<br>Culture: General audience | Explicitly measured eeriness (How eerie is this head?) | Measured believability (How believable is this head?) | - Head Movement: Reduced eeriness significantly and increased believability.<br>- Facial Expressions: Increased eeriness but also believability.<br>- Eye Movement: No significant effect on | - Small, gender-skewed sample may limit generalizability<br>- No dynamic interactions or speech included; findings were restricted to static, visual stimuli.<br>- Lack of ecological validity as stimuli were |
| | | | - Eye Movement: Naturalistic (tracking) vs. fixed. | | | | | | | | eeriness or believability<br>- The inclusion of head movement and facial expressions reduced the uncanny valley effect (lowered eeriness) while simultaneously improving believability (However, facial expressions increased eeriness slightly, suggesting a balance between emotional engagement and uncanniness) | simplified for experimental control.<br>-Possibility of different emotional responses in another context |
| 37. (Reuten et al., 2018)<br><br>"Pupillary Responses to Robotic and Human Emotions: The Uncanny Valley and Media Equation Confirmed" | How do emotional facial expressions of robots and humans elicit physiological responses (e.g., pupil dilation), and how do these responses correlate with perceived uncanniness and emotional recognition? | Within-subjects repeated-measures experimental design using pupillometry. | - Robotic faces varied in human-likeness (non-humanlike to highly human-like).<br>- Emotions expressed included happy, sad, angry, fearful, and neutral<br>- Backgrounds and colors were controlled for consistency across stimuli | Laboratory-based study | Static image of robots (e.g., iCat, Flobi, Sophia, Einstein) | 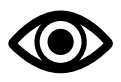 | Indirect | Sample Size: 40<br>---------------<br>Age: M=21.17<br>---------------<br>Gender: 30 females<br>---------------<br>Culture: Dutch university students | Explicitly measured eeriness (How eerie is the robot's appearance?), attractiveness, and pupillary responses as physiological markers of uncanniness | Not directly measured; Participants were asked to indicate how comfortable they would find it to interact with the robot in a variety of futuristic contexts (Housekeeping, Information, Conversation, Study help, Finance) to measure | - Robots rated as nearly human-like (e.g., Sophia, Diego-San) scored highest on uncanniness and lowest on naturalness and attractiveness, confirming the UVE<br>- Pupil dilation was weaker for uncanny robots, suggesting reduced | - Limited to static images; findings may not generalize to dynamic or interactive robot scenarios.<br>- Homogeneous sample (Dutch students) limits generalizability<br>- Emotional recognition and pupillary responses may be influenced by unmeasured individual differences (e.g., attention, |
| | | | - Pictures of humans were collected from the KDEF database | | | | | | | social acceptance. | emotional engagement.<br>- Recognition of emotions was significantly harder for robots compared to humans<br>- The UVE negatively impacts participants' willingness to engage in social interactions with uncanny robots (Robots in the UVE were rated as less comfortable for interaction across all tasks. | aesthetics, and to which observed persons are trusted can alter the amplitude of dilatory pupil responses.<br>- No measurement for uncanniness and human-likeness per robotic emotional expression |
| 38. (Rosenthal-von der Pütten et al., 2019)<br><br>"Neural Mechanisms for Accepting and Rejecting Artificial Social Partners in the Uncanny Valley" | How do neural mechanisms underlie human responses to artificial social partners, particularly in relation to the Uncanny Valley Effect? And what are their roles in decision-making? | Within-subjects design where participants evaluated all six stimulus categories under two tasks (rating and choice tasks) Repeated-measures fMRI study | - Stimuli included six categories: humans without physical impairments, humans with impairments, artificial (synthetic)humans, android robots, humanoid robots, and mechanoid robots.<br>- Stimuli varied systematically in human likeness based | Laboratory setting | Static image of robots | 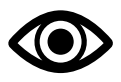 | Indirect | Sample Size: 21<br>---------------<br>Age: M=23.04<br>---------------<br>Gender: 14 females<br>---------------<br>Culture: recruited from general advertising on the Cambridge campus | *Behavioral measures:* likability, familiarity, and human likeness in both a rating task and a choice task using a nonlinear cubic polynomial to capture the depth of the UV.<br>*Neural correlates:* Neural responses in the ventromedial prefrontal | Indirectly measured through decision preferences in the choice task (e.g., preference for humans over robots for gift selection) | - Artificial humans and some android robots exhibited strong the UVE effects, with lower likability and nonlinear VMPFC activity (VMPFC activity increased with human-likeness, but showed a marked decrease for the highly human-like | - Limited to static images; findings may not generalize to interactive or dynamic settings.<br>- Small sample size and homogeneity (young adults).<br>- Stimuli categories were pre-selected and may not reflect broader artificial agent designs. |
| | | | on prior evaluations. | | | | | | cortex (VMPFC) also modeled UVE patterns. | | artificial humans)<br>- Decision preferences were influenced by likability and familiarity, with humans preferred over artificial agents.<br>- Neural activity in TPJ and FFG encoded linear and nonlinear human likeness signals, contributing to VMPFC's valuation of the UVE.<br>- Confirmed both behaviorally and neurologically, with highly humanlike artificial agents evoking the strongest UVE | |
| 39. (Schreibelmayr & Mara, 2022)<br><br>"Robot Voices in Daily Life: Vocal Human-Likeness and Application Context | RQ1. Voice realism and anthropomorphism: Are machines with more realistic voices actually more anthropomorphized than machines with less realistic voices?<br>RQ2. Human-likeness and the Uncanny Valley: | Between-subjects experimental design | Five types of female-sounding voices: real human, two synthetic voices (high human likeness), metallic voice (low human likeness), and comic voice (low human likeness). | Laboratory-based listening task focusing on social (care, companionship) and non-social (information, finance) applications. | Voice-only agents | 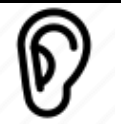 | Indirect | Sample Size: 163<br>---------------<br>Age: M=26.39<br>---------------<br>Gender: 99 females<br>---------------<br>Culture: German-speaking recruited from Johannes Kepler | Explicitly measured anthropomorphic attributions, pleasantness, and eeriness (e.g., How eerie did the voice sound?) adapted from Ho and MacDorman (2010) | Not directly measured; inferred through acceptance of application contexts. (How much would you agree with the use of the robot, you listened to in the following. | - Increasing vocal human-likeness does not elicit the UVE: higher vocal human-likeness correlated with lower eeriness and higher pleasantness.<br>- Real human voices were rated the least | - Limited to auditory-only stimuli; findings may not generalize to multimodal interactions<br>- Including only five stimulus voices<br>- No dynamic or interactive scenarios tested; findings restricted to |
| as Determinants of User Acceptance” | Is the degree of perceived human-likeness related to how eerie or pleasant users evaluate a given voice?<br>RQ3. Application context and acceptance of vocal human-likeness:<br>Does the acceptance of vocal human-likeness depend on the assumed application context, and more specifically on whether it is a social context?<br>RQ4. User personality and acceptance of vocal human-likeness:<br>Considering tolerance of ambiguity and the Big Five personality factors, do individuals differ in how. Positively they evaluate vocal human-likeness. | | | | | | | University in Austria | | areas? Care, Companionship, Information & navigation, Business & finance’ Entertainment, Customer service) | eerie and most pleasant.<br>- Social contexts (care, companionship) received lower acceptance scores compared to non-social contexts regardless of the degree of human realism in voices | passive listening<br>- All voices were female-sounding, which may bias the results<br>- Voices were German-speaking, limiting generalizability to non-German-speaking populations<br>- Participants' acceptance ratings, measured solely through self-reports, may not be independent across different contexts |
| 40. (Schreibelmayr et al., 2023)<br><br>“First Impressions of a Financial AI | How do initial perceptions of a financial AI assistant differ between people who report they (rather) trust and people who report they (rather) do not | Mixed methods: 2 x 1 Between-subjects experimental design and open-ended descriptions | Two videos differing in agency:<br>- Low Agency: Required human input for tasks<br>- High Agency: | Banking | AI-powered virtual assistant | 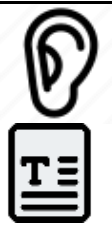 | Indirect | Sample Size: 127<br>---------------<br>Age: M=30.55<br>---------------<br>Gender: 62 females<br>---------------<br>Culture: | Explicitly measured human likeness and uncanniness (e.g., How eerie did the assistant feel?) | Measured competence, and intention-to-use, trust (e.g., I would trust the AI Banking assistant) | - Low Trust Group: Rated the AI assistant as significantly more uncanny and less human-like, competent, | - No pre-experience with financial AI systems was documented<br>- Spontaneous reactions (open-ended descriptions) were collected |
| Assistant: Differences Between High Trust and Low Trust Users” | trust the system, particularly in relation to uncanniness and trust-related attributes like competence and intention to use? | | Operated autonomously, connecting to accounts and managing tasks independently<br>- Voice was generated via text-to-speech software with consistent visualizations | | | | | Recruited from Johannes Kepler University in Austria using snowball sampling. | | | and trustworthy.<br>- High Trust Group: Found the assistant more competent, and rated high in trust and intention to use, but except uncanniness<br>- No significant effect of agency (low vs. high) on trust or intention to use contrary to the UVE<br>-Competence as a main factor in Technology Acceptance Model was proved to be a significant factor in intention to use (strong correlations between competence and trust) | at the end, potentially biasing results<br>- Findings may not generalize to other cultural or contexts |
| 41. (Seymour et al., 2021)<br>“Have We Crossed the Uncanny Valley? Understanding Affinity, Trustworth | RQ1: Is it possible with current technology to create. human-realistic avatars that cross the uncanny valley?<br>RQ2: How do user perceptions of affinity, | Mixed-methods field study: qualitative observations, qualitative interviews, and a quantitative survey<br>Social and professional interactions in VR and traditional 2D screens, with | A highly human-realistic avatar ("Digital MIKE") and caricature avatars | Immersive environments | Virtual avatars | 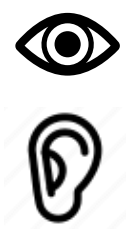 | Direct and indirect interactions: Guests interacted directly with the human-realistic avatar, while observers indirectly evaluated | Sample Size: 157: Survey 33: interviewees, 18: guest expert speakers<br>-----------------<br>Age: M=36<br>-----------------<br>Gender: 66% male<br>-----------------<br>Culture: | Assessed perceived humanness and uncanniness through qualitative indicators (e.g., aversive responses) and affinity measures. | Measured directly trustworthiness | - The human-realistic avatar was perceived as more trustworthy and preferred over caricature avatars.<br>- Participants wearing VR headsets rated affinity for the human-realistic avatar | - Fail to establish causality due to a lack of random assignment of participants to different conditions (no group control)<br>-Lack of generalizability due to specialized |
| iness, and Preference for Realistic Digital Humans in Immersive Environments” | trustworthiness and preferences differ between avatars with different levels of human realism?<br>RQ3: How does the virtual environment (i.e., immersive 3D or traditional 2D) impact user perceptions of, and preferences for, avatars with different levels of human realism? | avatars used as hosts and guests in a simulated interview at the SIGGRAPH conference (a major event for graphics design professionals) | | | | | interactions through VR headsets or 2D displays | Majority Caucasian | | | significantly higher than those viewing on 2D screens<br>-Guests interacting with the human-realistic avatar did not exhibit any aversive responses. They also made eye contact, interacted naturally, and matched verbal cues.<br>-Interviews: Observers reported positive affinity for the human-realistic avatar, preferring it over the caricature avatars, and generally felt that the avatars enabled meaningful discussion<br>-Participants did not find the caricature avatars repulsive or off-putting (indicating that caricature-style avatars are located on the left side of the UVE | designer samples (SIGGRAPH attendees).<br>- Observations relied on pre-defined avatars without varying levels of realism (only one level of realism for the digital MIKE avatar) |
| | | | | | | | | | | | -High affinity and trustworthiness ratings suggest that realistic avatars rendered in real-time can overcome the UVE | |
| 42.<br>(Sharma & Vemuri, 2022)<br><br>"Accepting Human-like Avatars in Social and Professional Roles" | How does the perception of human-like avatars affect their acceptance and proficiency ratings in social and professional roles? | Mixed methods: quantitative and qualitative (open-ended feedback) with three separate experiments:<br>- *Experiment 1*: Between-subjects comparative study of live-action vs. animated characters, using eye-tracking<br>- *Experiment 2*: Within-subjects eye-tracking study to measure pupil responses and gaze fixation to CGI human avatars. Also, one open-ended question was included. Participants split into two groups: Technically aware: Students familiar with CGI/mocap. Technically naïve: Staff members with no technical background.<br>- *Experiment 3*: Between-subjects | - Experiment 1: Clips from animated vs. live-action movies (Tintin, The Christmas Carol, and Tarzan)<br>- Experiment 2: A single CGI video clip (Dexter project)<br>- Experiment 3: Six computer-generated animations of avatars with varying degrees of human-likeness (e.g., Dexter, Siren, Andy)<br>Also, | General human and agent interaction | Static images and videos of film-based characters | 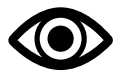 | Indirect | Sample Size:<br>Study1=22<br>Study2=21<br>Study3=65<br>-----------------<br>Gender females<br>Study1=4<br>Study2=4<br>Study3=27<br>-----------------<br>Age<br>Study1= M=20.31<br>Study2= M=20.18<br>Study3= M=21.42<br>-----------------<br>Culture Indian<br>Study1= Undergraduate students in Computer Science<br>Study2= 11 students, 10 staff | Measured through perceived humanness, eeriness, and familiarity. Also, emotional reactions are measured via. pupil size changes | Measured indirectly through comfort, and acceptance ratings in specific roles (friends, colleagues, and assistants) | - Human-like avatars (Dexter and Siren) were rated higher in comfort and acceptance but showed pupillary responses indicating unease during human-like movements<br>- CGI avatars (MoH and Andy) evoked significantly higher eeriness than more cartoonish (Tintin) or familiar avatars (Siren, Dexter)<br>- Eeriness was highest for Andy (male avatar with angry expressions), highlighting the influence of emotional displays on UVE.<br>- Acceptance was higher for | - Limited sample diversity (mostly students and staff).<br>- No live-action versions for some stimuli (e.g., Dexter, Siren).<br>- Lack of dynamic interactions or real-world context limits ecological validity<br>- testing with commercially available animated clips does not allow control over cinematographic confounds like lighting, background, and balance in facial and body movement.<br>- Gender bias reflected in role acceptance (e.g., lower ratings for female avatars in military roles). |
| | | questionnaire-based study evaluating comfort, eeriness, and proficiency of avatars in contextual roles: -Social (e.g., friends, colleagues, assistants). -Professional (e.g., military, IT professionals, artists, judges). | | | | | | | | | assistants, and lower for friends or judges, reflecting hierarchical preferences and UVE effects. - Familiarity mitigated UVE effects; avatars from familiar contexts (Tintin) evoked less discomfort despite cartoonish designs -Acceptance was highly role-dependent, with less acceptance in emotionally sensitive (Avatars with higher perceived human-likeness were accepted for roles requiring cognitive skills (e.g., IT) but resisted for roles demanding empathy or creativity (e.g., judges, artists)) | |
| 43. (Shin et al., 2019) | RQ1. Do the realism and animacy of avatars, rendered through a recent | 2 × 2 (*realism*: cartoonish vs. hyperrealistic) an avatar (*animacy*: still vs. animate) | - Avatar Realism: Cartoonish vs. hyperrealistic. | Virtual social networking services | Static image of virtual avatars | 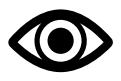 | Indirect | Sample Size: 134 ---------------- Age: M=33.7 | Explicitly measured eeriness (e.g., Reassuring–Eerie) | Measured PT (e.g., I can freely share my ideas, feelings, and | - UVE Confirmed: Hyperrealistic and animated avatars | - Only female avatars were tested; findings may not |
| "Uncanny Valley Effects on Friendship Decisions in Virtual Social Networking Services" | 3D scanning technology, influence a feeling of eeriness (i.e., UVE)?<br>RQ2. How does the uncanny valley effect impact the perceived trustworthiness (PT) of an avatar user?<br>RQ3. Does PT mediate the relationship between UVE and friendship decisions in virtual social networking services (SNS)? | factorial between-subjects online experiment | - Animacy: Still vs. animated.<br>- Stimuli included 20-second video clips showing avatars rendered through 3D scanning technology (CrazyTalk 8 Pro) | | | | | ----------------<br>Gender: 54% females<br>-----------------<br>Culture: Mostly American | | hopes with the person behind the avatar) Also, Friendship Decision was measured with a single item: Will you accept the friend request sent from the avatar user? | elicited significantly higher feelings of eeriness compared to cartoonish and still avatars.<br>- Mediation: Eeriness negatively impacted PT, which in turn reduced the likelihood of accepting friend requests.<br>- Animacy amplified eeriness for hyperrealistic avatars but not for cartoonish ones | generalize to male avatars.<br>- Participants' prior SNS experience was not measured; heavy SNS users may respond differently.<br>- The study did not examine how additional features (e.g., emotional expressions) might mitigate UVE |
| 44. (Song & Shin, 2022)<br><br>"Uncanny Valley Effects on Chatbot Trust, Purchase Intention, and Adoption Intention in the Context of E-Commerce: The Moderating Role of Avatar Familiarity" | Does the Uncanny Valley Effect influence trust, and subsequently affect purchase intention, and adoption intention in e-commerce chatbot interactions, and is this moderated by avatar familiarity? | Between-subjects design: 2 (*Human likeness*: hyperrealistic-animated chatbot vs. cartoonish-still chatbot) × 2 (*familiarity:* celebrity vs. unknown person). Participants simulated purchasing a laptop and interacting with a chatbot assistant | Four types of avatars:<br>*-Realism:* Hyperrealistic-animated and Cartoonish-still<br>*-animacy:* High familiarity (Depicted a celebrity, Tom Cruise) and Low familiarity (Depicted as a random person) | E-commerce setting: | Chatbots with anthropomorphic avatar-representations | 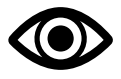 | Direct: | *Sample Size:* 185<br>-----------------<br>*Mean Age:* 24.21 years (SD = 6.22)<br>-----------------<br>*Gender:* 64% female<br>-----------------<br>*Culture:* University students in the Netherlands | Eeriness | Measured trust, purchase intention, and willingness to reuse the chatbot | Hyperrealistic-animated avatars induced more eeriness than cartoonish-still avatars.<br>- Eeriness negatively affected trust.<br>- Trust positively influenced purchase intention and reuse intention.<br>- Familiarity with the avatar (celebrity) moderated the UVE, reducing eeriness | - Celebrity Influence: The use of celebrity avatars may introduce confounding factors (e.g., attractiveness, expertise).<br>- Sample Bias: University students may not generalize to all e-commerce users.<br>- Scenario Specific: Focused on laptop purchases; results may differ for other products. |
| 45. (Stein & Ohler, 2018)<br><br>"Uncanny...But Convincing? Inconsistency Between a Virtual Agent's Facial Proportions and Vocal Realism Reduces Its Credibility and Attractiveness, but Not Its Persuasive Success" | How does inconsistency between a virtual agent's facial proportions and vocal realism influence perceived eeriness, credibility, attractiveness, and persuasive success? | 2 × 2 between-subjects experimental design (*agent's voice*: human voice vs. text-to-speech) vs. (*the design of its eye region*: humanlike proportions vs. exaggerated eyes) | - Facial Proportions: Humanlike (proportional eyes) vs. exaggerated (disproportional eyes).<br>- Voice Realism: Human voice vs. artificial text-to-speech voice.<br>- Four videos to show the conditions | Persuasive video about green genetic engineering presented by a virtual agent | Videos of virtual avatars | 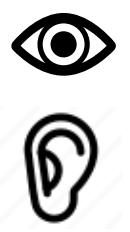 | Indirect | Sample Size: 107<br>-----------------<br>Age: M=24.8<br>-----------------<br>Gender: 70% female<br>-----------------<br>Culture: University-educated recruited from German-speaking regions. | Explicitly measured using a 7-item eeriness scale | Measured the speaker credibility index.<br>Also, Persuasive success was measured (e.g., unconvincing/ convincing) | - Consistency Matters for perceived argument quality, credibility, and attractiveness (Human-like proportional face + realistic voice: Highest credibility)<br>-Feature inconsistency (e.g., humanlike face + artificial voice) significantly reduced credibility and attractiveness but did not increase eeriness.<br>- Eeriness ratings were low across all conditions, showing no significant effects from manipulations.<br>- Persuasive success (attitude change) was unaffected by facial or vocal manipulations<br>-No significant UVE was detected, but design inconsistency (e.g., proportional | - Small, homogeneous sample (German-speaking, university-educated).<br>- Findings were limited to scripted persuasive settings; no real-world interactions or unscripted agents were tested.<br>- Low overall eeriness ratings may not generalize to agents with dynamic, real-time interactions<br>- A single topic for the persuasive message could influence the results |
| | | | | | | | | | | | face + artificial voice) negatively influenced credibility | |
| 46. (Stein et al., 2020)<br><br>“Matter over mind? How the acceptance of digital entities depends on their appearance, mental prowess, and the interaction between both” | How do digital agents’ appearance (human-like embodiment vs. text interface) and mental capabilities (simple algorithms vs. complex AI) interact to influence perceptions of eeriness and aversion? | 2 × 2 factorial between-subjects experimental design | - Portrayed agent Embodiment: Text interface (Cleverbot) vs. human-like rendering (Zoe).<br>- Portrayed agent Mental Prowess: Simple algorithms (limited responses) vs. complex AI (emotional and autonomous behavior)<br>- videos and vignettes | General human-agent interaction | Videos and vignettes of virtual avatars | 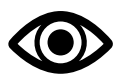<br>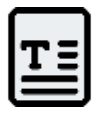 | Indirect | Sample Size: 134<br>-----------------<br>Age: M=23.17<br>-----------------<br>Gender: 84.3% females<br>-----------------<br>Culture: recruited through social media and mailing lists in Germany | Explicitly measured using eeriness scales (e.g., dull–freaky, predictable–eerie) and emotional warmth ratings. Also, perceived mental and visual human likeness were measured. | Indirectly measured through interest in future interactions (e.g., How likely are you to interact with this agent again?) | - Agents with human-like embodiment elicited significantly higher eeriness and emotional warmth compared to text-based interfaces.<br>- Complex AI was perceived as more eerie than simple algorithms, regardless of embodiment.<br>- The combination of human-like embodiment and complex AI triggered the highest eeriness ratings, consistent with the UVE.<br>-Neither the agent's mind complexity nor embodiment had a significant main effect on participants' interest in future interactions<br>- A general concern about | - Limited to passive observation; findings may not generalize to interactive settings.<br>- Small sample dominated by young female participants.<br>- Only one human-like female rendering (Zoe) was tested; results may not generalize to other designs.<br>- Eeriness and emotional warmth were positively correlated, which may indicate complex affective responses inconsistent with UVE predictions. |
| | | | | | | | | | | | autonomous technology was a strong predictor of lower interest in future interactions with digital agents<br>-Eeriness from mismatched designs (e.g., complex AI with human rendering) negatively influenced the intention to use<br>- Despite higher eeriness ratings, agents with human-like renderings were perceived as more useful | |
| 47. (Stein et al., 2022)<br><br>"Power-Posing Robots: The Influence of a Humanoid Robot's Posture and Size on its Perceived Dominanc e, Competen ce, Eeriness, | How do a robot's posture (expansive vs. constrictive) and size (child-sized vs. adult-sized) influence perceptions of dominance, competence, eeriness, and threat? | 2 × 2 between-subjects factorial design (*posture*: expansive vs. constrictive; *size*: child-sized vs. adult-sized) | - Posture Manipulation: Expansive poses (e.g., wide stance, arms on hips) vs. constrictive poses (e.g., arms crossed, head lowered).<br>- Size Manipulation: The robot is depicted as either child-sized (55 cm) or adult-sized (160 cm) using image | Laboratory -based study | Static image of Humanoid robot (NAO V5) | 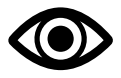 | Indirect | Sample Size: 204<br>----------------<br>Age: M=34.28<br>-----------------<br>Gender: 114 females<br>-----------------<br>Culture: recruited via social media and university mailing lists | Explicitly measured eeriness (e.g., How uncanny does this robot appear?) and threat. | Measured through competence (e.g., How capable is this robot?) | - Expansive poses significantly increased perceived dominance and competence but had no significant effect on eeriness or threat.<br>- Size manipulation did not significantly impact any measured variables.<br>- Null findings for eeriness | - Static images only; findings may not generalize to dynamic or real-life interactions.<br>- Use of a single robot design (NAO V5) limits applicability to diverse robot types.<br>- Limited to pre-designed poses; more nuanced behaviors were not tested. |
| and Threat” | | | scaling techniques. | | | | | | | | and threat suggest they do not depend on the posture or dimensions of a social robot (these perceptions may require dynamic interaction to emerge) | |
| 48. (Tastemirova et al., 2022)<br><br>“Microexpressions in Digital Humans: Perceived Affect, Sincerity, and Trustworthiness” | RQ1. How do people perceive digital humans with respect to affect, sincerity, and trustworthiness, as well as overall perception, when microexpressions are added?<br>RQ2. Is decision-making affected when microexpressions are added, and in which way?<br>RQ3. Can a designer use the almost infinite design options in a digital world to create more recognizable microexpressions and have a stronger impact on decision-making while still appearing natural? | 2 × 2 between-subjects experiment factorial repeated-measures design across two experiments | Four types of microexpressions: two *emotions* (happiness/anger) × two different *intensity levels* (normal/extreme) generated using Unreal Engine and the MetaHuman project.<br>- Experiment 1: Customer support setting with text ("I’d be happy to help") and video stimuli<br>- Experiment 2: Hiring scenario comparing two digital human candidates to check participants’ decision-making | General Social and professional settings | Videos of virtual avatars | | Indirect | Sample Size: 292<br>----------------<br>Age: M=34.2<br>-----------------<br>Gender: 112 females<br>-----------------<br>Culture: The majority of the USA | Explicitly measured overall perception in four dimensions: Humanness, eeriness, spine-tingling, and attractiveness | Measured directly perceived trustworthiness (e.g., I would trust the digital human). Also, for experiment 2, intention to hire was measured. | - Extreme microexpressions (both happiness and anger) increased eeriness compared to normal microexpressions (extreme anger elicited the highest eeriness and happiness, especially normal happiness, was rated as less eerie)<br>- Happiness was rated significantly higher in trustworthiness than anger.<br>- Intensity did not have a significant main effect on trustworthiness, but extreme happiness was perceived as the most trustworthy, and extreme anger as the | - No dynamic interactions were tested; findings were limited to passive viewing scenarios.<br>- Limited to two emotion types (happiness and anger); findings may not generalize to other emotions.<br>- Static settings may not replicate real-world interactions where microexpressions are contextual |
| | | | | | | | | | | | least trustworthy<br>- Microexpressions consistent with verbal communication (e.g., happy face + positive text) increased trust and reduced feelings of eeriness<br>-Happiness (both normal and extreme) increased participants' intention to hire compared to anger<br>-Extreme microexpressions (especially negative: anger) increased eeriness, reducing trust and overall perception of the digital human | |
| 49. (Thimm et al., 2024)<br><br>"Trust, (Dis)Comfort, and Voice Quality: Anthropomorphism in Verbal Interactions with NAO" | How does the type of robotic voice (synthetic vs. human) influence anthropomorphization, trust, and discomfort in human-robot interactions, particularly when discussing emotionally charged topics? | Mixed methods: Between-group lab experiment (two conditions: synthetic vs. human voice), quantitative study using pre- and post-experiment interviews, coded analysis, and observations for participants' interaction with the NAO robot | Two voice conditions:<br>- Synthetic voice: Pre-programmed robotic voice.<br>- Human voice: Female voice created from pre-recorded samples.<br>- NAO's physical appearance remained | General human-robot interaction | NAO humanoid robot | 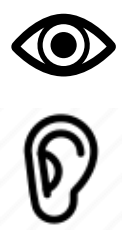 | Direct | Sample Size: 8<br>-----------------<br>Age: 18-24<br>-----------------<br>Gender: 5 females<br>-----------------<br>Culture: Students recruited from Germany | Observed through participants' post-interaction reports for eeriness and discomfort caused by voice-embodiment mismatch and human-likeness.<br>-e.g., "I think that's kind of | Explored qualitatively with two codings (trust-distrust) through post-experiment<br>-e.g., "Would you trust a robot with a robot voice more or less? Less. Because I did not have the feeling that people | - Participants interacting with a human voice NAO anthropomorphized the robot more than those interacting with a synthetic voice NAO.<br>- Human voice increased trust for participants | - Small sample size (N=8), limiting generalizability<br>- Limited demographic diversity (all students from Germany).<br>- Focused only on two voice types and one conversational topic (depression) |
| | | for 15 minutes, discussing depression | constant across conditions, including pre-programmed movements and gestures | | | | | | weird" and "I'd find that creepy" | had bothered to program it or put less thought and effort into programming it so that it was tailored to me." | who did not perceive the embodiment-voice mismatch as discomfort, but for participants who experienced the mismatch, trust decreased<br>- Synthetic voice felt neutral and expected for a robot, but less suited for emotional topics<br>- Voice-based anthropomorphism is highly contextual, and disruptions in any factors, such as embodiment or topic, can trigger cognitive conflict, leading to the UVE | |
| 50. (Troshani et al., 2021)<br><br>"Do We Trust in AI? Role of Anthropomorphism and Intelligence" | How does the humanness of AI applications affect consumers' trust in these applications in the service domain, and how do anthropomorphism and intelligence influence this trust?? | Exploratory qualitative study: three focus groups | No experimental manipulation. Participants discussed their experiences with various AI applications, such as chatbots, voice assistants, and | Service | Various AI systems, including voice assistants (e.g., Siri, Alexa), chatbots, and driverless cars, as described by participants | 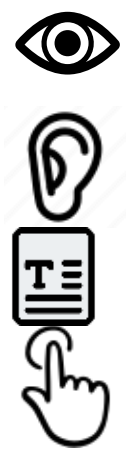 | Indirect | Sample Size: 17<br>----------------<br>Age: 18-51<br>-----------------<br>Gender: 11 females<br>-----------------<br>Culture: International | Explored through participant descriptions of discomfort, uneasiness, and cognitive conflict arising from anthropomorphism and intelligence, and analyzed thematically. There is no | Assessed through participants' perceptions of AI reliability, responsiveness, and socio-relational behaviors.<br>-e.g., I feel like the more human it sounds, the more I trust it | - Both anthropomorphism and intelligence influenced trust in AI, but their impact depended on context and perceived control (e.g., "I don't have confidence in a [AI] machine. At | - Small sample size and reliance on self-reported experiences with active users of AI applications with different experiences<br>- Potential bias due to groupthink in focus group discussions. |
| | | | driverless cars. | | | | | | UVE and trust as a core theme.<br>- Participants described AI as "creepy," and "weird," when too human-like -e.g., If it [AI] were something that was visually more human, that would be creepy | | the end of the day, I cannot control the machine.")<br>- Excessive anthropomorphism triggered cognitive dissonance, leading to the UVE<br>- Familiarity and predictability enhanced trust, while unpredictability and lack of control diminished it<br>-Higher anthropomorphism increased trust when AI behavior was predictable. | - Contexts and applications discussed were participant-driven, limiting systematic comparisons<br>-Moderator Bias: All three focus groups had the same moderator, |
| 51. (Tu et al., 2020)<br><br>"Age-Related Differences in the Uncanny Valley Effect" | Does the Uncanny Valley Effect vary across different age groups, and how does robot appearance affect acceptance of both robot service and companionship? | Within-subjects experimental design with a survey Participants randomly rated 16 out of 83 robot pictures for likability, humanness, and acceptance of service and companionship functions | 83 robot faces (Humanlike and non-humanlike designs) | Service: caregiving and companionship settings | Static image of robots | 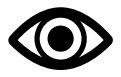 | Indirect | Sample Size: 255<br>-77 younger adults- M=21.87, 36 males<br>-87 middle-aged adults- M=50.33, 23 males<br>-91 older adults- M=65.30, 39 males<br>-----------------<br>Culture: majority Taiwanese | Measured through likability, disgust, and humanness | Assessed via participants' ratings of perceived trustworthiness and reliability of robot faces | - UVE was observed in younger and middle-aged adults, but not in older adults.<br>- Older adults showed a linear preference for humanlike robots, with no UVE<br>- Younger adults preferred non-humanlike robots, while middle-aged adults showed mixed preferences but still | - Used static images, limiting conclusions about dynamic robot interactions.<br>- Did not account for psychometric traits like anxiety or familiarity with technology.<br>- Small demographic scope (predominantly Taiwanese participants)<br>- Did not define terms like humanlike |
| | | | | | | | | | | | exhibited the UVE<br>- For trustworthiness/reliability, the results were similar to the patterns observed for likability (Younger and middle-aged adults showed the UVE for trust, but Older showed a linear positive relationship between humanness and trust)<br>- The higher the likability toward robots, the higher the acceptance of both robot companionship and service | and non-humanlike for participants, so they rate humanness based on their own judgment.<br>- Features of robot images, such as facial expressions, might affect the UVE |
| 52. (Weisman & Peña, 2021)<br><br>"Face the Uncanny: The Effects of Doppelganger Talking Head Avatars on Affect-Based Trust Toward Artificial | How do doppelganger talking head avatars, created using a participant's face, influence affect-based trust in AI-generated messages, and how is this relationship mediated by Uncanny Valley perceptions? | between-subjects design<br>3 (participant's face vs. stranger's face vs. audio-only) × 2 (pro-AI vs. anti-AI message pitch) factorial experiment | Talking head avatars (doppelgangers) generated by Reallusion CrazyTalk8 from participant photos, stranger photos, or no visual<br>Avatars delivered pro- or anti-AI persuasive pitches | Laboratory setting | Virtual avatars | 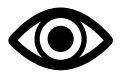<br>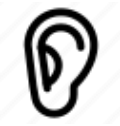 | Indirect | Sample Size: 228<br>-----------------<br>Age: M=19.6<br>-----------------<br>Gender: 86.2% females<br>-----------------<br>Culture: Undergraduate in the United States | Assessed Uncanny Valley perceptions (e.g., eerie/normal, not human-like/human-like, spine-tingling/numbing) | Measured Affect-based trust to assess security, comfort, and willingness to rely on AI-generated messages (e.g., I feel I can trust AI-generated messages to form my opinions on AIs) | - Doppelganger talking heads featuring the participant's own face increased the UVE perceptions compared to stranger faces and audio-only conditions<br>- The message type itself did not have a significant direct effect | - Imbalanced sample skewed toward female participants.<br>- Conducted in a controlled lab setting; results may not generalize to other contexts (e.g., virtual reality or robotic avatars).<br>- Limited exploration of cognitive trust or technical expertise attributions. |
| Intelligence” | | | | | | | | | | | on either trust or UVE<br>- UV perceptions negatively mediated affect-based trust, reducing trust in AI messages when self-resembling doppelganger avatars were used (the least trustworthy compared to other conditions, but no direct effect) | |
| 53. (Williams et al., 2014)<br><br>“Is Robot Telepathy Acceptable? Investigating Effects of Nonverbal Robot-Robot Communication on Human-Robot Interaction” | Does nonverbal robot-robot communication (robot telepathy) influence human perceptions of trust, cooperation, creepiness, and efficiency in human-robot interactions? | Mixed experimental design with between-subjects (*communication strategy*: verbal vs. nonverbal) and within-subjects (robot-specific perceptions) | Communication strategy:<br>- Verbal: Robots communicated audibly with each other in front of participants.<br>- Nonverbal: Robots used silent "telepathic" communication, leaving participants out of the loop | Laboratory setting: A simulated disaster relief task | Two robots: VGo (humanoid telepresence robot) and Roompi (iRobot Create with sensors). | 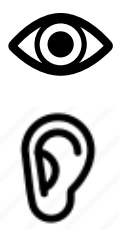 | Direct | Sample Size: 28<br>----------------<br>Gender: 14 females<br>-----------------<br>Age: 18-65<br>-----------------<br>Culture: Native English speakers from the USA | Creepiness | Assessed trustworthiness, cooperativity, and desire to interact with the robot again | - No significant differences in creepiness, cooperation, or efficiency, but marginal effects were observed for trustworthiness between verbal and nonverbal communication strategies<br>- Roompi was rated as slightly creepier than VGo (VGo was rated as marginally more trustworthy)<br>- Nonverbal communication did not increase | - Small sample size limits generalizability<br>- Wizard-of-Oz setup may not fully replicate autonomous robot behaviors.<br>- Results specific to the disaster relief context and robot design used in the study. |
| | | | | | | | | | | | perceptions of creepiness<br>- Robots may be able to communicate nonverbally without decreasing the trust of their human partners<br>- VGo was rated as more helpful, capable, and humanlike than Roompi. | |

***Note:*** The modality column uses icons to represent different study stimuli, while the type of interaction column explains the interaction between participants and the agent. Below are the details for these categories:

***Modality:**

- Visual: Covers studies relying on images, animations, or visual interactions.
- Auditory: Includes studies with spoken interactions, sounds, or auditory cues.
- Tactile/Physical: Encompasses studies with physical interactions, such as touch or haptic feedback.
- Textual: Captures studies using interaction based on text

***Type of Interaction:**

- Direct interaction: Participants interact directly with the agent, whether physically, virtually, or through conversation.

  *Example:* Talking to a robot, using a chatbot, or engaging with a virtual assistant in real-time

- Indirect encounters: Participants evaluate the agent without direct communication.

  *Example*: Observing a video of a robot, rating images of avatars, or reading a transcript of a chatbot's interaction.